\documentclass[twocolumn]{aastex61}
\usepackage[utf8]{inputenc}
\usepackage{graphicx}	% Including figure files
\usepackage{amsmath}	% Advanced maths commands
\usepackage{amssymb}	% Extra maths symbols
\usepackage{natbib}
\usepackage{tensor}
\usepackage{color}
\usepackage{tabularx}
\usepackage{bm} % fold math fonts

\usepackage{listingsutf8}

\newcommand{\arcmancer}{\textsc{Arcmancer}}
\newcommand{\am}{\arcmancer{}}
\newcommand{\odeint}{\textsc{Odeint}}
\newcommand{\grtrans}{\textsc{grtrans}}
\newcommand{\gyoto}{\textsc{GYOTO}}
\newcommand{\kertap}{\textsc{KERTAP}}
\newcommand{\GRay}{\textsc{GRay}}
\newcommand{\ASTRORAY}{\textsc{ASTRORAY}}

\newcommand{\amurl}{https://bitbucket.org/popiha/arcmancer}

\newcommand{\cpp}{\mbox{C\texttt{++}}}

\newcommand{\fromto}{\rightarrow} % kuvaus joukosta joukkoon
\newcommand{\chr}{\Gamma}% christoffel
\newcommand{\derfrac}[2]{\frac{\ud #1}{\ud #2}}
\newcommand{\parfrac}[2]{\frac{\partial #1}{\partial #2}}
\newcommand{\ud}{\mathrm{d}} % differentiaalin kursivoimaton d
\newcommand{\fR}{\mathbb{R}}
\newcommand{\bigO}{\mathcal{O}} % kertaluvun kauno-O
\newcommand{\atp}[2]{\left.#1\right|_{#2}} % something at point p
\newcommand{\defi}{:=}

\newcommand{\iprod}[2]{\left<#1,\,#2\right>}
\newcommand{\norm}[1]{\left\|#1\right\|}
\newcommand{\abs}[1]{\left|#1\right|}

\newcommand{\ix}[1]{\indices{#1}}

\DeclareMathOperator{\cond}{cond}

\newcommand{\sgra}{\mbox{Sgr A$^*$}} % sagittarius a*

\newcommand{\kronecker}{\delta}

\newcommand{\rsfac}{\mathcal{G}}

\newcommand{\polframe}{\mathcal{P}}

\newcommand{\vct}[1]{\bm{\mathit{#1}}}

\newcommand{\mat}[1]{\bm{\mathrm{#1}}}

\newcommand{\mulmat}{\mat{M}} % muller matrix
\newcommand{\svec}{\vct{I}} % stokes vector
\newcommand{\evec}{\vct{J}} % emission vector
\newcommand{\cmulmat}{\bm{\mathcal{M}}} % muller matrix
\newcommand{\csvec}{\bm{\mathcal{I}}} % stokes vector
\newcommand{\cevec}{\bm{\mathcal{J}}} % emission vector

\newcommand{\obs}{\mathcal{O}}

\newcommand{\Msun}{M_\odot}
\newcommand{\Medd}{\dot{M}_\text{Edd}}

\newcommand{\codeil}[1]{\lstinline!#1!}

\newcommand{\barr}{\bar{r}}

\turnoffeditone % remove referee changes indicators

\received{N.A.}
\revised{N.A.}
\accepted{N.A.}
\submitjournal{ApJ}

\begin{document}

\title{
General purpose ray-tracing and polarized radiative transfer in General Relativity
}

\newcommand{\affHelsinki}{
Department of Physics, University of Helsinki \\
Gustaf Hällströmin katu 2a\\
P.O. Box 64, FI-00014 University of Helsinki, Finland}

\newcommand{\affNordita}{Nordita, KTH Royal Institute of Technology and Stockholm University \\
Roslagstullsbacken 23\\
SE-10691 Stockholm, Sweden}

% TODO: Add Orcids
\correspondingauthor{Pauli Pihajoki}
\email{pauli.pihajoki@iki.fi}

\author[0000-0003-1758-1908]{Pauli Pihajoki}
\affiliation{\affHelsinki}

\author{Matias Mannerkoski}
\affiliation{\affHelsinki}

\author[0000-0002-3226-4575]{Joonas Nättilä}
\affiliation{\affNordita}

\author[0000-0001-8741-8263]{Peter H.~Johansson}
\affiliation{\affHelsinki}

%% AASTeX 6.1 has the new \collaboration and \nocollaboration commands to
%% provide the collaboration status of a group of authors. These commands 
%% can be used either before or after the list of corresponding authors. The
%% argument for \collaboration is the collaboration identifier. Authors are
%% encouraged to surround collaboration identifiers with ()s. The 
%% \nocollaboration command takes no argument and exists to indicate that
%% the nearby authors are not part of surrounding collaborations.

%% Mark off the abstract in the ``abstract'' environment. 
\begin{abstract}
    % XXX: Abstraktin koko max 250 sanaa. Tällä hetkellä 249
    % sanaa.
    Ray-tracing is a central tool for constructing mock observations of
    compact object emission and for comparing physical emission models
    with observations. We present \am{}, a publicly available general
    ray-tracing and tensor algebra library, written in \cpp{} and providing
    a Python interface.
    \am{} supports Riemannian and semi-Riemannian spaces of any
    dimension and metric, and has novel features such as support for multiple
    simultaneous coordinate charts, embedded geometric shapes, local coordinate
    systems and automatic parallel propagation. The \am{} interface is
    extensively documented and user-friendly.
    While these capabilities make the library well suited for a large variety
    of problems in numerical geometry, the main focus of this paper is in
    general relativistic polarized radiative transfer.
    The accuracy of the code is demonstrated in
    several code tests and in a comparison with \grtrans{}, an existing ray-tracing
    code. We then use the library in several scenarios as a way to showcase the wide applicability
    of the code. We study a thin variable-geometry accretion disk model,
    and find that polarization carries information of the inner disk opening
    angle. Next, we study rotating neutron stars and determine that to obtain
    polarized light curves at better than $\sim1\%$ level of accuracy, the rotation needs
    to be taken into account both in the space-time metric as well as in the
    shape of the star.
    Finally, we investigate the observational signatures of an accreting black
    hole lensed by an orbiting black hole. We find that these systems
    exhibit a characteristic asymmetric twin-peak profile both in flux and
    polarization properties.
\end{abstract}

%% Keywords should appear after the \end{abstract} command. 
%% See the online documentation for the full list of available subject
%% keywords and the rules for their use.
\keywords{
    methods: numerical
    --- gravitation
    --- gravitational lensing: strong
    --- radiative transfer
    --- polarization
    --- accretion, accretion disks
}

%% From the front matter, we move on to the body of the paper.
%% Sections are demarcated by \section and \subsection, respectively.
%% Observe the use of the LaTeX \label
%% command after the \subsection to give a symbolic KEY to the
%% subsection for cross-referencing in a \ref command.
%% You can use LaTeX's \ref and \label commands to keep track of
%% cross-references to sections, equations, tables, and figures.
%% That way, if you change the order of any elements, LaTeX will
%% automatically renumber them.

%% We recommend that authors also use the natbib \citep
%% and \citet commands to identify citations.  The citations are
%% tied to the reference list via symbolic KEYs. The KEY corresponds
%% to the KEY in the \bibitem in the reference list below. 

\section{Introduction} \label{sc:intro}

Fully covariant radiative transfer in General Relativity (GR) presents
distinct complications. Due to gravity, the path of a wave front of
radiation is curved even in vacuum. This leads to gravitational lensing,
which causes measurable effects all the way from the scales of the
Cosmic Microwave Background \citep{weinberg2013} and galaxy clusters
\citep{treu2010} to Supermassive Black Holes (SMBHs) in
centers of galaxies \citep{luminet1979}, down to single neutron stars
\citep{pechenick1983}.
Similarly, the rotation of the space-time
itself, such as around rotating Kerr black holes,
can cause an observable rotation of the direction
of polarization of light. This phenomenon is known as (gravitational) Faraday
rotation \citep{stark1977,connors1977,ishihara1988}.
Finally, the observed intensity is
also dependent on the relative position and velocity of the observer with
respect to the elements of the emitting, absorbing and scattering
medium -- typically an astrophysical plasma --
through which the light has propagated \citep[e.g.][]{gammie2012}.
This dependence is responsible for such effects as the Doppler
(de-)boosting, via velocities of the emitter and observer, and
gravitational and cosmological redshifts, via relative positions in
the space-time, respectively.

A full (classical) solution of the polarized radiative transfer problem
in GR requires solving the Einstein field equations, the
magnetohydrodynamic equations of motion of the radiating and
interacting matter, and the curved-space Maxwell equations simultaneously.
This is a formidable undertaking, also in terms of
computational resources, and significant progress has been made only
relatively lately \citep[see][and the references therein]{kelly2017}.
The problem becomes less taxing by assuming that the
radiation field makes a negligible contribution to both the space-time
curvature and the motion of the interacting medium.
In this case, the underlying space-time structure and the state of the
interacting medium can either be specified by analytic means, or by
a separate numerical computation. However, even in this case, the full
curved-space Maxwell equations need to be solved in the entire
computational domain, which is still a computationally demanding
task.

%Using the solution of the Maxwell equations, a mock observation
%can be constructed for any observer in the computational domain.
%However, the problem can be further simplified by noting that typically we
%are interested in constructing mock observations only for certain
%special observers, typically situated far from the emitting
%region and with a specific orientation with respect to it.
\edit1{The situation is considerably simplified by the fact that
the exact time-dependent behaviour of the
electromagnetic $\vct{E}$ and $\vct{B}$ fields is not usually required,
and knowledge of the radiative energy flow, i.e.\ specific intensity,
is enough. In this case, it suffices to solve the usual radiative
transfer equation, while taking into account General Relativity.
There are numerous approaches with different limitations to solving the
radiative transfer equation, such as Monte Carlo (MC) methods or the
method of characteristics \citep[see e.g.][and the references
therein]{baron2004}.
One of the perhaps conceptually simplest approaches is
to use \emph{ray-tracing}.
}
In \edit1{ray-tracing}
a mock observation can be constructed by connecting the observer to the
emitting region through null geodesics (when plasma effects are
unimportant, otherwise see e.g. \edit1{\citealt{gedalin2001}} and \citealt{broderick2003}),
through either analytic or numerical means.
The bending of these geodesics captures the lensing effects of the
gravitational field.
The relativistic polarized radiative transfer equation can
then be solved along these geodesics to capture the remaining
relativistic effects. This process, called ray-tracing,
is computationally efficient and naively parallelizable, enabling high
resolution mock observations to be computed in seconds or minutes on a
standard desktop computer. \edit1{However, straightforward ray-tracing methods are
limited to problems where scattering is not dominant. This is due to the
fact that strong scattering couples all directions and spatial locations
of the solution, whereas simple ray-tracing only samples the rays reaching
the observer. In addition, if the radiative eigenmodes propagate
differently due to strong plasma effects, then all eigenmodes must be
propagated separately and similarly radiative transfer must be computed multiple
times, increasing the workload considerably.}

\edit1{Despite these limitations,} using ray-tracing to compute mock observations of highly relativistic
objects has a relatively long history.
Already in \citet{cunningham1972}, the light curve of a star orbiting around
a black hole was computed, followed by studies of the effects of gravity
on the observed accretion disk spectra
\citep{cunningham1975,cunningham1976}.
Polarization effects of relativistic motion and strong gravity in the
Kerr solution were studied using ray-tracing in \citet{stark1977},
\citet{connors1977} and \citet{connors1980}.
The first resolved mock observation of an accretion flow around a black
hole was computed via ray-tracing remarkably early as well, in \citet{luminet1979}.
Following these pioneering studies,
the ray-tracing approach was quickly adopted to investigations of a great variety of
relativistic phenomena, including but not limited to:
hot spots and accretion columns on rotating neutron stars
\citep{pechenick1983, riffert1988},
the general mock observation problem in the Kerr space-time
\citep{viergutz1993},
details of the resolved black hole accretion disk structure
\citep{fukue1988,bromley2001}, %,watarai2005,bambi2012},
accretion disk hot spots \citep{karas1992},
accretion disk microlensing \citep{rauch1991,jaroszynski1992},
accretion disk line profiles \citep{chen1989,ebisawa1991}, %,fanton1997,beckwith2004},
optical caustics \citep{rauch1994} %,bozza2008},
and the shadow cast by the black hole event horizon
\citep{falcke2000}.%,broderick2003}.

In particular, the topics of the black hole shadow and accretion flow as well as
the observable polarization
properties of neutron stars are currently especially relevant.
The interest in black hole shadows and accretion flows is warranted by 
the recent progress in programs for interferometric observations
at the event horizon scales of \sgra{}, the Milky Way supermassive black hole, and the SMBH
in M87, the dominant galaxy of the Virgo cluster.
The event horizon is approached both in the
sub-mm wavelengths, via the Event Horizon Telescope (EHT) VLBI program
\citep{doeleman2009}, and in optical wavelengths via the VLTI GRAVITY
instrument \citep{eisenhauer2008}. The surging interest is evident also in
the number of recent studies focusing on the
black hole shadow and accretion flow modeling using ray-tracing, especially in the
context of \sgra{}
\citep[e.g.][]{%
dexter2016,garcia2016,broderick2016,atamurotov2016,chael2016,vincent2016,gold2017,porth2017,moscibrodzka2018}.

Likewise, accurate modeling of the
observable properties of neutron stars is timely due to the
current and near-future increase in X-ray sensitive space
missions such as NICER \citep{gendreau2012} and eXTP \citep{zhang2016},
of which the latter is also sensitive to
polarization. In anticipation, a number of recent papers have applied the ray-tracing
approach to model observations of neutron stars
\citep[e.g.][]{baubock2015a,baubock2015b,miller2015,ludlam2016,gonzalez2016,defalco2016,nattila2017,vincent2017}.

It is evident even from the short review above that ray-tracing is an
important numerical tool, especially for general relativistic radiative
transfer in a variety of astrophysical situations. However,
the numerical means to compute curves has an even wider applicability
in the sense that in addition to the path of light, curves also
represent the timelines of massive particles and observers in a space-time.
Furthermore, it is often convenient to have various tensorial quantities such as
local Lorentz frames parallel transported (or more generally, Fermi--Walker
transported) along curves. It is also necessary to perform various algebraic
computations involving tensor quantities, often mixing different coordinate
systems.

To help facilitate numerical studies requiring curve and tensor
manipulations in any (Semi-)Riemannian context, which naturally includes GR,
we have implemented \am{} \edit1{\citep[][]{code_zenodo}}%
\footnote{%
    \edit1{Codebase:} \url{\amurl}%
}, a publicly available
general ray-tracing and tensor algebra library. 
From an astrophysical point of view, \am{} is
useful for such varied tasks as radiative transfer and mock
observations, computing the paths of massive charged particles in curved space-times
or calculating the orbits of extreme mass-ratio inspirals (EMRIs).
However, the \am{} library offers capabilities beyond purely physically
motivated applications. It can compute all kinds of curves, 
both geodesic and externally
forced, on Riemannian and semi-Riemannian
manifolds of any dimension and metric, using multiple simultaneously
defined coordinate charts to circumnavigate coordinate singularities and
to facilitate easy input and output of data in any preferred coordinate
system. \am{} can also be used to define tensors of any rank, and to perform
tensor algebra, as well as for example automatically parallel propagate
tensorial quantities along curves. This last feature is particularly
useful for problems of radiative transfer, on which we will mainly focus
in this paper.

In this paper, we present an overview of the \am{} library and its
implementation.
We show the results of various code tests to establish
the accuracy of the code, and present several astrophysical
applications using the \am{} library.
In this paper, the main focus of the tests and applications is in general relativistic
polarized radiative transfer using ray-tracing exclusively.
\edit1{The generality of the \am{} library makes it
straightforward to use for more general purpose radiative transfer
methods such as Monte Carlo (MC) radiative transfer or hybrid
MC--ray-tracing schemes, and the application of these will be
demonstrated in future works.}
% paperin rakenne
The paper is organized as follows. In Section~\ref{sc:code_overview}
we present an overview of the \am{} library and its capabilities.
In Section~\ref{sc:implementation} we
discuss how the various mathematical objects and functionalities
provided by the \am{} library are implemented.
For convenience, these differential geometric concepts
are briefly reviewed in Appendix~\ref{app:diffgeo}, to
which Section~\ref{sc:implementation} cross-references to.
Section~\ref{sc:radtrans} describes the implementation details of the radiative
transfer scheme implemented in \am{}.
In Section~\ref{sc:code_tests}, we present a series of numerical tests,
measuring the accuracy of the numerics implemented in \am{}. These
include a test of the radiative transfer features, where the
results obtained with \am{} are compared to another recent general relativistic
code \grtrans{} \citep{dexter2016}.
In Section~\ref{sc:applications}, the \am{} code is applied to
various astrophysical phenomena in order to showcase the versatility of
the code.
Finally, in Section~\ref{sc:conclusions} we give concluding remarks, and
discuss some future prospects concerning the \am{} library and the
ray-tracing approach in astrophysics.
The paper comes with several Appendixes. Appendix~\ref{app:diffgeo}
presents a highly condensed review of the various differential geometric
concepts used in the code. Appendix~\ref{app:radtrans} presents the
general relativistic polarized radiative transfer equation used in
\am{}, and how it relates to the usual flat-space equation.
Appendix~\ref{app:builtins} presents the built-in manifolds and coordinate
systems available in \am{}.

The reader interested mainly in a broad overview of the code and its
astrophysical applications is urged to browse Sections
\ref{sc:code_overview}, \ref{sc:grtrans_comp} and \ref{sc:applications}.
Those interested in technical details may want to read through Sections
\ref{sc:implementation}, \ref{sc:radtrans} and \ref{sc:code_tests}, and the
Appendixes as well.

Throughout the paper, we use a system of units where $G=c=1$, unless
explicitly otherwise specified. For Lorentzian space-times, we use a
metric signature $(+---)$ in the paper, although \am{} supports other
signatures as well. The abstract index notation (see
Appendix~\ref{app:tensors}) is assumed throughout.

\section{Overview of the code}\label{sc:code_overview}

The \am{} library consists of a core library, written in modern
\cpp{}, a Python interface and a suite of example \cpp{}-programs and Python
scripts. The core library code, Python interface and example programs
are all thoroughly documented.
The code, the examples and instructions for installation and getting started are
all freely available at the code repository, \url{\amurl}.\footnote{\edit1{
The version of the code used to produce the results in this paper is
version \mbox{0.2.0}, which corresponds to the
commit identifier \texttt{38b0879909990746f28e09fb1f94167063608be9} at
the master branch of the repository.}}

The underlying idea behind the \am{} library is to provide all the
mathematical tools needed to perform a large variety of relativistic
computations that require numerical tensor algebra and curve
propagation. In addition, the library and the Python interface are designed with
easy extensibility in mind.
These design decisions make it possible to use \am{} for a wide variety of
astrophysical problems, including for example particle dynamics and
radiative transfer, as well as for problems in applied mathematics.

These design goals give \am{} some distinct advantages compared to existing
`pure' ray-tracing codes such as 
\grtrans{} \citep{dexter2016},
\gyoto{} \citep{vincent2011},
\kertap{} \citep{chen2015},
\GRay{} \citep{chan2013,chan2017}
or
\ASTRORAY{} \citep{shcherbakov2013}.
Namely, \am{} can work in any dimension and with metric spaces that are
either Lorentzian, as in GR, or purely Riemannian. For Lorentzian
geometry, all types of geodesics -- null, spacelike and timelike -- are
supported, as well as general curves of indeterminate classification.
\am{} can also work
with spaces for which the geometry, through the metric, is available
only numerically, such as from a numerical relativity simulation.
In addition, \am{} supports any number of
simultaneous coordinate systems with automatic conversion of all
quantities between coordinate systems.
The use of multiple coordinate systems
makes it possible to input and output data in whatever coordinates are
most convenient for the given problem. 
Furthermore, simultaneous use of multiple coordinates
makes it possible for \am{} to avoid coordinate singularities, and to
automatically choose the numerically most optimal coordinate system for
propagating a curve (see Section~\ref{sc:chartsel}).

\am{} provides full support for tensorial quantities
of any contra- or covariant rank (see Section~\ref{sc:tensors}).
This support is built on top of the
Eigen Linear Algebra Library and 
includes all the usual tensor operations such as sums, products,
contractions of indices and raising and lowering of indices with the
metric. All these operations are checked at compile-time so that
mathematically malformed operations, such as mixing points and vectors
or contracting two similar indices are automatically detected. In
addition, \am{} can automatically parallel transport all tensor
quantities along curves, so that e.g.\ smooth local coordinates can be
constructed for an observer undergoing arbitrary geodesic motion. This
functionality also supports Fermi--Walker transport for accelerating observers,
and fully general transport for e.g.\ accelerating and rotating observers.

\am{} also provides support for including user-defined embedded
geometry (see Section~\ref{sc:surfaces}).
This feature can be used, for example, to model surfaces of
optically thick or solid astrophysical objects, such as planets,
photospheres of stars or neutron stars or optically thick accretion
disks. The surfaces are easy to define through level sets, and can be
given tangential vector fields, which represent movement
along the surface, such as in the case of a rotating surface of a neutron star
or an optically thick accretion disk.

Finally, while \am{} comes with a suite of built-in space-times, coordinates, geometries,
and radiation models, the library is designed to be easily
extensible by the user. Several examples showcasing this easy
extensibility are bundled together with the \am{} library.
These examples include such programs as
simple black hole and neutron star imagers, as well as a full
postprocessor for two-dimensional data produced by the GR
magnetohydrodynamics (GRMHD) code HARM
\citep{gammie2003,noble2006}.

In the following, we will discuss in more detail how the \cpp{} library
implements the mathematical concepts required for the wide variety of
applications described above.

\section{Implementation of differential geometry and ray-tracing}\label{sc:implementation}

The main aim of the \am{} implementation is to provide the user with
\cpp{}/Python objects that match the mathematical objects of differential
geometry (see Appendix~\ref{app:diffgeo}) as closely as possible.
This approach makes converting mathematical formulae to code
straightforward. It also has the additional benefit of eliminating
errors stemming from code that expresses mathematically invalid
operations. These include, for example, assigning to the components of a
point from the components of a vector or a one-form, since all can be
expressed as a tuple of $n$ numbers, or assigning to the components of a
vector from the components of a vector defined at a different point, in
a different chart, or even defined on a different manifold.  Likewise,
for tensorial quantities, an error such as contracting two similar
indices is easily made if working in terms of pure components.

The implementation in \am{} guarantees that all programmed operations
correspond to mathematically valid
statements. This feature eliminates a large set of logical errors of the kind
described above -- a major benefit, since currently there are no
codebase analysis tools able to identify errors of this
kind.

\begin{figure*}[hbtp]
    \includegraphics[width=\textwidth]{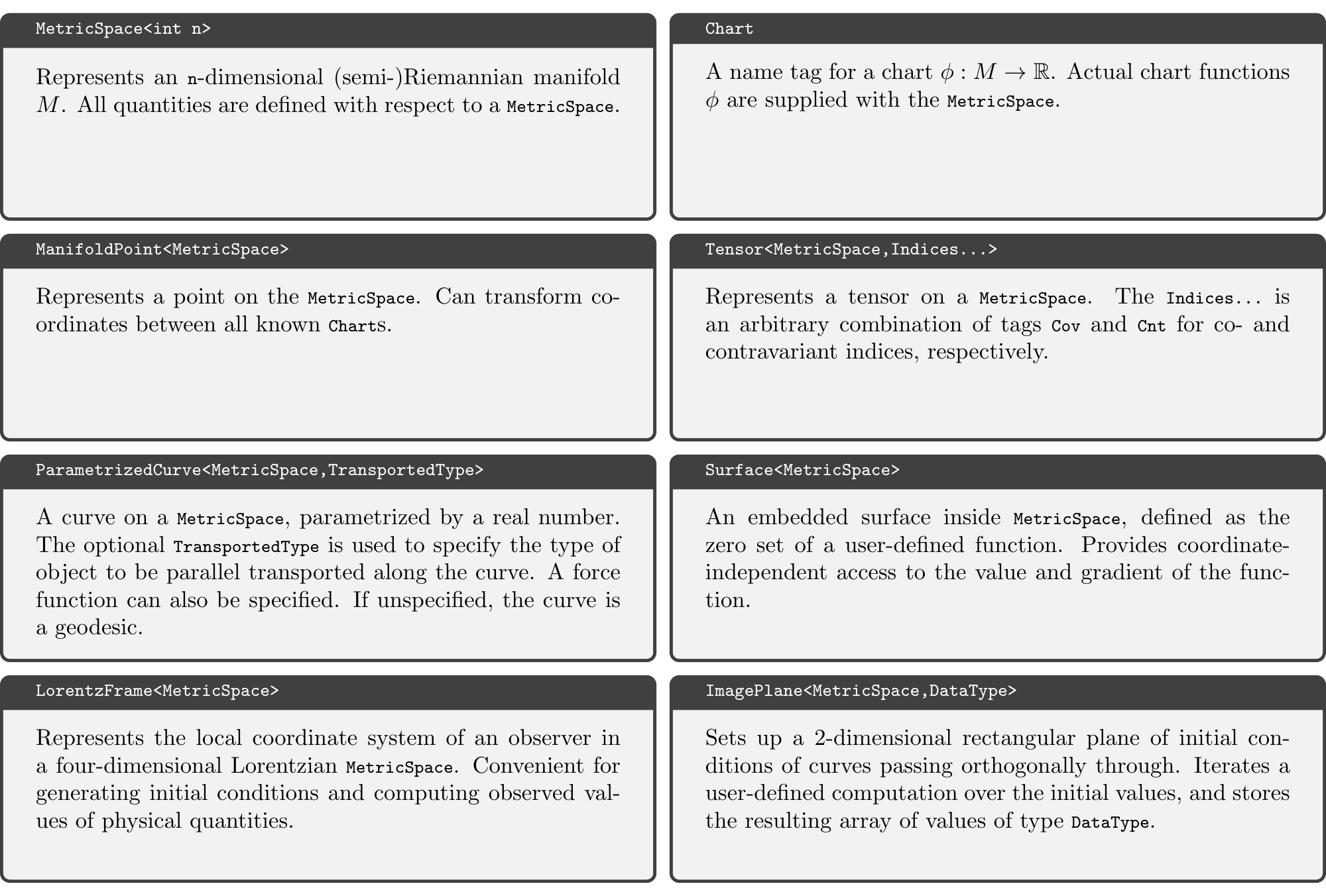}
    \caption{Short descriptions of the most often-used classes in the \am{} library.}
    \label{fig:classlist}
\end{figure*}

In the following, we describe how the mathematical objects are
implemented in the \am{} code. To make the exposition
easier to follow, we have provided a list of the most important \cpp{} classes of the
\am{} library together with their descriptions in
Figure~\ref{fig:classlist}. The Python
interface provides corresponding counterparts to these classes, together with
some additional convenience classes. A listing of these can be found in
the documentation accompanying the code.

\subsection{Manifolds and charts}\label{sc:manifolds_and_charts}

The most fundamental object in the \am{} library is \codeil{MetricSpace<n,Signature>},
representing an $n$-dimensional (semi-)Riemannian manifold of a given signature.
Defining a new
\codeil{MetricSpace} requires the specification of dimensionality $n$,
one or more charts, and functions returning the components of the metric
tensor field and its derivatives in each chart.
For four-dimensional semi-Riemannian spaces, the metric signature must
also be specified. \am{} supports both timelike $(+---)$ and spacelike
$(-+++)$ signatures.

A chart is
represented as a class \codeil{Chart} that in the current implementation
only contains a description and serves to give meaning to a tuple of
coordinate numbers.
The points on the manifold are implemented as a
class \codeil{ManifoldPoint<MetricSpace>}. These can be constructed by specifying $n$
coordinates and the corresponding chart. After this, the components of
the point can be requested in any available chart, and the object itself
behaves much like the mathematical idea of a point on a
manifold (see Appendix~\ref{app:manifolds}).

For transforming the components of tensorial objects, the transition
functions and their Jacobians between the charts must also be specified.
For $N$ different charts this would naively require $N(N-1)$ transition
functions and Jacobians to be implemented. The amount of work increases
quadratically. However, when the domains of
charts $\phi_i$, $\phi_j$ and $\phi_k$ overlap suitably, the transition
function from $i$-coordinates to $j$-coordinates fulfills
\begin{equation}\label{eq:trans-decomp}
\phi_j\circ\phi^{-1}_i = \phi_j \circ \phi^{-1}_k \circ \phi_k \circ
\phi^{-1}_i,
\end{equation}
and the Jacobian $J_{i\fromto j}=d(\phi_j\circ\phi^{-1}_i)$ decomposes similarly,
\begin{equation}\label{eq:jacobi-decomp}
    %d(\phi_j\circ\phi^{-1}_i) = d(\phi_j \circ \phi^{-1}_k) \circ d(\phi_k \circ \phi^{-1}_i).
    J_{i\fromto j} = J_{k\fromto j}  J_{i\fromto k}.
\end{equation}

The \am{} library uses the properties \eqref{eq:trans-decomp} and
\eqref{eq:jacobi-decomp} to build a directed \emph{graph of charts},
wherein each chart is a node, and the Jacobians and transition functions
define the edges. This makes it possible to introduce $N$ charts while
supplying only the minimum number of $2(N-1)$ transition functions and
Jacobians to make the graph connected. Then, when the components of a
point or a tensorial quantity are requested in a different chart, the
code walks through the graph building the transition function and
Jacobian piece by piece using equations \eqref{eq:trans-decomp} and
\eqref{eq:jacobi-decomp}.

For a listing of the built-in metric spaces and chart implementations provided
with \am{}, see Appendix~\ref{app:builtins}.

\subsection{Tensors}\label{sc:tensors}

Tensor algebra and calculus for tensors of arbitrary rank (see
Appendix~\ref{app:tensors}) is provided by
the \codeil{Tensor<MetricSpace,Indices...>} template class. Here \codeil{MetricSpace} is
the base manifold and \codeil{Indices...} is an
arbitrary combination of index tags \codeil{Cov} and \codeil{Cnt}, for
covariant or contravariant index, respectively. The implementation is
pointwise, using a set of $n^{k+l}$ components in a given chart to
specify a tensor of rank $(k,l)$ on a manifold $M$ with $\dim(M) = n$ at
a given \codeil{ManifoldPoint}.

As such, similarly to a \codeil{ManifoldPoint}, defining a tensor at a
point requires the input of $n^{k+l}$ components and the corresponding
chart. After this, the chart is abstracted away in the sense that
algebraic operations between tensors defined at the same point
can be performed irrespective of the chart the tensors were originally
defined in. The \codeil{Tensor} class provides all the usual algebraic tensor
operations: sum of tensors of same rank, tensor product, contraction and
additionally raising and lowering of the indices using the underlying
\codeil{MetricSpace} structure. The implementation checks all
operations for index correctness at compile time, so that e.g.\ no
contraction between indexes of same type is allowed. In addition, during
runtime, all operands are inspected to ensure that they are defined at
the same base point. These checks guarantee that operations
expressed in code correspond to mathematical operations that are well
defined.

The \codeil{Tensor} class also provides some elements of tensor
calculus. Namely, the class automatically computes the derivatives
required for parallel transporting a tensor along a general curve.
Given a curve tangent vector $u^a$, the class can compute the
contractions with $\chr^a_{bc}u^c$ required in the parallel transport
equation~\eqref{eq:T-partrans}.

\subsection{Curves}\label{sc:curves}

Functionality for working with curves $\gamma$, including geodesics, is
provided by the class \codeil{ParametrizedCurve<MetricSpace,TransportedType>} 
along with a convenience subclass \codeil{Geodesic}.
Curves are implemented as sequences of points $(\lambda, p)$ on a
manifold, where $\lambda$ is the curve parameter and $p\in M$.
More concretely, the implementation is based on an ordered
queue of objects of type \codeil{ParametrizedPoint<MetricSpace,TransportedType>}, 
which combine a \codeil{ManifoldPoint} with a real value $\lambda$ specifying the
position along the curve. 
In addition, the \codeil{ParametrizedPoint}
can include any arbitrary object $A$ of type \codeil{TransportedType}
to be parallel transported along a geodesic or, for example, Fermi--Walker transported along a
forced curve.
The only requirement is that the object be representable as a (chart-dependent) tuple of real
numbers, and that a function $D_A(u^a,\chr^a_{bc}u^c, f^a)$ yielding the derivatives
$\derfrac{A(\gamma(\lambda))}{\lambda}$ is provided. The function $D_A$
\edit1{depends} externally on the current tangent vector of
the curve $u^a$, the contractions $\chr^a_{bc}u^c$ and optionally the force
$f^a$.
As mentioned above, \codeil{Tensor} class provides the derivative
function automatically, and as such arbitrary tensors can be parallel
transported along all generic curves without any extra programming effort.

In practice, a curve is computed by specifying the initial conditions in
some given chart. These consist of the initial point
$(\lambda_0,p_0)\in \fR \times M$, the
components of the curve tangent vector $u^a(p_0)\in T_{p_0}M$, the
components $A(p_0)$ of the
possible parallel transported object, and an optional force function
$f^a$. The \am{} library then computes points along the curve for the
desired interval $I\subset\fR$ containing $\lambda_0$ by solving the
set of equations (see Appendix~\ref{app:geodesics})
\begin{align}
    \derfrac{\gamma(\lambda)}{\lambda} &= u^a 
        \label{eq:gam-der} \\
    \derfrac{u^a(\lambda)}{\lambda} &= -\chr^a_{bc}u^bu^c + f^a
        \label{eq:u-der} \\
    \derfrac{A(\lambda)}{\lambda} &= D_A(u^a, \chr^a_{bc}u^c, f^a)
        \label{eq:A-der}
\end{align}
in a suitable chart (see Section~\ref{sc:chartsel} for details on the
chart selection).

\am{} computes the solution using the integration methods
offered by the \odeint{} \cpp{} library \citep{odeint}. The default method is the
Dormand--Prince 5th order Runge--Kutta method \citep{dormand1980}, which
offers error estimation and automatic stepsize adjustment, as well as a fair
numerical performance in most cases. The absolute and relative error
tolerances and stepsize and iteration limits are fully
user-configurable.
After the computation is finished, the
\codeil{ParametrizedCurve} class provides access to the solution in any
chart and for \emph{any} $\lambda\in I$. Internally this is achieved
through a cubic spline interpolation.

\subsection{Surfaces}\label{sc:surfaces}

An interface for implementing hypersurfaces is available through the class
\codeil{Surface<MetricSpace>}. Surfaces are useful for representing solid or
highly optically thick objects, or regions of interest. Examples include
the surfaces of neutron stars, white dwarfs or
planets but also black hole event horizons, optically thick accretion
disks or the limits of computational domains. The \am{} implementation of
surfaces is based on the concept of level hypersurfaces (see
Appendix~\ref{app:surfaces}).

A new surface is implemented by supplying a real valued function
$S$ taking a \codeil{ManifoldPoint} as an argument, as well as the
gradient $\partial_a S$. The surface is then defined as the set of
points $\{p\in M | S(p)=0\}$.
In addition, a tangent vector field $t^a$ on
the surface must be defined. This field is primarily used to represent
the four-velocity field of observers fixed on the surface, and is
required for e.g.\ computations involving rotating neutron stars (see
Section~\ref{sc:neutron_stars}).

The \am{} library automatically detects intersections of curves with
surfaces, and numerically finds the exact (to within tolerance)
intersection point. The intersections are found by examining the sign of
the product $S(p_{k+1})S(p_k)$ for two successive points $p_{k+1}$ and
$p_{k}$ on a curve. If the product is negative, the two points must lie
in different regions bounded by the surface. The exact intersection
point is then found using the so-called Hénon's trick \citep{henon1982}.
The `trick' consists of changing the independent variable $\gamma$, the
curve parameter, in equations~\eqref{eq:gam-der}--\eqref{eq:u-der} to
$S$, or the value of the surface function. The transformed equations
read
\begin{align}
    \derfrac{\lambda}{S} &= (u^b\partial_b S)^{-1}
        \label{eq:H-lambda-der} \\
    \derfrac{\gamma(S)}{S} &=  (u^b\partial_b S)^{-1}\, u^a
        \label{eq:H-gam-der} \\
    \derfrac{u^a(S)}{S} &= (u^b\partial_b S)^{-1}\,
    \left(-\chr^a_{bc}u^bu^c + f^a\right)
        \label{eq:H-u-der} \\
    \derfrac{A(S)}{S} &=  (u^b\partial_b S)^{-1}\, D_A(u^a, \chr^a_{bc}u^c, f^a)
        \label{eq:H-A-der}.
\end{align}
These equations can then be numerically propagated for a single step of
length $-S(p_{k+1})$ starting from the point $p_{k+1}$ to yield the
intersection point to within numerical tolerance.

\subsection{Automatic chart selection}\label{sc:chartsel}

\begin{figure}
    \includegraphics[width=\columnwidth]{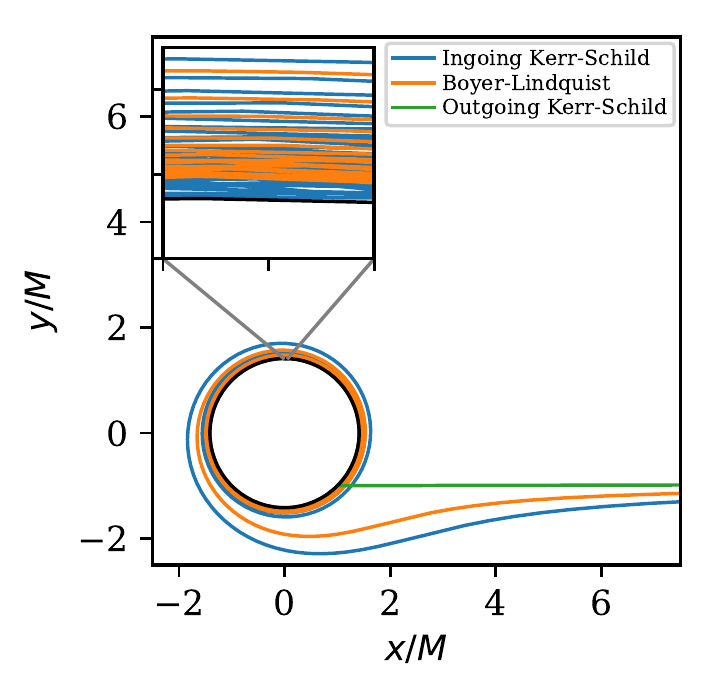}
    \caption{A null geodesic emanating from near the event horizon
    of an extremal Kerr black hole, shown in the outgoing and ingoing
    Kerr--Schild coordinates as well as Boyer-Lindquist (BL) coordinates,
    in the $xy$-projection.
    For BL coordinates, the transformation 
    $(x,y,z) = \sqrt{M^2+a^2}(\sin\theta\cos\phi, \sin\theta\sin\phi,
    \cos\theta)$ was used. The black line shows the location of the
    event horizon. The inset shows a zoomed-in region from near the event
    horizon.
    }
    \label{fig:kerr_geo_charts}
\end{figure}

Perhaps the most novel and interesting feature of \am{} is the
possibility to use multiple coordinate charts simultaneously and seamlessly. The most
immediate benefit is that objects can be input and output in any
available chart, with all transformations handled automatically by
\am{}. However, there are important computational benefits to free
selection of coordinate charts as well. The most obvious benefit is
the fact that a given problem may be much easier to solve numerically in
some specific coordinates compared to others.
This is illustrated in Figure~\ref{fig:kerr_geo_charts}, where the same
null geodesic in an extremal Kerr space-time is shown in the
outgoing Kerr--Schild coordinates, the ingoing Kerr--Schild coordinates
and the Boyer--Lindquist coordinates (see Appendix~\ref{sc:app-kerr-space}). 
From the figure, it is easy to
appreciate how in the outgoing Kerr--Schild coordinates the geodesic is
essentially straight, and long integration steps can be
taken. On the other hand, in the ingoing Kerr--Schild coordinates and the
Boyer-Lindquist coordinates the geodesic twists around the event horizon
at an increasing rate as the event horizon is approached.
The magnitudes of the derivatives with respect to
the curve parameter increase correspondingly, making the problem
eventually numerically impossible to solve.

The possibility to simultaneously use multiple charts
makes it possible to avoid the coordinate singularities present
in any single chart, such as the pole singularity in any spherical
coordinate system, or the coordinate singularity at the event horizon
present in the usual Schwarzschild coordinates. In addition,
using multiple charts makes it possible to switch the chart used for solving
the equations of motion for a curve on the fly, useful for situations
such as the one depicted in Figure~\ref{fig:kerr_geo_charts}.
It is not obvious which chart is to be preferred, which is why \am{}
currently implements several heuristics for automatically choosing the
numerically optimal chart.

The first heuristic consists of finding a chart $\phi_i$ where the matrix of
the components of the metric $\mat{G} = (g_{ab})$ has the largest 
inverse condition number, defined as the ratio of the smallest and
largest singular value of the matrix, i.e.\
$\cond^{-1}(\mat{G}) = \sigma_\text{min}/\sigma_\text{max}$.
This is based on two key observations. Firstly, floating point
addition and substraction between numbers of different
magnitude causes a loss of precision.
Secondly, the equations of motion for a curve and for parallel transport
along it, \eqref{eq:gam-der}--\eqref{eq:A-der}, contain a mix of the
components of the metric and its derivatives on the right-hand side. As
such, it would be intuitively advantageous to perform the computations
in a chart where the matrix formed by the metric has eigenvalues that
span as small a range as possible. This is achieved by maximizing the
inverse condition number.

In some cases the condition number of the metric is not enough to detect
a computationally awkward chart. For example, in the case of a Kerr
black hole, the condition number cannot differentiate between the
ingoing and outgoing Kerr--Schild charts. However, as is seen in
Section~\ref{sc:kerr-tests}, using one over the other can cause a
large difference in computation time and accuracy for radial geodesics,
depending on whether they are falling towards or emanating from the
event horizon. As such, a further heuristic is needed.

If the condition number heuristic does not separate two promising charts,
the \am{} code next tries to minimize the maximal absolute value of the intrinsic
derivatives, $-\chr^{a}_{bc} u^b u^c$, of the curve tangent vector $u^a$. As
such, this heuristic needs to know the current curve tangent vector
$u^a$, unlike the condition number test, for which only the current point is required.
For Cartesian coordinates in a Euclidean or Minkowskian space
$\chr^{a}_{bc}\equiv 0$,
so in effect this procedure looks for the most
Cartesian-like chart in which the metric looks most Euclidean (or
Minkowskian) in the direction of the current curve tangent vector $u^a$.

Formal proofs of the performance of these heuristics are beyond the scope of
this work, but the numerical results in Section~\ref{sc:code_tests}
indicate that they work reasonably well.

\subsection{Local Lorentz frames}\label{sc:lorentz_frames}

For four-dimensional Lorentzian manifolds, \am{} provides a
functionality to construct local Lorentz frames (see
Appendix~\ref{app:local-frames}) through the class
\codeil{LorentzFrame<MetricSpace>}. The user supplies a timelike vector $e_t$ and two
spacelike vectors $e_z$ and $e_x$. From these, a complete Lorentz frame
$\{E_t, E_x, E_y, E_z\}$ is constructed by first normalizing $e_t$ to
yield $E_t$ and then orthonormalizing $e_z$ and $e_x$ sequentially.
Finally, $E_y$ is defined by the remaining orthogonal direction through
%$E_y = \pm E_y\times E_z$, 
$E_y^a = \pm \epsilon\ix{^a_{bcd}} E_t^b E_z^c E_x^d$, where
$\epsilon_{abcd}$ is the Levi--Civita tensor,
with sign depending on the desired handedness (positive
for a right-handed frame).

The \codeil{LorentzFrame} object can be automatically parallel transported
along a \codeil{ParametrizedCurve}. In addition, \codeil{Tensor} objects
can be constructed from components given with respect to a
\codeil{LorentzFrame}. Likewise, the components of any
\codeil{Tensor} can be extracted in a given \codeil{LorentzFrame} as
well, see equation~\eqref{eq:app-tenscoordtolocal}.

\subsection{Image plane generation}

To produce mock observations, an observational instrument must be
emulated somehow. For ray-tracing purposes, this usually means
using an \emph{image plane}. The image
plane is positioned near the object of interest, and only the rays
intersecting the plane orthogonally are considered. These rays are
then assumed to propagate in vacuum all the way to the distant observer.
This approximation neglects atmospheric and instrumental effects, but
these can be modeled afterwards using dedicated tools if necessary.

There are three main sources of error when generating the image plane: 
required deviations from
perpendicularity, perturbations caused by the curvature of the space and
the assumption of vacuum propagation.
The assumption of perpendicularity is typically excellent. For
distant objects, the maximum deviation from perpendicularity
$\Delta\theta$ is approximately equal to the observed angular size of
the object, or $\Delta \theta \sim L/(2D)$, where $L$ is the linear
extent of the source perpendicular to the line of sight and $D$ the
distance. For example, in the case of \sgra{} (Sagittarius A$^*$), we
have $\Delta\theta\sim 10^{-11}$, and for a typical galactic neutron star
$\Delta\theta \sim 10^{-17}$.
The effects of remaining space-time curvature at the image plane location
can be estimated by looking at the bending angle $\beta$ that the
image plane rays will make when propagated to infinity. Sufficiently far away
from the object so that the Schwarzschild metric can be used, this angle turns out to be
\citep[e.g.][]{beloborodov2002} $\beta\sim 2GM/(c^2 R)$, where $M$ is
the total mass of the observed object and $R$ is the radial distance of
the image plane from the object. Thus, for $R \gtrsim 10^4 \, GM/c^2$ we have
$\beta \lesssim 2\times 10^{-4}$, and so the
effects of residual curvature are negligible.
The assumption of propagation in vacuum is typically valid for objects
that are not situated at cosmological distances as far as the light
bending is concerned. However, corrections for effects
such as extinction, frequency dispersion or Faraday rotation may need to
be added in further postprocessing.

In many ray-tracing codes, the construction of image planes is achieved by a
assuming a flat space and explicitly constructing the starting points and
tangent vectors for a planar configuration of geodesics
\citep{broderick2004b,cadeau2007,dexter2009,vincent2011,dexter2016,chan2017}.
\am{} provides a general-purpose tool for constructing plane-parallel
initial conditions for Lorentzian space-times in class
\codeil{ImagePlane<MetricSpace,DataType>}.
The user specifies a \codeil{LorentzFrame} at the
center of the plane and the extent and the resolution (number of grid points)
of the plane in the local $E_x$ and $E_y$ directions. The local Lorentz
frame is then parallel transported to the desired grid points via
spacelike geodesics, using the \am{} curve propagation functionality.
Initial conditions for curves passing through the plane are set up by
assigning the tangent vectors $u^a(0)$ to be spatially parallel to the
parallel transported $E_z$ vector.
The collection of parallel transported frames defines a best local
approximation to a flat plane that is threaded by orthogonal curves, and
\emph{corrects} the effect of the bending $\beta$ caused by the curvature to
first order. Thus, the \am{} \codeil{ImagePlane} can safely be used in
regions where the curvature is small but non-negligible.
The method is also general purpose in the sense that it
works similarly in any coordinate system and only requires specifying a
local Lorentz frame at one point.

\section{Implementation of radiative transfer}\label{sc:radtrans}

\subsection{Fluid and radiation models}

Radiative transfer functionality in \am{} is built with flexibility in
mind. For this purpose, the interface declares two types of functions.
The first type is a
\codeil{FluidFunction<MetricSpace,FluidData>} which maps points on the base manifold
\codeil{MetricSpace} to a user-defined set of fluid variables
\codeil{FluidData}, which represent local material properties such as
temperature or density. The only restriction is that \codeil{FluidData}
must include a single bulk fluid four-velocity $w^a$ and a single
reference direction (often magnetic field) $t^a$ orthogonal to $w^a$.

The second type of
function is \codeil{RadiationFunction<FluidData>}, which
computes the Stokes emissivity vector $\evec$ and the response matrix
$\mulmat$ (see Appendix~\ref{app:radtrans}) from the given
\codeil{FluidData}, local fluid rest frame frequency $\nu$, and the
rest frame angle $\theta$ between the reference direction $t^a$ and the
current direction of the light ray (the tangent vector $k^a$).

This approach makes implementing different fluid and radiation models
rather straightforward. For example,
the fluid variables for a given point can be
obtained from a GR magnetohydrodynamics (GRMHD) simulation or from
an analytic model. The \am{} suite includes an example application which
reads outputs from the HARM GRMHD code \citep{gammie2003,noble2006} and computes
mock observations using a thermal synchrotron radiation model based on
the results in \citet{dexter2016}.
See Section~\ref{sc:grtrans_comp} for computational results.

\subsection{Solving the radiative transfer equation}

With \am{}, a radiative transfer problem (see
Appendix~\ref{app:radtrans}) is solved by first propagating
a set of curves $\gamma_i$ (typically geodesics, unless plasma effects
are significant) along which the radiative transfer
equation, eq.~\eqref{eq:app-pol-scalar}, is to be solved as a curve integral.
%While the current
%implementation of \am{} supports curve integrals over arbitrary curves,
%solving the radiation transfer equation is only supported along null
%geodesics.
Usually, the most
convenient approach is to use an \codeil{ImagePlane} and let \am{}
propagate the set of initial conditions backwards in time through the region
of interest. Each propagated curve must
include a parallel transported \codeil{PolarizationFrame}, a pair of two
orthogonal spacelike vectors $\polframe=\{v^a, h^a\}$,
also orthogonal to the geodesic and the four-velocity of the observer,
representing the vertical and
horizontal linear polarization basis vectors of the observer at one end
$\gamma_i(\lambda_\text{obs}) = p_\text{obs}$ of the curve. If using an
\codeil{ImagePlane}, these can be conveniently obtained from the $E_x$
and $E_y$ vectors of the local Lorentz frame at each point.

The four-velocity $u^a(p_\text{obs})$ of the observer $\obs$ at $p_\text{obs}$, the
four-velocity $w^a(p)$ of the fluid at each point $p=\gamma_i(\lambda)$
and the curve tangents $k^a(p)$ and $k^a(p_\text{obs})$ define a connection between the 
photon frequency $\nu_0$ observed by $\obs$ at $p_\text{obs}$ and the
corresponding photon frequency $\nu$ in the local rest frame of the fluid at
$p$. This is given by the \emph{redshift factor}
\begin{equation}
    \rsfac = \frac{u_a(p_\text{obs}) k^a(p_\text{obs})}{w_a(p) k^a(p)} =
    \frac{\nu_0}{\nu}.
\end{equation}

\begin{figure}
    \includegraphics[width=\columnwidth]{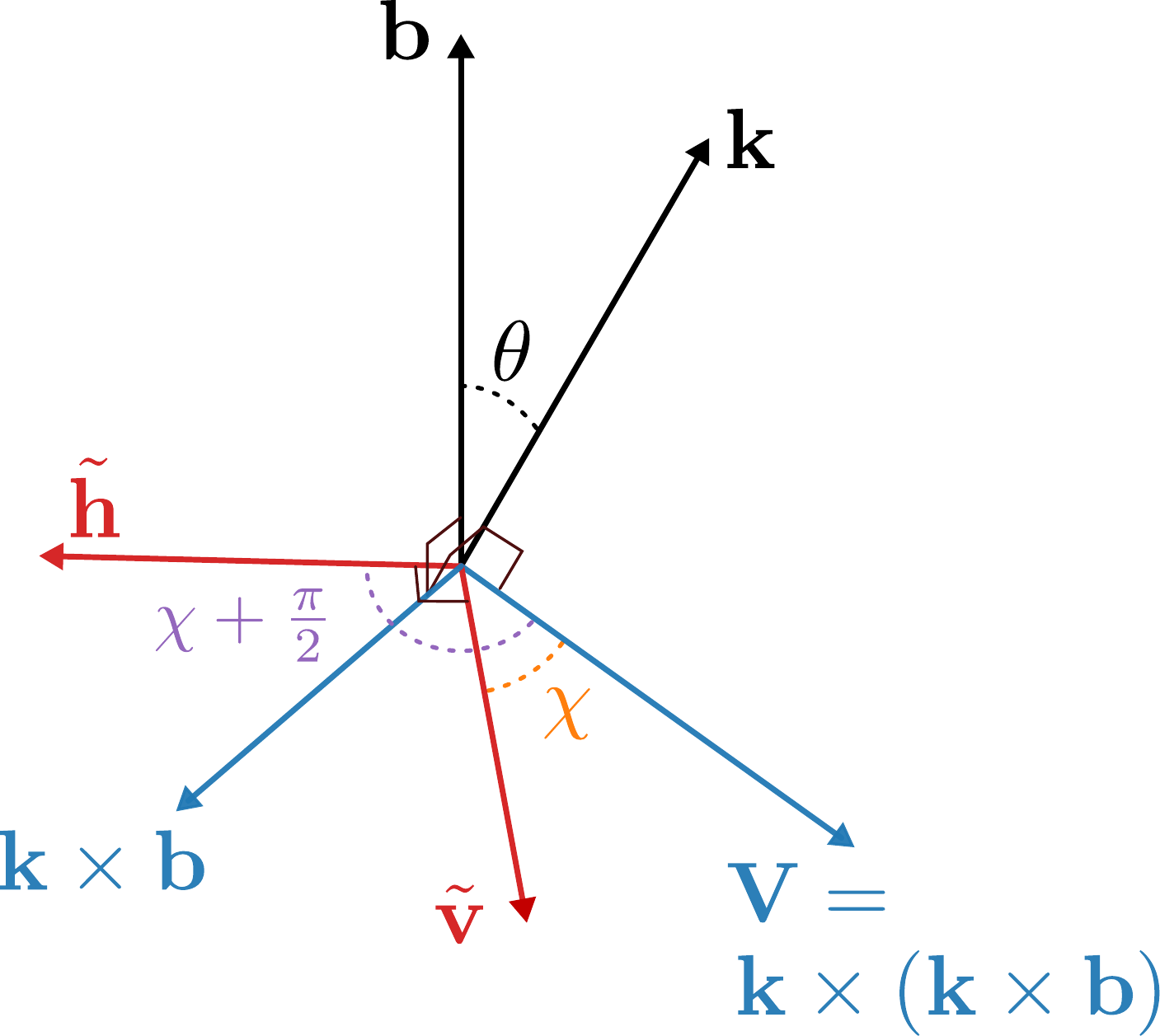}
    \caption{Definition of the angles $\theta$ and $\chi$, in the
    three-dimensional rest frame of the fluid. Also shown are the local
    reference direction $\vct{b}$, the direction of the geodesic $\vct{k}$ and
    the local vertical polarization direction $\vct{V}$.
    The vectors
    $\tilde{\vct{v}}$ and $\tilde{\vct{h}}$ are the spatial parts of $\tilde{v}^a$
    and $\tilde{h}^a$.
    }
    \label{fig:pol_angles}
\end{figure}

The initial conditions are set by defining initial invariant specific
intensities $\{ \csvec_{\nu_{0,i}/\rsfac} =
\svec_{\nu_0/\rsfac}(\nu_0/\rsfac)^{-3} \}$ at the
other end $p_\text{start}$ of the curve, one for each observed frequency
$\nu_{0,i}$ of interest. Often these can be set to zero, but for example
in the case of radiation emanating from optically thick or solid surfaces,
the initial intensity can be non-zero. Solving the radiative transfer
equation itself proceeds in a manner following \citep{shcherbakov2011}.
See Figure~\ref{fig:pol_angles} for a diagram of all
the vectors and angles.

At each point $p\in M$ during the calculation, the \am{} library
evaluates the given \codeil{FluidFunction} to obtain the fluid
four-velocity $w^a$ and the rest of the fluid parameters in the rest frame of
the fluid. This includes the local reference direction $b^a$, which typically
is the direction of the local magnetic field.
From these, the angle $\theta(w;b,k)$
between the reference direction $b^a$ and the light ray tangent $k^a$ as seen in
the fluid rest frame is computed using equation~\eqref{eq:vecangle}.
This angle is required by some radiation models, such as synchrotron emission
models.
The
reference direction also defines the local vertical direction of polarization
$\vct{V} = \vct{k}\times(\vct{k}\times\vct{b})$, where $\vct{k}$ and $\vct{b}$
are the spatial parts of $k^a$ and $b^a$, respectively.

The next step is to project the
parallel transported polarization frame $\polframe$ to the fluid rest frame
using the screen projection operator, equation~\eqref{eq:app-screen-proj},
yielding $\tilde\polframe=\{\tilde{v}^a,\tilde{h}^a\}$, where 
\begin{align}
    \tilde{v}^a &= P_\perp(w,k)^a_b v^b \\
    \tilde{h}^a &= P_\perp(w,k)^a_b h^b.
\end{align}
Now we can compute the angle $\chi$ between the projected parallel transported polarization frame
$\{\tilde{v}^a,\tilde{h}^a\}$ and the polarization frame of the fluid, defined
by $\vct{V}$, from
\begin{equation}
    \tan\chi = \frac{-V_a \tilde{h}^a }{V_a \tilde{v}^a},
\end{equation}
where $V^a = (0,\vct{V})$.

Next, the angle $\theta$ and the fluid parameters are passed to the 
\codeil{RadiationFunction} to obtain the Stokes emissivity  and
the response (Müller) matrix $\mulmat_\nu$ in the fluid rest frame.
These are related to the parallel transported and projected polarization frame
$\tilde\polframe$
using the angle $\chi$ and the transformation properties of the Stokes
components under rotation \citep[e.g.][]{chandrasekhar1960}.
The emissivity vector $\evec_\nu$ and response matrix $\mulmat_\nu$ are
transformed via $\evec_\nu \mapsto R(\chi)
\evec_\nu$ and $\mulmat_\nu \mapsto R(\chi) \mulmat R(-\chi)$, where
\begin{equation}
    R(\chi) = \begin{pmatrix}
        0 & 0           & 0            & 0 \\
        0 & \cos(2\chi) & -\sin(2\chi) & 0 \\
        0 & \sin(2\chi) & \cos(2\chi)  & 0 \\
        0 & 0           & 0            & 0
    \end{pmatrix},
\end{equation}
gives the transformation of Stokes vectors under rotations of the
polarization plane. Finally, it can be shown that the Stokes components in any
two polarization frames $\polframe$ and $\tilde\polframe$ related by a screen
projection are equal, so that the
radiative transfer equation to be solved along the geodesic is 
\begin{equation}\label{eq:final_radtrans}
    \derfrac{\csvec_{\nu}}{\lambda} = 
    L \frac{u_a(p_\text{obs}) k^a(p_\text{obs})}{\nu_0} \left(
    \cevec_\nu - \csvec_\nu \cmulmat_\nu
    \right),
\end{equation}
where
\begin{align}
    \cevec_\nu &= \nu^{-2} R(\chi) \evec_\nu \\
    \cmulmat_\nu &= \nu R(\chi) \mulmat_\nu R(-\chi) \\
    \nu &= \nu_0/\rsfac,
\end{align}
and $L$ is the unit of length. For example, in problems related
to black holes, a typical choice is $L = GM/c^2$, where $M$ is the black
hole mass.
Internally, equation~\eqref{eq:final_radtrans} is solved using the
\odeint{} Runge--Kutta--Fehlberg 8th-order method. However, for problems
where the optical thickness is large, the
equation~\eqref{eq:final_radtrans} can become stiff, and an implicit method
would provide better performance.

\section{Code tests}\label{sc:code_tests}

\subsection{Curves, parallel transport and chart selection}

\subsubsection{Geodesic propagation}

The accuracy of the basic curve propagation functionality
(Section~\ref{sc:curves}) was
verified by investigating curves on a two-dimensional spherical
surface. The computations were performed both in two dimensions,
using the intrinsic spherical coordinate chart
$(\theta,\phi)$, equation~\eqref{eq:app-two-sphere-metric}, and
in a three-dimensional Euclidean slice at $t=0$ of the Minkowski space
using the spherical coordinates $(0,r,\theta,\phi)$,
equation~\eqref{eq:app-minkowski-sph}.
To force the curve to stay on the surface of a sphere in the
three-dimensional case,
a constraint force $f(u^a) = (0, -u^a u_a, 0, 0)$ was specified.
Here $u^a$ is the curve tangent, in three-dimensional
spherical coordinates.

\begin{figure}
    \includegraphics[width=\columnwidth]{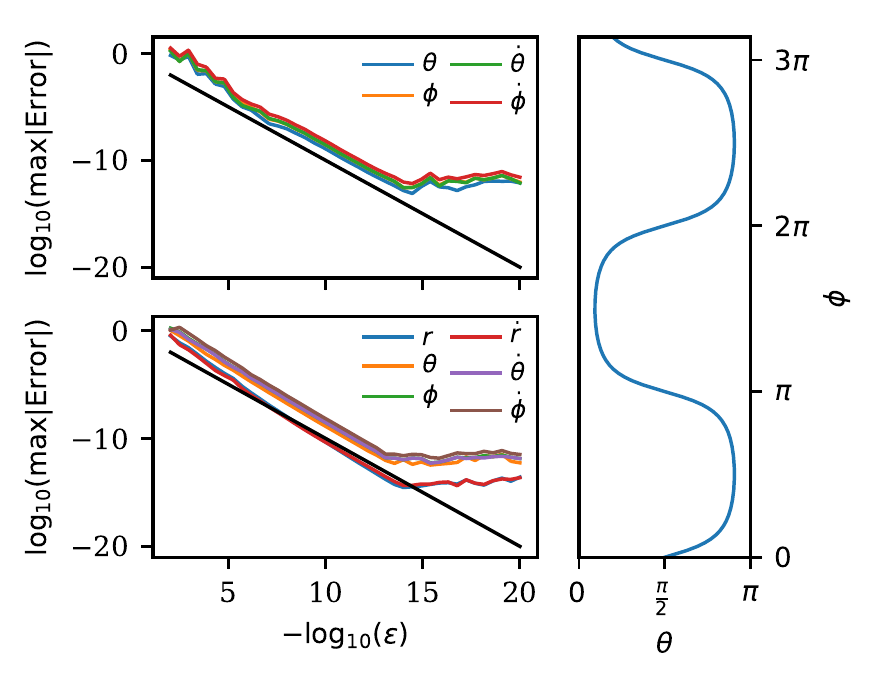}
    \caption{Differences between the analytic and the numerically computed geodesic 
    on a spherical surface, fulfilling $(\theta(0),\phi(0)) = (\pi/2, 0)$ and
    $(\dot{\theta}(0),\dot{\phi}(0))=(1.0, 0.3)$. 
    \emph{Top left panel:} Results computed in using the intrinsic
    two-dimensional metric. Shows maximal numerical errors along
    the curve in the coordinate position $(\theta,\phi)$ and the tangent
    vector $(\dot{\theta},\dot{\phi})$ as a function of the tolerance
    $\epsilon_\text{abs}=\epsilon_\text{rel}=\epsilon$.
    \emph{Bottom left panel:}
    Results computed in a three-dimensional space using a constraint force.
    Shows maximal numerical errors along
    the curve in the coordinate position $(r,\theta,\phi)$ and the tangent
    vector $(\dot{r},\dot{\theta},\dot{\phi})$ as a function of the
    tolerance.
    \emph{Right panel:} The orbit of the curve in the two-dimensional
    spherical coordinate chart.
    The solid black line segments in top and bottom left-hand panels are
    guides to the eye and show the identity function $f(\epsilon)=\epsilon$.
    }
    \label{fig:two_sphere_conv}
\end{figure}

Numerical convergence was estimated using 
a single geodesic curve $\gamma(\lambda)$ passing through $(\theta,\phi) = (\pi/2, 0)$ at
$\lambda=0$ with a
tangent vector $u^a=(\dot{\theta},\dot{\phi})=(1.0, 0.3)$. The initial values
were chosen so as to avoid a purely polar or equatorial geodesic, but were
otherwise chosen arbitrarily.
The geodesic was computed several times using a range
of equal relative and absolute numerical tolerances $\epsilon_\text{rel}$ and
$\epsilon_\text{abs}$ from $10^{-20}$ to $10^{-2}$ in 40 steps.
The differences between the numerical results and
the known analytical solution are shown in
Figure~\ref{fig:two_sphere_conv}. We see that both in the intrinsic
two-dimensional and the constrained three-dimensional case the numerical
curves converge towards the analytical solution linearly with the
tolerance parameters.
The convergence saturates at tolerance parameters $\sim 10^{-15}$ when
the relative precision floor of the double precision floating point
numbers is reached.

\subsubsection{Parallel propagation in the Kerr space-time}\label{sc:kerr-tests}

The functionality for parallel transporting tensorial quantities (see
Sections \ref{sc:tensors} \& \ref{sc:curves}) along a
curve was assessed in the context of a Kerr space-time (see
Section~\ref{sc:app-kerr-space})
with a near-extremal non-dimensional spin parameter $\chi = 0.99$ and
mass $M=1$.
First, initial conditions $\gamma(0) = (t_0,r_0,\theta_0,\phi_0) = (0, 10, 1, 0)$
and $u^a(0) = (1, 0, 0.01, 0.03)$
were fixed in the Boyer--Lindquist coordinates
(see equation~\eqref{eq:app-kerr-bl-metric}).
These initial values were chosen to yield a generic timelike geodesic, and to
avoid special cases such as equatorial geodesics, but were otherwise chosen
arbitrarily.
The geodesic was then augmented by including the metric
$g_{ab}$ and a Lorentz frame $\{E_t, E_x, E_y, E_z\}$ as quantities to
be parallel transported. The geodesic was then computed until
$\lambda=990$ to yield several complete orbits around the black hole, using tolerances
$\epsilon_\text{abs}=\epsilon_\text{rel}=10^{-10}$.
Finally, the parallel transported values were evaluated for accuracy by
comparing to analytic expectations.

\begin{figure}
    \includegraphics[width=\columnwidth]{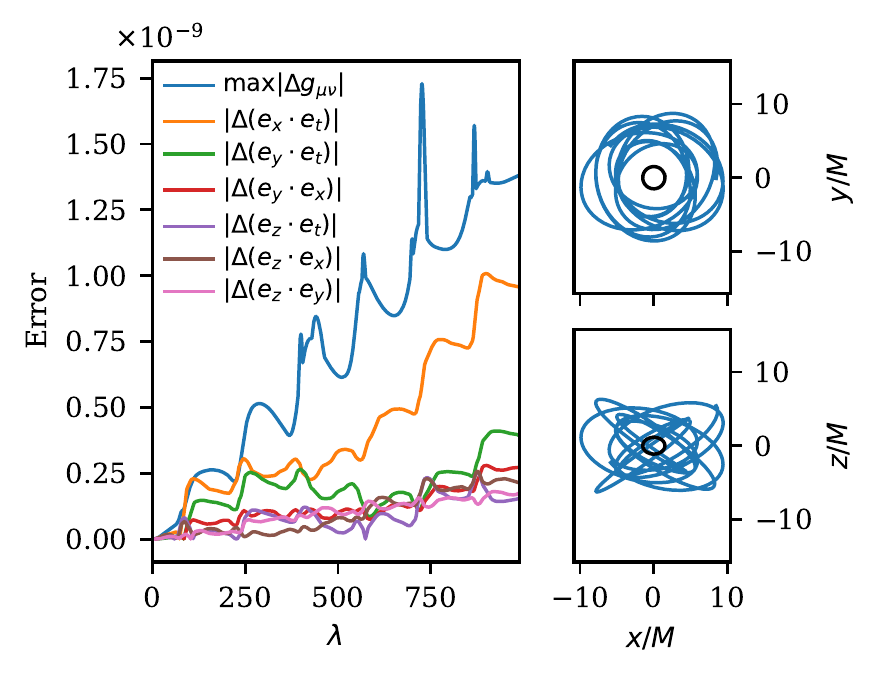}
    \caption{
        \emph{Left panel:}
        The absolute values of the errors accumulated during the
        parallel transport of the
        metric $g_{ab}$ and a local Lorentz frame $\{E_t, E_x, E_y,
        E_z\}$, computed in the ingoing Kerr--Schild coordinates.
        \emph{Right panels:} The orbit of the geodesic along which the
        parallel transport was computed, shown in ingoing Kerr--Schild
        coordinates using $xy$ (top) and $xz$ projections (bottom).
        The black circle shows the location of the event
        horizon.
        }
    \label{fig:kerr_partrans}
\end{figure}

Figure~\ref{fig:kerr_partrans} shows the orbit of the geodesic.
It also depicts magnitudes of the maximum difference $\max\abs{\Delta
g_{ab}}$ of the components of the parallel transported metric with
respect to the analytic expression, both computed in the ingoing
Kerr--Schild chart. Also shown are the
absolute values of all the pairwise inner products of the parallel
propagated Lorentz frame which should be identically zero. From the
figure we see that the errors in all of these conserved quantities
increase in a secular fashion, while
the single step errors are below the set numerical tolerance.
This is an expected and well-known behavior for
non-symplectic numerical integration methods, such as the 5th order
Dormand--Prince scheme used in \am{}, which do not respect the
geometric structure of the phase space \citep{hairer2008}.
Symplectic methods for the inseparable Hamiltonians occurring in geodesic
propagation have been discovered recently
\citep{pihajoki2015}, but these are not yet available in \odeint{}.
In general, the secular accumulation of integration error poses no
problem for the applications we demonstrate in this paper.  However, for
integrations over long periods of time, such as for computing dynamics
of massive particles orbiting a black hole, a symplectic method for
inseparable Hamiltonians might need to be implemented.

\begin{figure*}
    \includegraphics[width=1.0\textwidth,height=0.8\textheight,keepaspectratio]{%
        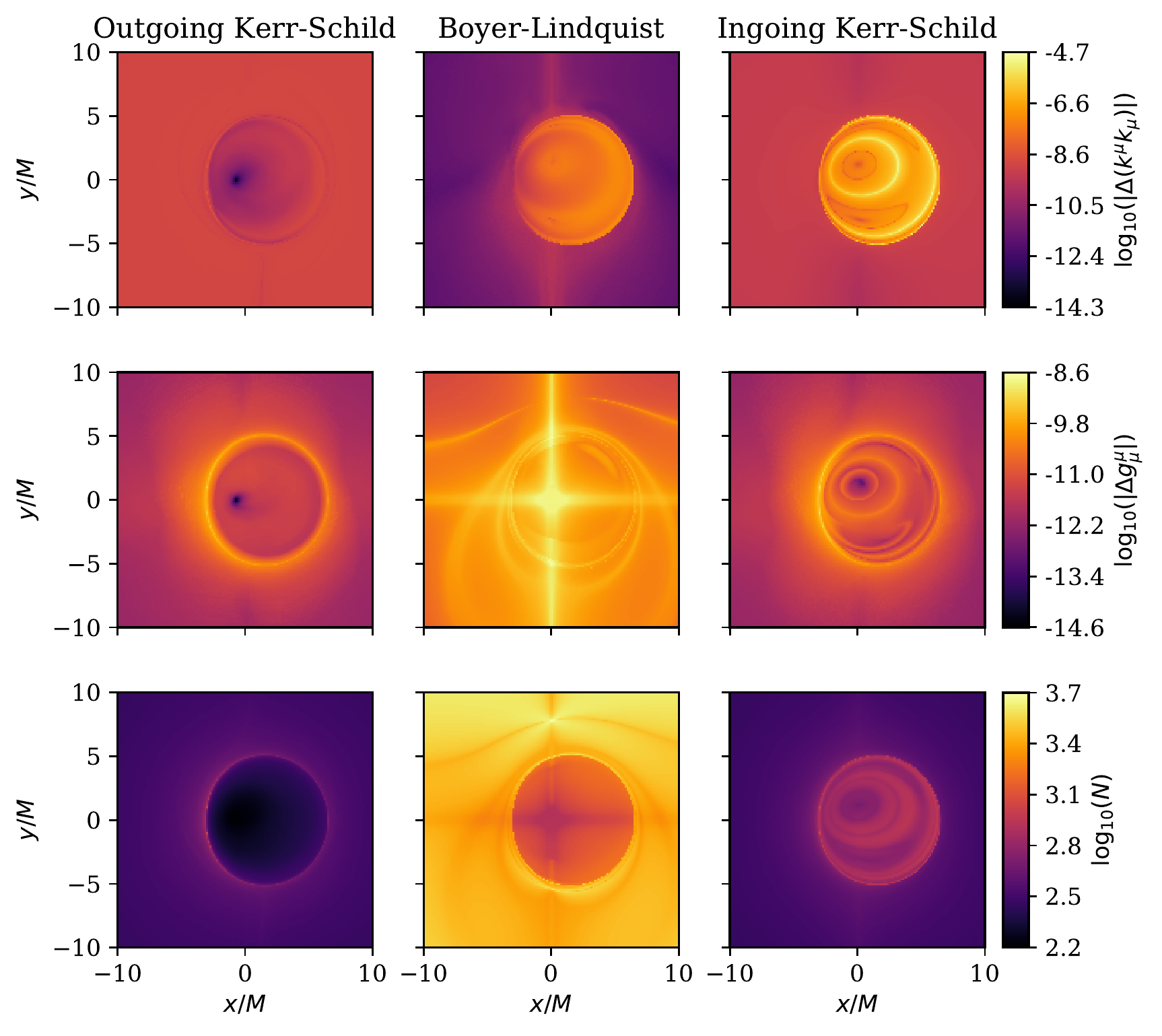}
    \caption{
        Maximum absolute errors and the number of computational
        steps taken along null geodesics computed in a Kerr
        space-time with a dimensionless spin parameter $\chi=0.95$. The
        errors are shown on an image plane situated at $r_0=10^5 M$ and
        an inclination of $\theta_0=50^\circ$, in Boyer-Lindquist
        coordinates. Geodesics were computed from $\lambda=0$ to
        $\lambda=-2 r_0$ or until an intersection with a 
        surface at $r=1.03 r_H$, where $r_H$ is the Kerr event horizon radius.
        The computation was performed three times with the chart fixed to
        either the outgoing
        Kerr--Schild (left column), Boyer--Lindquist (middle column) or
        ingoing Kerr--Schild coordinates (right column).
        \emph{Top row:} Maximum error in the value of the Hamiltonian
        $H=k_ak^a$, where $k^a$ is the tangent of the geodesic. 
        \emph{Middle row:} Maximum error in the 
        trace $g\ix{^a_a}$ of the parallel transported metric tensor.
        \emph{Bottom row:} Number of steps, $N$, taken by the curve
        integration routine.
        }
    \label{fig:kerr_imgplane_cons}
\end{figure*}

The accuracy and performance of both the curve propagation and
parallel transport functionality was also assessed
as a function of the geodesic and the coordinate chart.
To this end, we set up an image plane at $(r_0=10^5,
\theta_0=50^\circ)$ in the Boyer--Lindquist (BL) coordinates of a Kerr
space-time with $\chi=0.95$ and $M=1$. From the image plane, null geodesics were
propagated backwards from $\lambda=0$ to $\lambda=-2 r_0$ or until
intersection with a surface slightly outside the event horizon,
defined by $r=1.03 r_H$, where $r_H$ is the
event horizon radius. This radius was chosen since the computation in
the Boyer--Lindquist and ingoing Kerr--Schild coordinates must be
terminated before the event horizon itself (see
Figure~\ref{fig:kerr_geo_charts}).
The geodesics were computed three times, each time
fixing the chart (automatic chart selection disabled) to either ingoing
Kerr--Schild (KS), outgoing KS or the Boyer--Lindquist chart.
Standard tolerances of $\epsilon_\text{rel}=\epsilon_\text{abs}=10^{-10}$
were used.
We then computed the maximal absolute errors in
the value of the curve Hamiltonian, $H(x,k) = k_a k^a = 0$, and
the trace $g\ix{^a_a}$ of the parallel transported metric along the
geodesics, and plotted these on the image plane, in addition to the number of
integration steps $N$. The results are shown in
Figure~\ref{fig:kerr_imgplane_cons}.

From the figure, it is evident that the outgoing KS
coordinates offer significantly better numerical performance than the
ingoing KS coordinates or the BL coordinates.
This is not surprising, since the outgoing KS chart is adapted
to radially outgoing null geodesics. As a consequence, the more radial the geodesic
is, the more nearly a straight line it is in the outgoing KS chart. In
the figure, this can be seen as the remarkable decrease in the maximal error
and the number of computational steps for the geodesics starting near the
origin of the image plane (see also Figure~\ref{fig:kerr_geo_charts}).
On the other hand, the BL coordinates are seen to perform significantly worse.
This is related to both the fact that the geodesic `wraps around' the
black hole near the event horizon (see
Figure~\ref{fig:kerr_geo_charts}), but also the fact that the condition
number (the ratio of the maximum to minimum singular value) of the
matrix of the metric components scales as $\bigO(r^2)$ (see
Section~\ref{sc:chartsel}).
In addition, the
coordinates are singular at the poles. All these factors combine to make
the BL chart the most numerically disadvantageous of the three.
Finally, the ingoing KS chart fares worse than the BL chart for
the conservation of the Hamiltonian, but better for the trace of the metric
tensor and number of steps taken.
This is
understandable, since these coordinates are adapted to radially ingoing
null geodesics, and outgoing geodesics `wrap around' the
black hole near the event horizon twice as fast compared to the BL
coordinates. This is partly offset by the fact that the condition number
of the metric components is better behaved than for the BL coordinates.
The `wrap-around' behavior is suppressed near the poles of the black
hole, which in the figure can be seen as the slight decrease in the
error of the metric trace around the `North' pole of the black hole for
the BL and the ingoing KS coordinates.

The accuracy in general is seen to be consistent with the given
numerical tolerances. The outgoing KS chart in particular provides
excellent accuracy, with results much better than even the set
tolerances for nearly radial geodesics. In addition, there is a factor of
$\sim 10$ difference in the number of steps taken between the outgoing
KS chart and the BL chart, which was also directly reflected in the
computational time. The results strongly suggest that the outgoing
KS metric should be preferred in all codes computing mock observations
using geodesics emanating from the vicinity of a Kerr black hole.
Likewise, for studies of radiation scattering from a black hole, ingoing
KS coordinates should be used for computing the incoming radiation and
outgoing KS coordinates for the scattered, outgoing radiation.

\subsection{Radiation tests}

We assessed the accuracy and convergence properties of the \am{} radiative
transfer functionality by postprocessing a general relativistic
magnetohydrodynamics (GRMHD) simulation, and
by comparing to an existing polarized radiative transfer code \grtrans{}
\citep{dexter2016}.
To facilitate an easy comparison, we used the same simulation data as
was used to test \grtrans{}, as the data
is conveniently distributed with the \grtrans{}
code.\footnote{The data is found in the file \texttt{dump040}, found online at
\url{https://github.com/jadexter/grtrans/blob/master/dump040}.}
The simulation data used was computed with the GRMHD code HARM
\citep{gammie2003,noble2006}, and describes an axisymmetric
optically thin accretion flow around a Kerr black hole with a
dimensionless angular momentum of $\chi=0.9375$.
The black hole mass $M$
and its accretion rate $\dot{M}$ were set to the \grtrans{} defaults for HARM,
$M=4\times 10^6\,\Msun$ and $\dot{M} =
1.57\times 10^{15}\,\text{g}\,\text{s}^{-1} =
2.49\times10^{-11}\,\Msun/\text{yr}$. Assuming in
addition that the source is at a distance of $D=8\,\text{kpc}$,
these values approximate \sgra{} \edit1{\citep{dexter2016}}, although for this
particular data set at the observed frequency
of $\nu=230\,\text{Ghz}$ the total computed flux of $\sim13\,\text{Jy}$ (see
below) is roughly three times too large compared to the observed value of
$\sim3\text{--}4\,\text{Jy}$ \citep[e.g.][]{bower2015}.
The radiative model was taken to be relativistic
thermal synchrotron radiation, using the updated formulae in
\citet{dexter2016}. The electron and proton
temperatures in the plasma were assumed equal, and the
ideal gas equation of state was assumed, also corresponding to \grtrans{}.

All mock observations were computed using
a square image plane with physical dimensions $x,y\in[-L, L]$, where
$L=13\,M$, in
the local Lorentz frame of a stationary observer with $E_t=(1,0,0,0)$ (see
Section~\ref{sc:lorentz_frames} and Appendix~\ref{app:lorentz_frames}).
The negative $z$-axis is pointed towards the origin of the
Boyer-Lindquist (BL) coordinates.
The image plane was set at a distance $r=10^4\,M$ and
an inclination of $\theta=50^\circ$ (in BL coordinates), as in
\citet{dexter2016}.
Geodesics from the image plane were then computed backwards in time from
the image plane and the radiative transfer computed at the observed frequency of
$\nu=230\,\text{GHz}$ along these geodesics to
form the final image. Numerical tolerances for both the geodesic computation
and the radiative transfer computation were set to $10^{-10}$. These tolerances
guarantee that the accuracy during radiation transfer is
governed by the chosen sampling rate $\Delta\lambda/\lambda_\text{max}$,
where $\lambda_\text{max}$ is the total range of the affine parameter over which
the radiation transfer is computed. This ensures that the convergence
and comparison results are not affected by the characteristics of
sampling induced by timestep control.

\subsubsection{Flux convergence}\label{sc:flux_convergence}

First, we investigated the convergence of the total flux in the Stokes
variables $\svec_\nu=(I_\nu,Q_\nu,U_\nu,V_\nu)$
as the maximum stepsize $\Delta\lambda$ in affine
parameter and the image size $P$ in pixels per side were varied.
For each pixel $i$, we computed the observed flux
\begin{equation}\label{eq:physflux}
    \vct{F}_{\nu,i} = \svec_\nu(x_i,y_i)\, \frac{\Delta x \Delta y}{D^2},
\end{equation}
where $\Delta x = \Delta y = 2L/P$ is the physical size of each pixel and $D$ is
the (non-cosmological) distance of the target.
From the pixel-by-pixel fluxes, the total integrated fluxes
$\vct{F}_\nu = \sum_i \vct{F}_{\nu,i}$ were then computed.

\begin{figure*}
    \includegraphics[width=\textwidth]{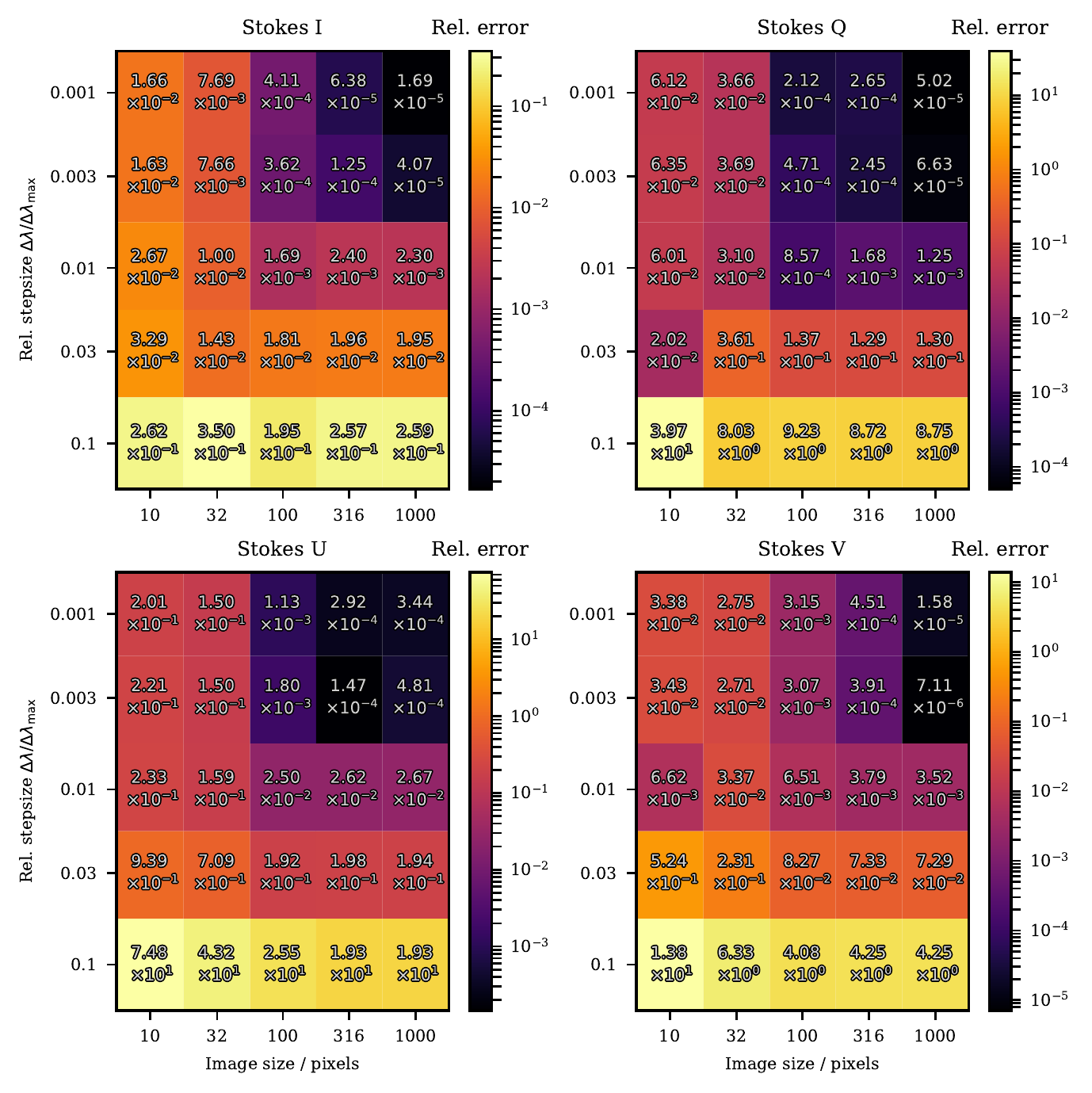}
    \caption{
    Convergence of total Stokes fluxes,
    see eq.~\eqref{eq:physflux},
    at $\nu=230\times 10^9\,\text{Hz}$  over a mock observation
    image (see text) when
    relative integration stepsize $\Delta\lambda/\lambda_\text{max}$ and mock image size $P$ (width and
    height, in pixels) are
    varied.
    All differences are relative to a data set computed with $P=1500$
    and $\Delta\lambda/\lambda_\text{max} = 10^{-4}$. Differences are shown as
    density maps with the numerical value inset, with one map for each Stokes
    flux component $I$, $Q$, $U$ and $V$.
    }\label{fig:convergence}
\end{figure*}

The convergence results are shown in Figure~\ref{fig:convergence}.
The general trend is that the benefits of a smaller stepsize saturate
quickly for smaller sized images, where the spatial sampling noise
dominates. Similarly, increasing the image size is only effective up to
the point where the noise from sampling of the small scale structures
starts to dominate. For this particular case, the benefit of increasing
the image size beyond $P=316$ pixels per side is already marginal. 
At this size, a $0.1\%$ convergence is achieved at a maximum relative step size
of $\Delta \lambda/\lambda_\text{max} = 3\times 10^{-3}$.

\subsubsection{Comparison to \grtrans{}}\label{sc:grtrans_comp}

In addition to ensuring the consistency and convergence of the \am{}
results, we performed a comparison to a publicly available radiative
transfer code \grtrans{} using the same HARM data set as above. Both
codes were used to compute a square image $400$ pixels wide, as above.
\grtrans{} was configured to take $2000$ steps, which according to
\citet{dexter2016} should net a relative accuracy for total flux at
$10^{-3}$ level. Similarly, \am{} was constrained to take steps of at
most $\Delta\lambda/\lambda_\text{max} = 3\times 10^{-3}$, which should
guarantee a relative accuracy of better than $10^{-3}$ by the convergence
results above.

\begin{figure*}
    \includegraphics[width=\textwidth]{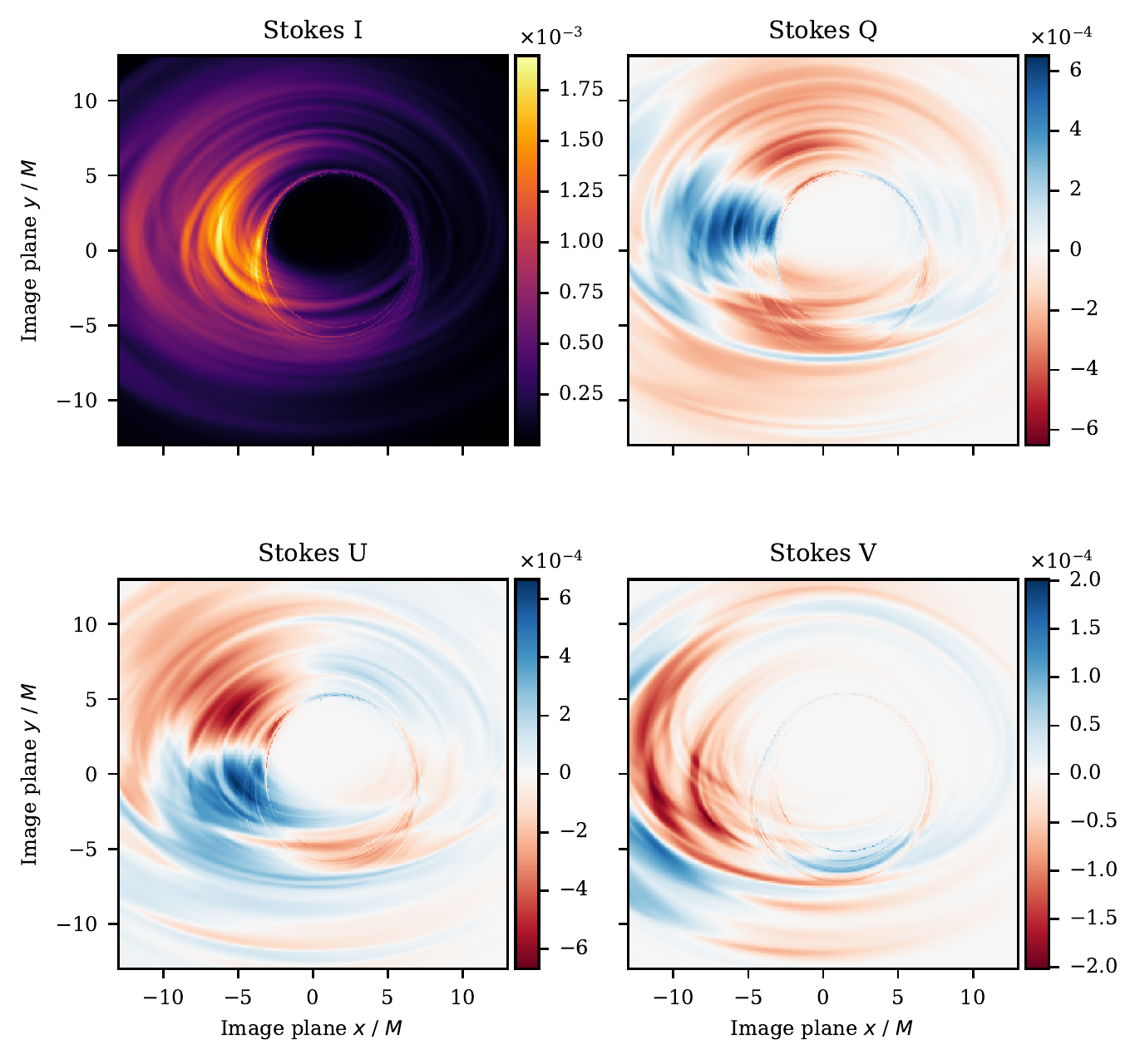}
    \caption{Observed specific Stokes intensities for an accretion flow around 
    a rotating black hole with a mass $M=4\times 10^6\,\Msun$, dimensionless
    spin parameter $\chi=0.9375$ and an accretion rate of
    $\dot{M} = 2.49\times10^{-11}\,\Msun/\text{yr}$.
    The mock observation was computed at $\nu=230\times 10^9\,\text{Hz}$,
    using \am{}.
    The $x$- and $y$-axis are in units of $GM/c^2$, whereas the intensity values
    are given in cgs units of
    $\text{erg}\,\text{s}^{-1}\,\text{Hz}^{-1}\,\text{cm}^{-2}\,\text{sr}^{-1}$.
    }\label{fig:arc_harm}
\end{figure*}

\begin{figure*}
    \center{\includegraphics[width=\textwidth,height=0.95\textheight,keepaspectratio]{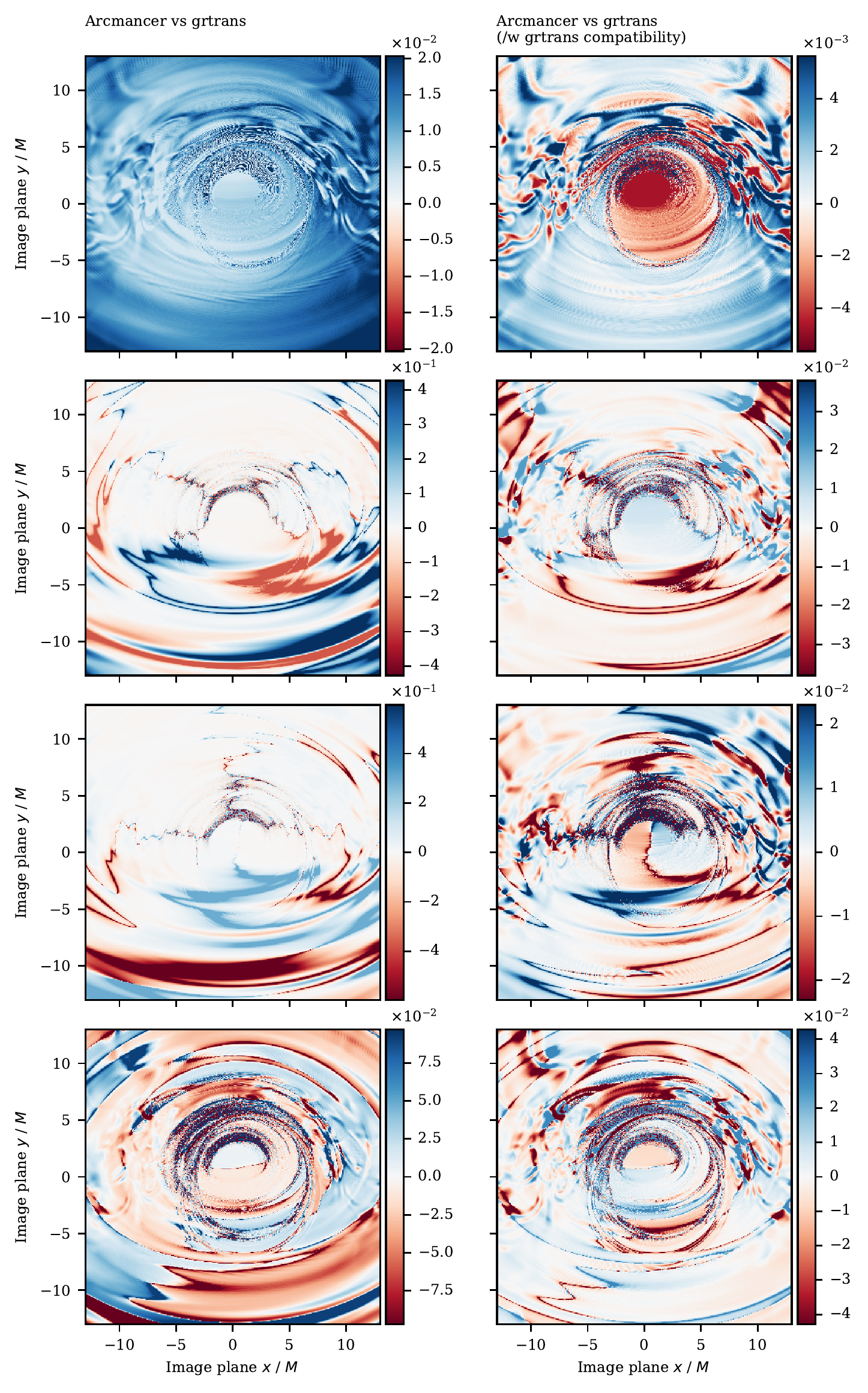}}
    \caption{\emph{Left column:} Pixel-by-pixel relative difference
    between the Stokes intensities computed using \am{} and \grtrans{}, winsorized to 95th
    percentile. \am{} results computed using exact Bessel functions (see text).
    \emph{Right column:} Same as above, but using the \grtrans{}
    approximations for the Bessel functions and fundamental constants in \am{}.
    \emph{Note the very different color bar normalization between left and
    right columns.}
    }\label{fig:grtrans-comp}
\end{figure*}

The resulting Stokes intensity maps
computed with \am{} are shown in
Figure~\ref{fig:arc_harm}. In addition,
Figure~\ref{fig:grtrans-comp} shows the relative differences in the Stokes
intensities as computed by \am{} versus \grtrans{}.
In the left panel of Figure~\ref{fig:grtrans-comp}, we see that
the unpolarized intensity predicted by \am{} is consistently higher, and there
is a clear difference in the polarized results, especially in the $Q$
and $U$ components.

The reason for this
discrepancy was traced to two separate numerical issues. Firstly, \grtrans{} uses
values for the gravitational constant $G$ and Boltzmann constant
$k_B$ that were truncated to three significant figures,
while in \am{} the CODATA 2014 \citep{codata} values are used up to the
known experimental precision.
Secondly, \grtrans{} uses an approximation for computing the cylindrical Bessel
functions used in the relativistic thermal synchrotron radiation model,
equations (B4) and (B14) in \citet{dexter2016}. The \grtrans{} code, as well as
some other codes, such as \citep{moscibrodzka2018}, use first order
approximations for the cylindrical Bessel functions, but at least in this
example case the approximations are not always valid throughout.

If the same physical
constants and Bessel function
approximations are used in \am{}, the agreement both in unpolarized and
polarized intensities is excellent, as can be seen in the right panel of
Figure~\ref{fig:grtrans-comp}.
The unpolarized and polarized total intensities agree with \grtrans{} to a
relative level of $\sim 10^{-3}$,
and the pixel-by-pixel errors are below percent level on average.
There are a small number of high difference outliers located either at regions
where the absolute intensity values are very small, or at the
strongly lensed rings of emissivity. The former outliers are caused by
numerical noise and the latter mainly by spatial sampling noise, since the small
scale structure in these rings is not resolved while simultaneously the emission
is highly boosted, amplifying the differences.
However, it can be seen from the results in Figures~\ref{fig:convergence} 
and the numerical total fluxes tabulated in Table~\ref{tb:total_fluxes} 
that these
pixels make no significant difference in the observed integrated fluxes.

\begin{table}
    \caption{Total integrated fluxes as computed with \am{} in the test
    scenario depicted in Figures~\ref{fig:arc_harm} and~\ref{fig:grtrans-comp},
    \emph{relative to the values obtained with \grtrans{}}.}
    \label{tb:total_fluxes}

    % Taulukossa pyöristämätön kommentoituna, ja näkyvissä viiteen merkitsevään
    % numeroon pyöristetty
    \begin{center}
    \begin{tabular}{cccc}
         $I$ & $Q$ & $U$ & $V$ \\ \hline\hline
         \multicolumn{4}{l}{\am{}} \\
         %1.00825765 &  0.98674832 &  1.16706022 &  1.00536923 \\
         1.0083 &  0.98675 &  1.1671 &  1.0054 \\ \hline
         \multicolumn{4}{l}{With \grtrans{} compatibility} \\
         %1.00089596 &  1.0029475  &  0.99690254 &  1.0023228 \\
         1.0009 &  1.0029 &  0.99690 &  1.0023 \\ \hline
         \multicolumn{4}{l}{With no polarization} \\
         %1.12642151 & -0.         & -0.         & -0.        \\
         1.1264 & 0         & 0         & 0        \\ \hline
    \end{tabular}
    \end{center}
\end{table}

For an interesting test case, we also ran the same test scenario with
all polarization effects disabled. That is, we set
$j_{\{Q,U,V\}}=\alpha_{\{Q,U,V\}}=r_{\{Q,U,V\}}=0$ in $\evec$ and $\mulmat$ so that
only the unpolarized degrees of freedom were propagated. As shown in
Table~\ref{tb:total_fluxes}, the resulting
total flux is $\sim 13\%$ higher than in the polarized case. This
suggests that creating mock observations of unpolarized flux can be
misleading if polarization effects are completely ignored.

\section{Applications}\label{sc:applications}

The capability of \am{} to compute radiation transfer through an
emitting and absorbing relativistic fluid (plasma) was showcased in
the previous section. In the following, we present further applications
of \am{} in different scenarios. The focus is on leveraging the
capability of \am{} to work with all kinds of emitting and absorbing
surfaces, both moving and stationary.

\subsection{Effects of thin accretion disk geometry}\label{sc:thinthick}

\am{} makes it easy to compute mock observations of emitting surfaces
with different user defined geometries. Here this feature is
demonstrated through a toy model by computing the changes on the observed
spectropolarimetric features caused by varying the opening half-angle $\beta$ of a
geometrically thin but optically thick accretion disk around a Kerr
black hole.

Often \citep[e.g.][]{vincent2011,psaltis2012,bambi2012,dexter2016} a thin disk is modeled
in mock observation simulations as an infinitely thin equatorial plane
around the black hole. For $\alpha$-disk models
\citep{shakura1973,novikov1973}, this is in many cases a satisfactory
approximation. This is because the maximal angle made by the disk photosphere and
the symmetry plane is $\beta \sim \arctan(0.2 \dot{m})$, computed in the
Schwarzschild coordinates, where $\dot{m} = \dot{M}/\Medd$ is
the black hole accretion rate in units of the Eddington accretion rate
$\Medd$.
However,
for an accretion rate of $\dot{m}=0.3$, this maximal angle is already
$\beta\sim 4^\circ$, which can be expected to have observable
consequences. This is since the maximal $\beta$ in the
Shakura--Sunyaev solution is found at $r=(27/2)M$, where $M$ is the black
hole mass, which is in the bright inner region of the disk.

Instead of the geometry of the $\alpha$-disk model, which has a
photospheric surface profile dependent on the accretion rate, we use a
disk defined by a hyperbolic surface in the outgoing Kerr--Schild
coordinates,
\begin{equation}
    S(t,x,y,z) = \frac{x^2+y^2-a^2}{r_\text{min}^2} -
    \frac{z^2}{r_\text{min}^2\tan^2\beta} - 1,
\end{equation}
where $a = M\chi$ is the normalized angular momentum of the black hole,
$\beta$ is the half-opening angle of the hyperboloid and $r_\text{min}$ sets the inner
boundary of the disk, here fixed to the innermost stable circular orbit
(ISCO) of the black hole. The choice of this surface is motivated by
the intention to investigate only the effects of geometry on the
observable properties, while keeping the emission properties of the
disk otherwise fixed.

To compute the mock observation, we first set up an image plane with physical dimensions $x,y\in[-40M,40M]$ at
a distance of $r=10^4 M$ (in BL coordinates). From this surface, null geodesics were
propagated backwards until they intersected the disk surface or the
event horizon. A \codeil{PolarizationFrame} was parallel transported
with the geodesic to enable Faraday rotation effects to be captured.
Points that hit the disk were given a blackbody spectrum $B_\nu(T)$ with a temperature $T$ matching
the Novikov--Thorne disk model \citep{novikov1973,page1974}, using a
mass $M=10\,\Msun$, accretion rate $\dot{m}=0.3\,\Medd$
and a dimensionless viscosity parameter $\alpha=0.1$.
The intensity and the linear polarization of the point were computed based on
the electron scattering atmosphere model given in \citet{chandrasekhar1960},
using the impact angle $\theta$ between the geodesic and the disk normal,
computed in the rest frame of the rotating disk surface. The exact solution
requires solving an integral equation. We instead used the Padé approximants
\begin{align}
    \frac{I_\nu(\mu)}{S_\nu} &= \frac{1+2.3\mu-0.3\mu^2}{2\times1.19167}
    \label{eq:es-int} \\
    P &= 0.117126\, \frac{1+14.9165\mu-15.8923\mu^2}{1+22.2420\mu+44.8893\mu^2}, \label{eq:es-pol}
\end{align}
where $\mu=\cos(\theta)$,
for the intensity $I_\nu$ normalized by the source function $S_\nu$ (in this example, 
$S_\nu=B_\nu(T)$), and polarization \edit1{fraction $P$}, respectively.
Both approximations are accurate to within $2\%$ over the
range $\mu\in[0,1]$.
It should be noted that the combination of a blackbody spectrum and a beamed
intensity profile is not fully self-consistent, since a genuine blackbody
emitter is isotropic. However, the combination serves to illustrate the effects
of an anisotropically emitting surface. In addition, the spectral shape for thin
accretion disks around stellar mass black holes is in any case well described
using a \emph{diluted} blackbody \citep{davis2005}.

\begin{figure*}
    %\mbox{\includegraphics[width=0.5\textwidth]{fig/thin_disk_image}%
    %\includegraphics[width=0.5\textwidth]{fig/thick_disk_image}}
    \includegraphics[width=\textwidth]{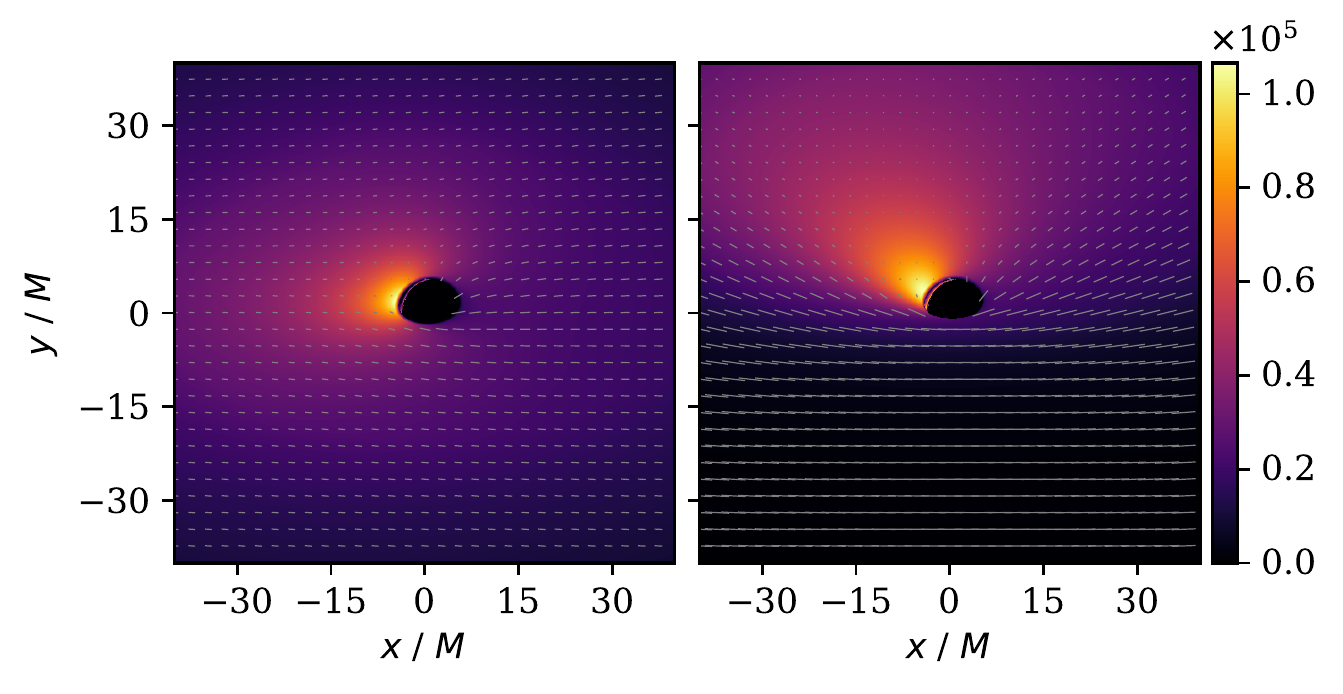}
    \caption{
        Specific intensity maps at $\nu=8\times10^{17}\,\text{Hz}$ of a thin
        Novikov--Thorne model around a $10\,\Msun$ Kerr black hole with a
        dimensionless spin parameter $\chi=0.7$. 
        The disk model is computed with $\alpha=0.1$, and
        $\dot{m}=0.3$, and the observer inclination is $i=60^\circ$ from
        the disk symmetry axis. The disk opening angles are $\beta=0.001^\circ$
        and $\beta=25^\circ$ for the left and the right panels, respectively.
        The direction of observed linear polarization is shown by the gray
        lines, with the degree of linear polarization proportional to
        the length of the line. Both panels are computed with a
        resolution of $600$ pixels per side.
        The intensities are in cgs units, i.e.\
        $\text{erg}\,\text{s}^{-1}\,\text{cm}^{-2}\,\text{Hz}^{-1}\,\text{sr}^{-1}$.
    }
    \label{fig:disk_angle_images}
\end{figure*}

\begin{figure*}
    \includegraphics[width=\textwidth]{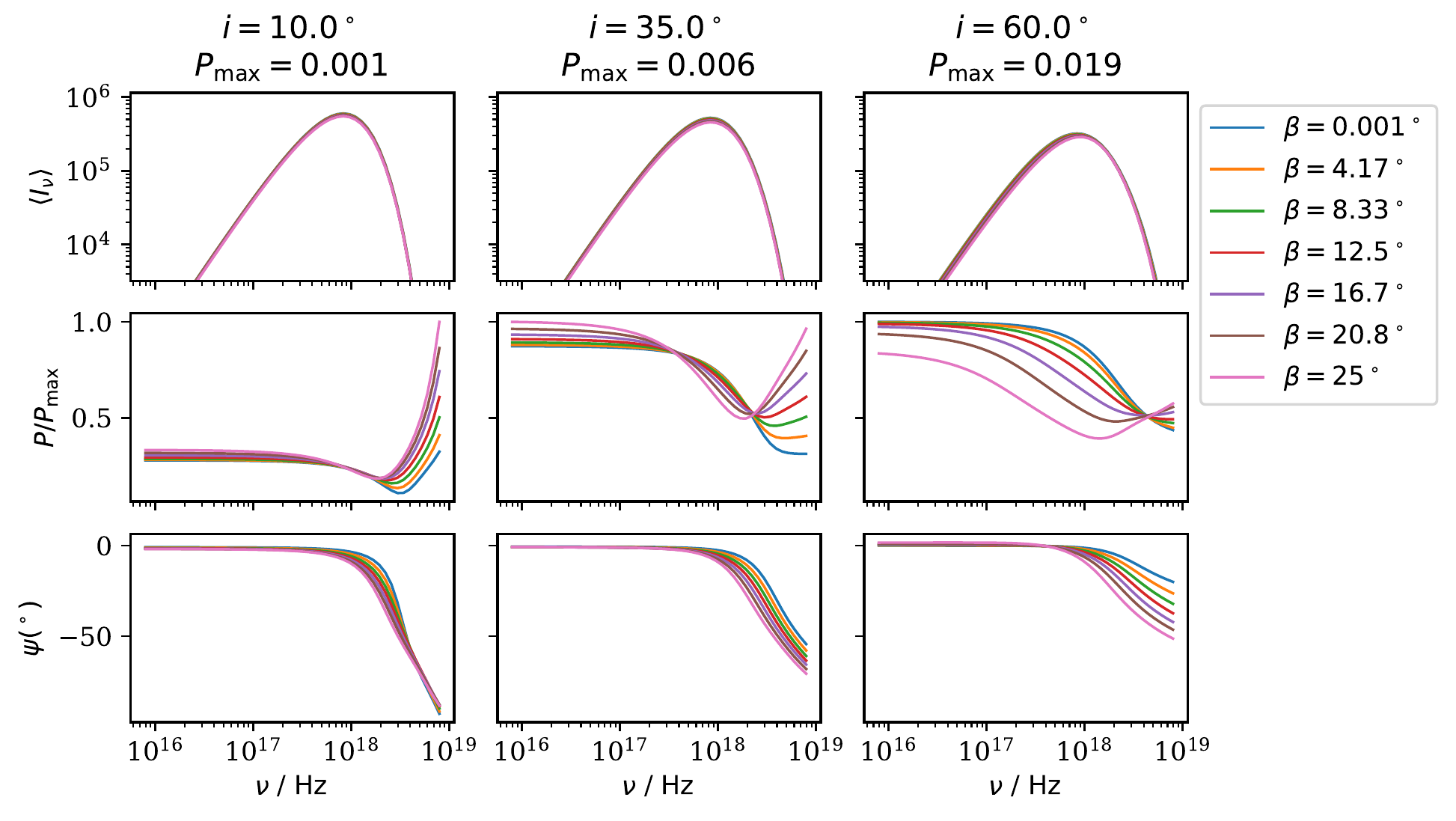}
    \caption{
        The mean intensity in cgs units (\emph{top panel}), normalized degree of polarization
        (\emph{middle panel}) and the polarization angle (\emph{bottom
        panel}) as a function of frequency,
        obtained from the integrated Stokes intensities of mock images of a Novikov--Thorne disk
        around a Kerr black hole, with parameters as in Figure~\ref{fig:disk_angle_images}.
        The lines correspond to the different disk opening angles $\beta$,
        shown in the legend.
    }
    \label{fig:disk_angle_spectrum}
\end{figure*}

To construct the image from these data, instead of running full radiation
transfer, the radiation was assumed to propagate in vacuum. This is not
a particularly good assumption physically, since the thin disks are
expected to have a tenuous, hot coronae
\citep[e.g.][]{liang1977,czerny1987}, but it was made so as to not add
additional uncertainties and keep the focus
on the effects of changing disk geometry.
The values of the
intensity and polarization were
directly transferred to the image plane after scaling the intensity by
the redshift factor and rotating the polarization to match the rotation
of the parallel transported \mbox{\codeil{PolarizationFrame}.} This computation
was repeated for seven values of the disk opening angle from
$\beta=0.001^\circ$ to $\beta=25^\circ$ and three observer inclination
angles $i=10^\circ$, $35^\circ$ and $60^\circ$.
The results are collected in Figures~\ref{fig:disk_angle_images} and
\ref{fig:disk_angle_spectrum}, which
show mock images of the two extreme cases ($\beta=0.001^\circ$,
$\beta=25^\circ$) and the polarization spectra for all the computation
runs.

The effect of the disk opening angle
is clearly seen in Figure~\ref{fig:disk_angle_images}, which shows
specific intensity maps as seen by an observer at an inclination of
$i=60^\circ$ for the extreme opening angles of $\beta=0.001^\circ$ and
$\beta=25^\circ$. The intensity patterns differ significantly, with most of
the emission coming from the opposite side of the disk for the
disk with the larger opening angle. In addition, the structure of the
ring caused by radiation that has traveled around the black hole once
is noticeably changed by the increased disk thickness
\citep[cf.][]{luminet1979}.
Despite the visual differences,
Figure~\ref{fig:disk_angle_spectrum}
shows that the shape of the spectra
obtained from the integrated emission is hardly changed at all, and as
such the shape of the observed spectrum is not very sensitive to the disk
geometry in this example.

Figure~\ref{fig:disk_angle_images} also shows a
significant difference in polarization patterns, with the large opening
angle disk exhibiting a large asymmetry between the upper and lower halves
of the mock observation image. This is caused by a purely geometrical
effect, wherein the geodesics emanating from the opposite side of the
disk from the observer's point of view are more closely aligned with the
local disk surface normal. For the geodesics coming from the observers
side of the disk, the situation is the opposite. The polarization
\edit1{fraction} of the electron scattering atmosphere model is strongly dependent
on the angle of the geodesic with respect to the disk normal, with
stronger polarization for lower incidence angles.
The graphs of the degree of polarization,
\begin{equation}\label{eq:poldegree}
    P = \frac{\sqrt{Q^2+U^2+V^2}}{I},
\end{equation}
and the polarization angle,
\begin{equation}\label{eq:polangle}
    \psi = \frac{1}{2}\arctan(U/Q),
\end{equation}
in
Figure~\ref{fig:disk_angle_spectrum} show that unlike for intensity,
the polarization asymmetry does not average out. Indeed, for the largest
observer inclination ($60^\circ$) shown in
Figure~\ref{fig:disk_angle_spectrum}, we see that there is a strong
dependency of the degree of net polarization on the disk opening angle
$\beta$. A similar but weaker effect is seen also for the observer inclinations
$i=10^\circ$ and $i=35^\circ$. Figure~\ref{fig:disk_angle_spectrum} also
shows the behavior of the net polarization angle $\psi$. With all
observer inclinations, a similar behavior of rotation of the
polarization angle at high photon energies is seen. However, for these
model parameters, the rotation mainly occurs at the high energy end of
the spectrum, where the exponential cutoff makes the effect hard to
observe in practice.

The changes in polarization with observation frequency described above, for
$\beta\sim 0$, are consistent with those of \citet{schnittman2009}, who
studied an infinitely thin disk using a Monte--Carlo approach.
However, for a physically more realistic result, the accretion disk
corona as well as the radiation returning and reflecting to the disk
need to be taken into account, as in \citet{schnittman2010}.
In addition, here we have shown that the geometry of the optically thick part
of the disk cannot be neglected, which is an assumption used in
\cite{schnittman2010}.
Combining the effects of the geometry with the effects of the corona and the
returning radiation is straightforward using \am{}, and will be
investigated in a future work.

\subsection{Neutron stars}\label{sc:neutron_stars}

Another natural application of user-definable surfaces is the imaging
of neutron stars. A solid surface is an excellent approximation for the
radiating atmosphere of a neutron star, since the atmospheric thickness
is on the order of $\sim 10\,\text{cm}$, whereas the radii of the neutron
stars are in the range $\sim 10\,\text{km}$
\citep[see e.g.][for a review]{potekhin2014}. Thus terminating geodesics on the top of
the atmosphere, and using a separate atmospheric model to provide the
(angle-dependent) specific intensity and polarization as initial conditions is
an attractive possibility.

The use of a numerical geodesic propagation code such as \am{} is further
warranted due to the fact that a rotating neutron star is not exactly spherical
but oblate, and the space-time near the star cannot be
exactly described by the Kerr metric
\citep{stergioulas2003,bradley2009,urbanec2013}. Both complications are difficult to
take into account when using fully analytic approaches, such as in e.g.\
%long list\citet{pechenick1983, strohmayer1992, page1995,  miller1998, weinberg2001, poutanen2003, viironen2004, poutanen2006, lamb2009b}, 
%short list:\citet{pechenick1983, strohmayer1992, miller1998, poutanen2003},
\citet{pechenick1983, strohmayer1992, miller1998, poutanen2003},
and \citet{lamb2009a},
where the neutron star is
modeled as a spherical surface in a Schwarzschild space-time. The reason is
two-fold: the intersections of geodesics with the oblate surface are much more
involved to compute (but not impossible, see \citealt{morsink2007, lo2013,
miller2015, stevens2016}),
and since the Carter's constant \citep{carter1968} of the Kerr solution is not available,
the geodesics themselves cannot be analytically solved even in
quadrature.  Another benefit of
using a fully covariant approach throughout is that the pitfalls of
trying to combine special relativistic and general relativistic effects
separately in an ad hoc way (as done in e.g.\ \citealt{lo2013})
are avoided. For example, see \citealt{nattila2017} and \citealt{lo2018}
for a thorough discussion of an error in the calculation of the observed
flux in the ad hoc approach that has gone undetected for years.
Finally, incorporating polarization in an analytic geodesic
propagator is only possible for Kerr (and Schwarzschild) space-times, but
even then it is not trivial \citep[see][]{viironen2004,dexter2016}. However, 
polarization data for this application is critical, since for small
hot spots there is a severe degeneracy in the unpolarized pulse profile
between the spot colatitude $\theta_s$ and the observer inclination $i$
\citep{poutanen2003}.

In this section, we use \am{} to assess the effects of the oblateness of the
neutron star surface and the deviation of the neutron star space-time from the
simple Schwarzschild space-time on the radiative transfer calculation.
For this purpose, we use the AlGendy--Morsink
(AGM) form of the Butterworth--Ipser space-time (\citealt{algendy2014}, and see
also Appendix~\ref{app:agm}).
The AGM space-time describes the surroundings of a rotating neutron star, taking
into account the oblate shape of the star. The space-time is parametrized by the
dimensionless rotational parameter $\bar{\Omega}=\Omega R_e^{3/2} M^{-1/2}$, and
the compactness parameter $x = M/R_e$,
where $\Omega$ is the angular velocity of the rotation as seen by a distant
observer, and $M$ and $R_e$ are the mass and the equatorial radius of the star,
respectively. The oblate shape of the star is obtained from
equation~\eqref{eq:app-oblate-shape}.

As an example case, we studied a rotating neutron star with a
mass $M=1.6\,\Msun$, equatorial radius $R_e=12\,\text{km}$ and a rotational
frequency of $\nu=700\,\text{Hz}$, with $\Omega = 2\pi\nu$.
The high value of the spin was chosen to
accentuate the effects of oblateness, yielding from
equation~\eqref{eq:app-flattening} a flattening of $f=1-R_p/R_e\sim0.09$, where
$R_p$ and $R_e$ are the polar and equatorial radii of the star, respectively.
However, the high spin value is still within the observed range for
neutron stars \citep{hessels2006}. Similarly, the mass and the radius are well
within the observed and inferred limits \citep{steiner2016,ozel2016,alsing2017}.
Using \am{}, we computed surface maps of flux and polarization
characteristics, again assuming that the emission originates from an electron
scattering atmosphere, using equations \eqref{eq:es-int} and \eqref{eq:es-pol}.
This is a good approximation for neutron stars where the emission originates from
thermonuclear outbursts on the surface
% long list:
%\citep{london1986, lapidus1986, pavlov1991, madej1991, madej2004, majczyna2005, suleimanov2011, suleimanov2012, nattila2015, medin2016}.
% short list:
%\citep{london1986,lapidus1986, pavlov1991, madej1991, majczyna2005, suleimanov2011, medin2016}.
% ultra-condensed:
\citep[see e.g.][and the references therein]{suleimanov2011}.
However, we note that the results can be extrapolated on a more qualitative
level to shock-heated accretion-powered hot spots as well \citep[see e.g.][]{basko1976, 
lyubarskii1982, viironen2004}.
Otherwise, the radiation transfer is computed as in Section~\ref{sc:thinthick}.

\begin{figure*}
    \includegraphics[width=\textwidth]{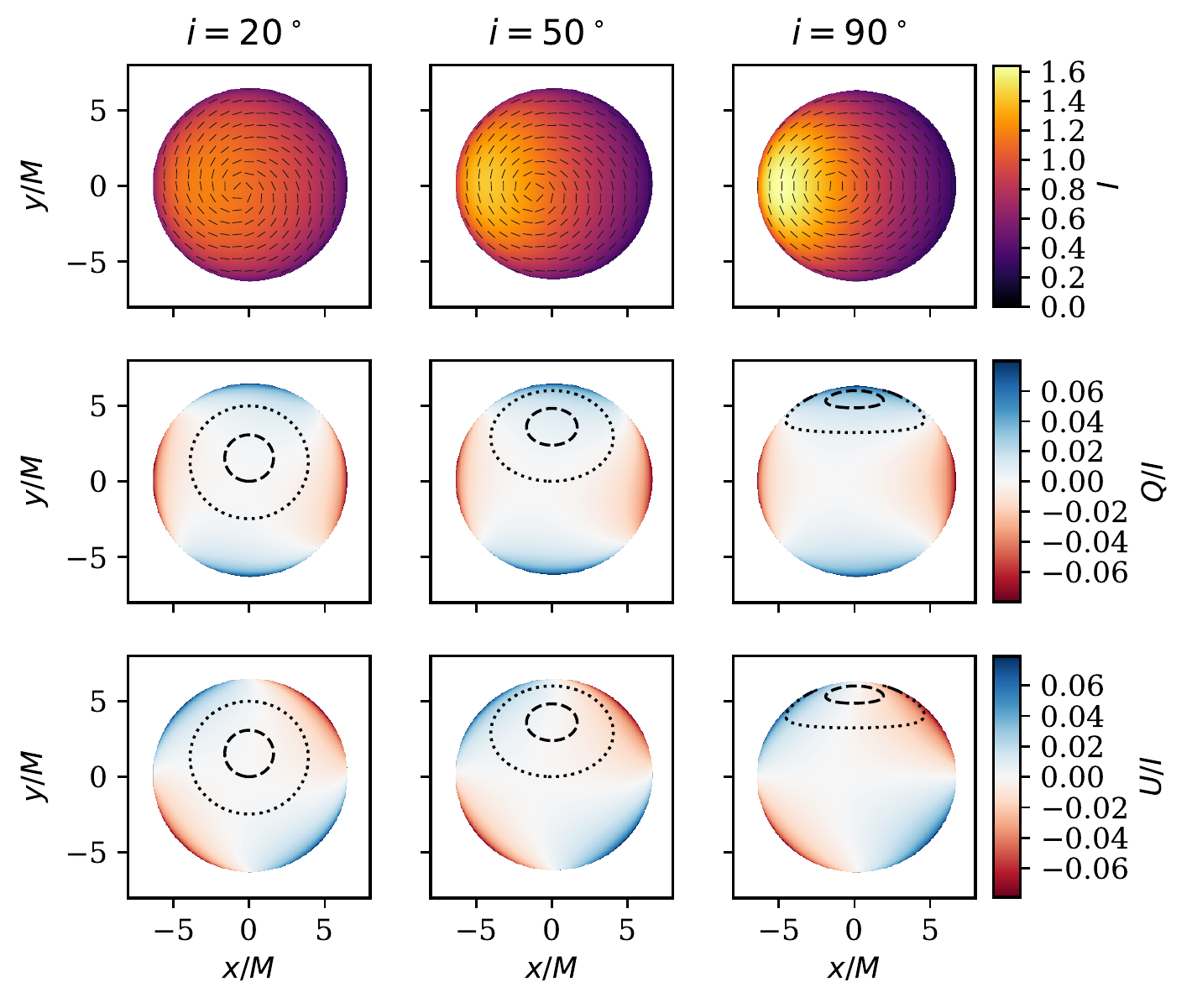}
    \caption{
    Surface maps of an oblate rotating neutron star with an equatorial
    radius $R_e=12\,\text{km}$, mass $M=1.6\,\Msun$ and rotational frequency
    $\nu=700\,\text{Hz}$, computed with \am{}. The top row shows
    $I_\nu/S_\nu$, or specific intensity normalized with the source function (see
    eq.~\ref{eq:es-int}) as a color map,
    while the direction of linear polarization is
    indicated by black lines.
    Middle and bottom row show the linear
    polarization fractions $Q/I$ and $U/I$, respectively.
    The dashed and dotted lines indicate contours of constant colatitude of
    $20^\circ$ and $50^\circ$, respectively (cf.\
    Figure~\ref{fig:pulse_profiles}).
    Columns from left to right show results with observer inclinations of
    $i=20^\circ$, $50^\circ$ and $90^\circ$ with respect to the rotational axis of the star.
    }
    \label{fig:agm_values}
\end{figure*}

\begin{figure*}
    \includegraphics[width=\textwidth]{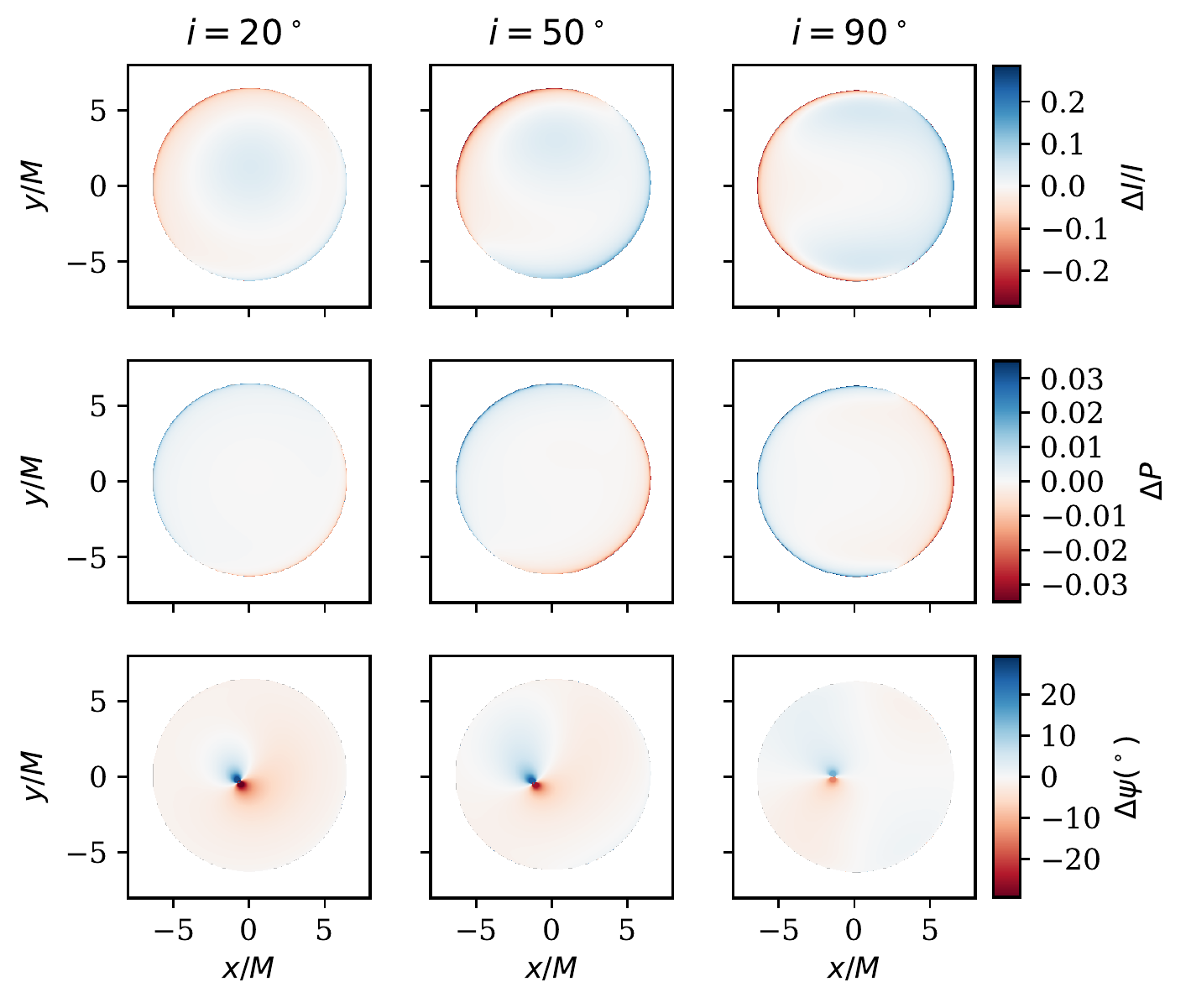}
    \caption{
        Surface maps for a rotating neutron star with $R_e=12\,\text{km}$,
        $M=1.6\,\Msun$ and $\nu=700\,\text{Hz}$, computed with \am{}. The maps indicate relative
        differences between a solution using the AlGendy--Morsink space-time versus using the
        Schwarzschild space-time. The same oblate shape of the neutron star is
        used for both metrics.
        The rows show the relative difference in normalized intensity
        (see eq.~\ref{eq:es-int}) $I_\nu/S_\nu$ (top
        row), degree of polarization $P$ (middle row) and
        polarization angle $\psi$ (bottom row).
        Columns from left to right show results with observer inclinations of
        $i=10^\circ$, $50^\circ$ and $90^\circ$.
    }
    \label{fig:agm_obl_sch_diff}
\end{figure*}

\begin{figure*}
    \includegraphics[width=\textwidth]{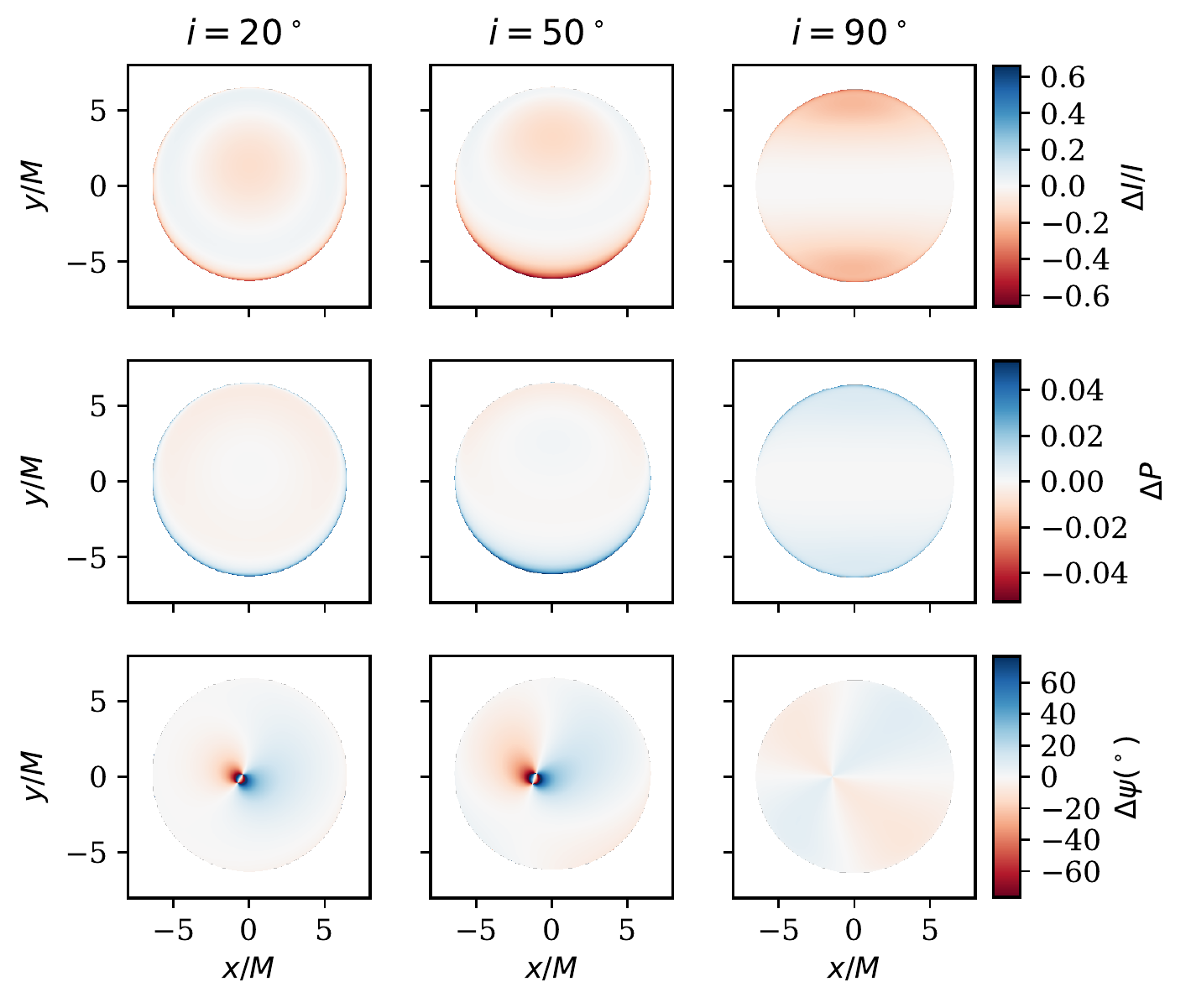}
    \caption{
        Same as Figure~\ref{fig:agm_obl_sch_diff}, but showing the relative
        differences between an oblate neutron star surface, corresponding to the
        rotational rate of $\nu=700\,\text{Hz}$,
        and a spherical surface, both in the Schwarzschild space-time. 
        The oblate surface has a flattening of $f\sim0.09$, so that the polar
        radius is $\sim91\%$ of the equatorial radius. The spherical surface
        has a radius equal to the equatorial radius of the oblate surface.
    }
    \label{fig:obl_sph_sch_diff}
\end{figure*}

\begin{figure*}
    %\center{
    %\includegraphics[width=\textwidth,height=0.45\textheight,keepaspectratio]{fig/pulse_profile_20}
    %\includegraphics[width=\textwidth,height=0.45\textheight,keepaspectratio]{fig/pulse_profile_50}
    %}
    \mbox{\includegraphics[width=0.5\textwidth]{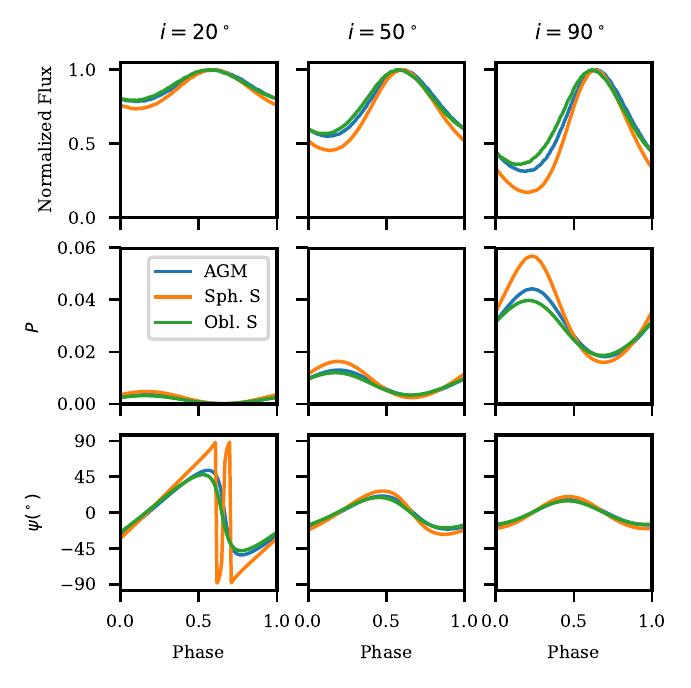}
    \includegraphics[width=0.5\textwidth]{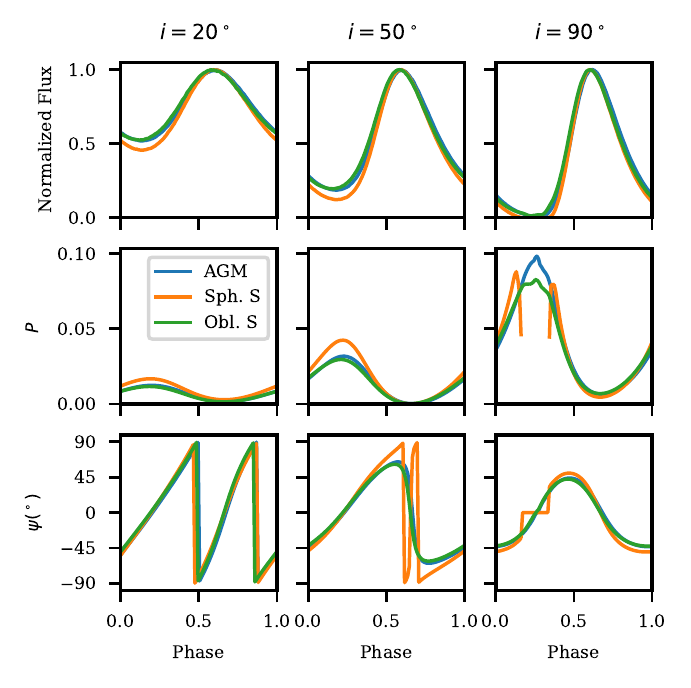}}
    \caption{
        Pulse profiles for thermonuclear-powered hot spots rotating with the
        surface of a neutron star with $R_e=12\,\text{km}$,
        $M=1.6\,\Msun$ and $\nu=700\,\text{Hz}$. The spot has an angular radius
        of $5^\circ$, and a constant colatitude of either $20^\circ$
        (\emph{left panel})
        or $50^\circ$
        (\emph{right panel}). In both panels, the top row shows
        the integrated flux normalized to the maximum value, assuming a constant spot temperature.
        The middle and bottom rows show the integrated polarization fraction and
        polarization angle, respectively.
        Three cases are shown: oblate star with the AGM space-time (blue curve),
        oblate star with the Schwarzschild space-time (green curve) and spherical
        star with the Schwarzschild space-time (orange curve).
    }
    \label{fig:pulse_profiles}
\end{figure*}

The computations were repeated three times: for an oblate star
using the AGM space-time (hereafter, AGM+Obl), for an oblate star using
the Schwarzschild space-time (Sch+Obl) and for a spherical star in the
Schwarzschild space-time (Sch+Sph).
The results are shown in Figures \ref{fig:agm_values},
\ref{fig:agm_obl_sch_diff}, \ref{fig:obl_sph_sch_diff} and
\ref{fig:pulse_profiles}.
Figure~\ref{fig:agm_values} shows the behavior of $I_\nu/S_\nu$, the specific intensity 
divided by the source function, and polarization over the star
surface, computed using the AGM space-time at observer inclinations of
$i=20^\circ$, $50^\circ$ and $90^\circ$. The combination of Doppler boosting
and the strong angular dependence of the electron scattering atmosphere yield an
intensity that varies significantly over the neutron star surface. The net polarization
is high only near the edges, where the impact angle is large.
Figure~\ref{fig:agm_values} also shows two possible paths of constant colatitude
hot spots, assuming that the star is rotating around the vertical axis. From the
figure it is then easy to appreciate that a rotating hot spot should exhibit
large periodic variation in the observed polarization angle. This variation can be
directly seen in Figure~\ref{fig:pulse_profiles}, which is consistent 
with the results in \citet{viironen2004}.

Figure~\ref{fig:agm_obl_sch_diff} shows the difference in normalized intensity
$I_\nu/S_\nu$, degree of
polarization and polarization angle when the computation is performed using the AGM
metric versus the Schwarzschild metric (i.e., AGM+Obl vs.\ Sch+Obl). The effects
of the rotation become significant only near the star, and consequently the
differences stay moderate for the most part, below $\sim 10\%$ for the intensity
and below $\sim 2\%$ for the degree of polarization. There are areas of larger
differences, but these are concentrated on the edges of the visible disk of the
neutron star, and their total area is small. The differences in the polarization
angle are larger, around $\sim 20^\circ$ overall. There are very
large differences near the point where the radiation was emitted towards the
zenith in the frame of the neutron star surface, but this area corresponds to
vanishing polarization, and as such these differences are unobservable.

In contrast, Figure~\ref{fig:obl_sph_sch_diff} displays the same differences but
between computations performed using an oblate star versus a spherical star, both in
Schwarzschild space-times (i.e., Sch+Obl vs.\ Sch+Sph). The spherical star was
given a radius equal to the equatorial radius of the oblate star.
In this case, the differences in all quantities are
much more pronounced. This is not a surprise, since a change in the shape of the
star affects the redshift distribution on the surface due to variations in local
surface gravity. These differences become even more evident when one looks at
Figure~\ref{fig:pulse_profiles}, which shows two examples of light curves and
the time varying degree of polarization and polarization angle for a rotating
hot spot. Firstly, the pulse and polarization profiles closely match those obtained by
\citet{viironen2004} for the Sch+Sph case, and confirm that the
observational degeneracy in unpolarized flux between observer inclination and spot
colatitude is lifted by the polarization measurements.
However, from the figure it can be seen that the approximation of a spherical
star produces results that differ significantly both in intensity and
polarization properties from the result obtained using an oblate surface.
In addition, there is a small but non-negligible difference between the results
obtained using the AGM metric versus a plain Schwarzschild metric. Similar
results for the unpolarized flux were obtained already in \citet{psaltis2014},
although for an isotropically emitting atmosphere.

Based on our preliminary study, we can conclude that the error introduced when
computing the polarization angle with the Schwarzschild space-time approximation is
largest when both the observer inclination and the spot colatitude are small.
Likewise, the error in the degree of polarization is largest when the spot is
near the equator, i.e.\ spot colatitude is close to $\sim 90^\circ$.
We conclude that to obtain polarized pulse profiles that are accurate to below
the $\sim1\%$ level, it is necessary that the rotation and the geometric shape of the
star are both accurately modeled.
In practice this means that the analytic
results based on the Schwarzschild space-time such as in e.g.\
%long list:
%\citet{pechenick1983, page1995, strohmayer1992, miller1998, weinberg2001, poutanen2003, viironen2004, poutanen2006, lamb2009b, lamb2009a, lo2013} 
%slightly shorter list: 
% \citet{miller1998, weinberg2001, poutanen2003, viironen2004, poutanen2006, lamb2009b, lamb2009a, lo2013} 
% even shorter list:
\citet{weinberg2001, viironen2004, lamb2009b, lo2013}
and \citet{miller2015}
should be used with caution.
However, to actually reach $\sim1\%$ level of accuracy, other systematic errors
in e.g.\ modeling the emission from the neutron star and its surrounding
environment would also need to be resolved.

\subsection{Binary black holes}

To further explore the possibility to use arbitrary metrics and multiple
surfaces which may also move, we consider a toy model of an accreting
black hole with a secondary black hole companion. To set up the problem,
we use an approximative metric, constructed using the outgoing
Kerr--Schild form of the Kerr metric, equations \eqref{eq:app-KS-sum} and
\eqref{eq:app-cart-KS-out}.
In the limit of zero spin, $a=0$, the metric is
\begin{equation}
    g_{ab} = \eta_{ab} + F(M,r) l_a(\vct{x}) l_b(\vct{x}),
\end{equation}
where $M$ is the mass of the black hole, $\vct{x}=(x,y,z)$,
\begin{align}
    F(M,r) &= -\frac{2M}{r} \\
    l_a(\vct{x}) &= \left(-1,\frac{x}{r}, \frac{y}{r}, \frac{z}{r}\right)
\end{align}
and $r^2 = x^2+y^2+z^2$.
To this form we add a perturbation representing
a distant second black hole moving at a slow coordinate velocity. Taking
the spatial position of the second black hole to be a function
$\vct{x}_2(t)$ of the coordinate time, we set
\begin{equation}\label{eq:bbh-metric}
    \begin{split}
        g_{ab} = \eta_{ab}
        &+ F(M_1,r) l_a(\vct{x}) l_b(\vct{x}) \\
        &+ F(M_2,r_2) l_a(\vct{x}-\vct{x}_2) l_b(\vct{x}-\vct{x}_2)
    \end{split}
\end{equation}
where $M_1$ and $M_2$ are the masses of the primary and secondary black holes, respectively,
$\vct{x}_2$ is the spatial position of the secondary black hole,
and $r_2^2 = (x-x_2)^2 + (y-y_2)^2 + (z-z_2)^2$.

The metric~\eqref{eq:bbh-metric} \emph{is not a solution} of the
vacuum Einstein field equations, for which no exact \emph{dynamic} binary black hole
solution is known.\footnote{However, there are a number of known \emph{static} 
solutions for multiple black holes. Examples include
\emph{any} number Schwarzschild black holes in a collinear
configuration \citep{israel1964}, or any number 
of maximally charged Reissner--Nordström
black holes in \emph{any} configuration \citep{papapetrou1945,majumdar1947}. 
}
For example, the metric~\eqref{eq:bbh-metric} does not contain the
gravitational wave component expected from the motion of multiple
gravitating bodies.
However, in the limit $M_2\sim 0$ and
$\ud\vct{x}_2(t)/\ud t \sim 0$ for all $t$, the perturbation caused by the
secondary is small and remains small, and the gravitational wave component
is negligible, and in this sense the approximation is reasonable.
For black hole binary systems with smaller separations and larger velocities,
a discretized metric from a full GR simulation should be used with \am{}.
More accurate analytical approximations, such as from \citet{mundim2014},
can also yield satisfactory accuracy \citep{sadiq2018}, but the approximative
analytical metrics are on the other hand algebraically complex.

We set \edit1{$M_1=5\times 10^6\,\Msun$}, representing a supermassive black hole (SMBH) and
$M_2=0.05 M_1$, which falls into the intermediate mass black hole (IMBH) range.
Otherwise we set up the system as in Section~\ref{sc:thinthick}, by
placing an image plane with physical dimensions $[-50 M_1,50 M_1]^2$ at
$R=10^7 M_1$ at an observer inclination of $i=60^\circ$.
The secondary black hole is set on rectilinear coordinate path 
$\vct{x}_2(t) = \vct{x}_{2,0} + t\vct{v}_2$,
where 
\begin{align}
    \begin{split}
        \vct{x}_{2,0} &= 
        (L\sin i-\delta \cos i \cos \phi_v,\,
        -\delta \sin \phi_v,\, \\
        &\quad\quad L\cos i + \delta \sin i \cos \phi_v)
    \end{split} \\
    \vct{v}_2 &= v_2 \left(
    -\cos i \sin \phi_v,\, \cos \phi_v,\, \sin i \sin \phi_v
    \right).
\end{align}
Here $\delta$ is the apparent offset of the secondary's path, $L =
\sqrt{r_{2,0}^2-\delta^2}$ is the orthogonal distance from the primary
to the image plane, $r_{2,0}$ is the minimum distance between
the black holes, $v_2$ is the velocity of the secondary and $\phi_v$ is
the angle between the path of the secondary and the image plane
$x$-axis. For this particular example, we set $r_{2,0} = 10^3 M_1$, $\delta=-4 M_1$,
$\phi_v = 25^\circ$, and $v=3.162\times 10^{-2}$.
These initial conditions approximately correspond to an IMBH on a
circular orbit around an SMBH, a situation that could possibly follow
a merger of a more massive galaxy with a dwarf galaxy \citep{graham2013}.
In order to have something to make a mock observation of,
the primary black hole was given an infinitely thin Novikov--Thorne
accretion disk, with $\alpha=0.1$ and an accretion rate $\dot{m}=0.01$
in units of the Eddington accretion rate.
Geodesics were then propagated backwards in time from the
image plane starting at $150$ different values of the coordinate time,
evenly distributed in $[-3000 M_1, 3000 M_1]$.
For each set of geodesics, mock images, integrated fluxes, and
polarization fraction and angle were computed.

\begin{figure*}
    \includegraphics[width=\textwidth]{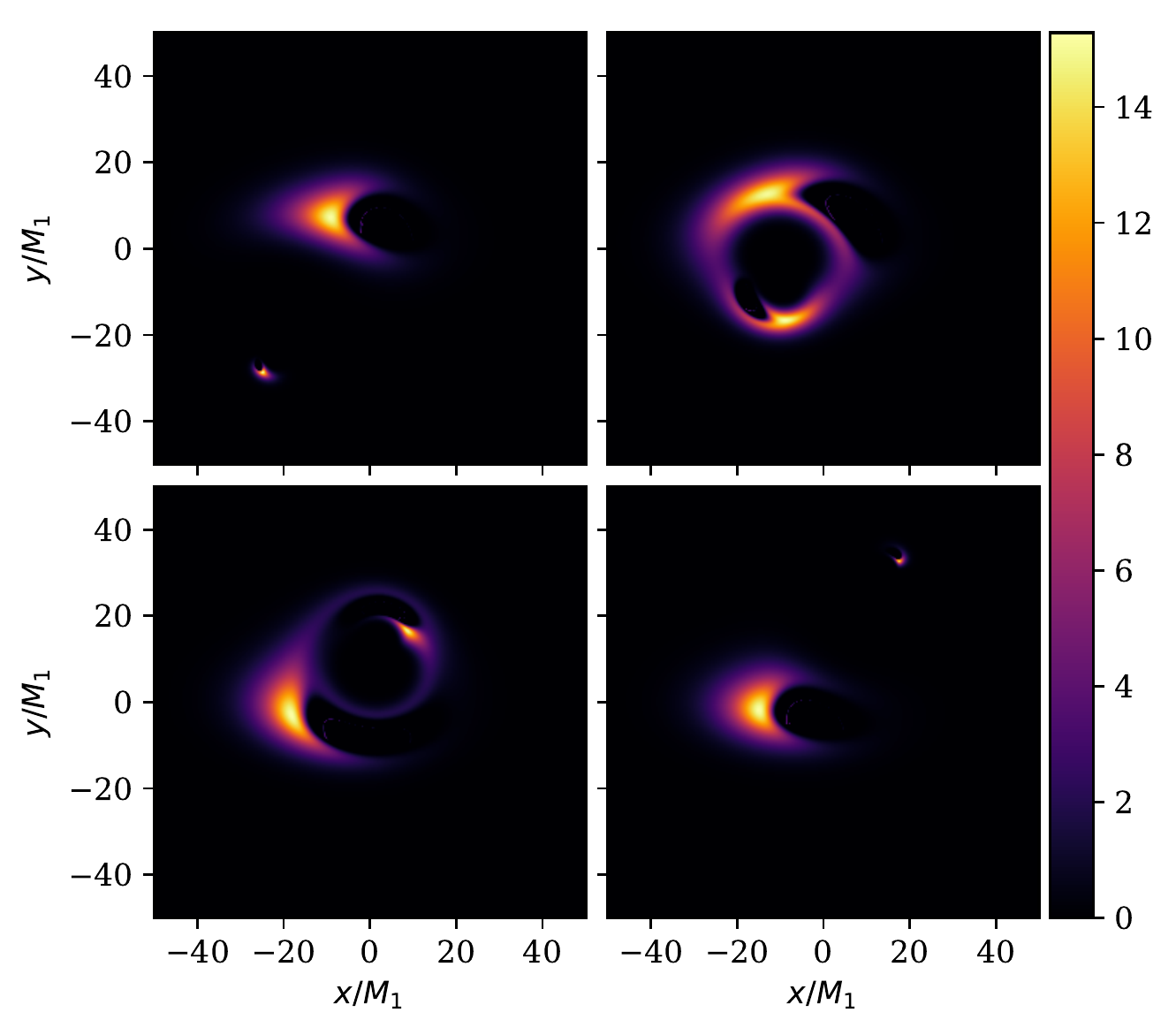}
    \caption{Resolved mock observations at $\nu=5\times 10^{16}\,\text{Hz}$ of a simulated accreting binary
    black hole system with primary of mass $M_1=5\times10^6\Msun$ and a
    secondary of $M_2 = 0.05M_1$ (see text for orbital parameters). The primary
    has an geometrically thin and optically thick Novikov--Thorne accretion
    disk. The images show intensity maps of the unpolarized Stokes $I$
    from the accretion disk, with the coordinate time increasing from
    left to right and top to bottom.
    The $x$- and $y$-axis are in units of $GM_1/c^2$, and the
    intensities are in cgs units.}
    \label{fig:bbh_frames}
\end{figure*}

\begin{figure*}
    \begin{center}
    %\mbox{%
    %\includegraphics[width=0.5\textwidth]{fig/binary_bh_lightcurve}%
    %\includegraphics[width=0.5\textwidth]{fig/binary_bh_lightcurve_lowinc}}
    \includegraphics[width=\textwidth,height=0.45\textheight,keepaspectratio]{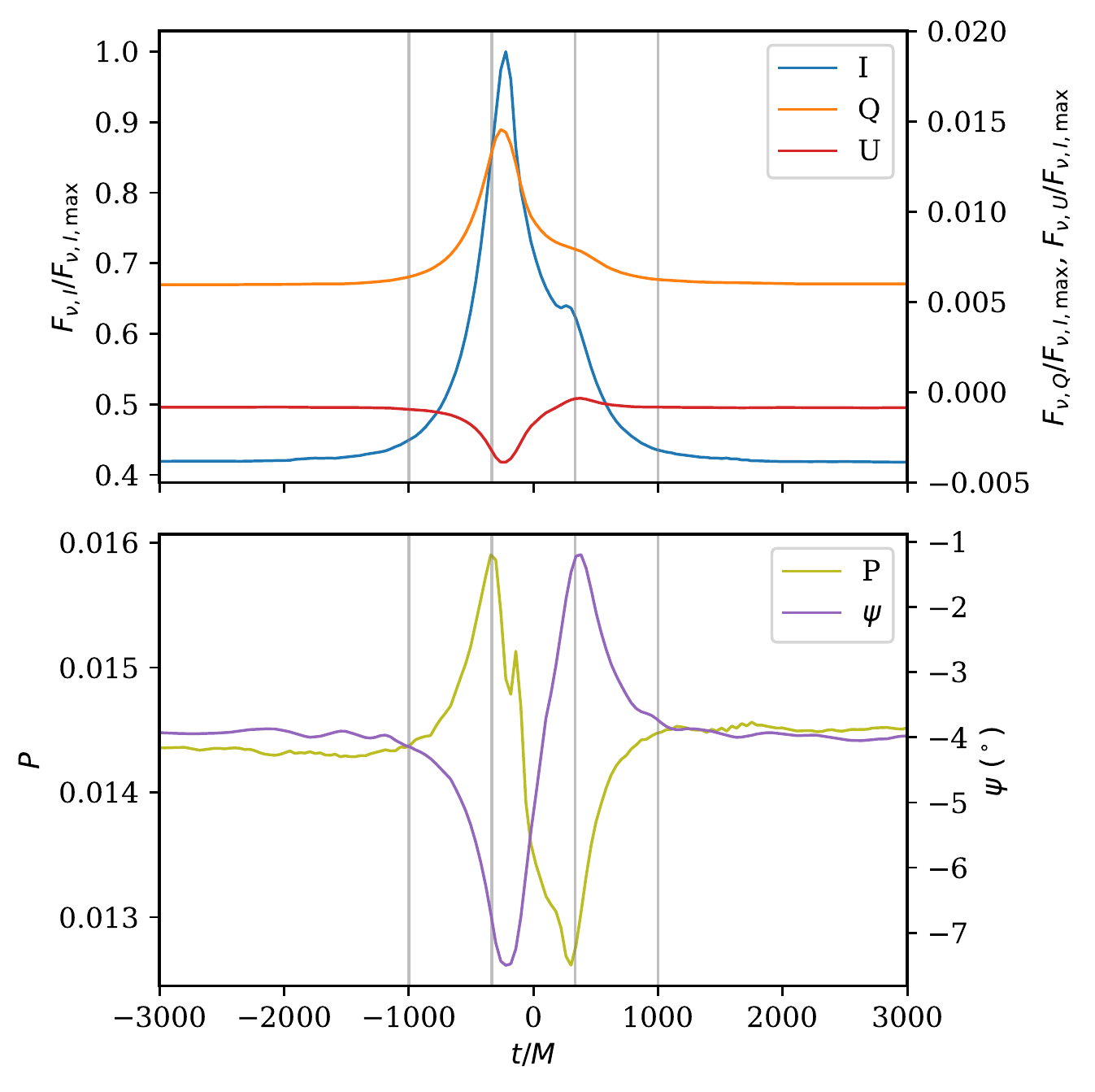}
    \includegraphics[width=\textwidth,height=0.45\textheight,keepaspectratio]{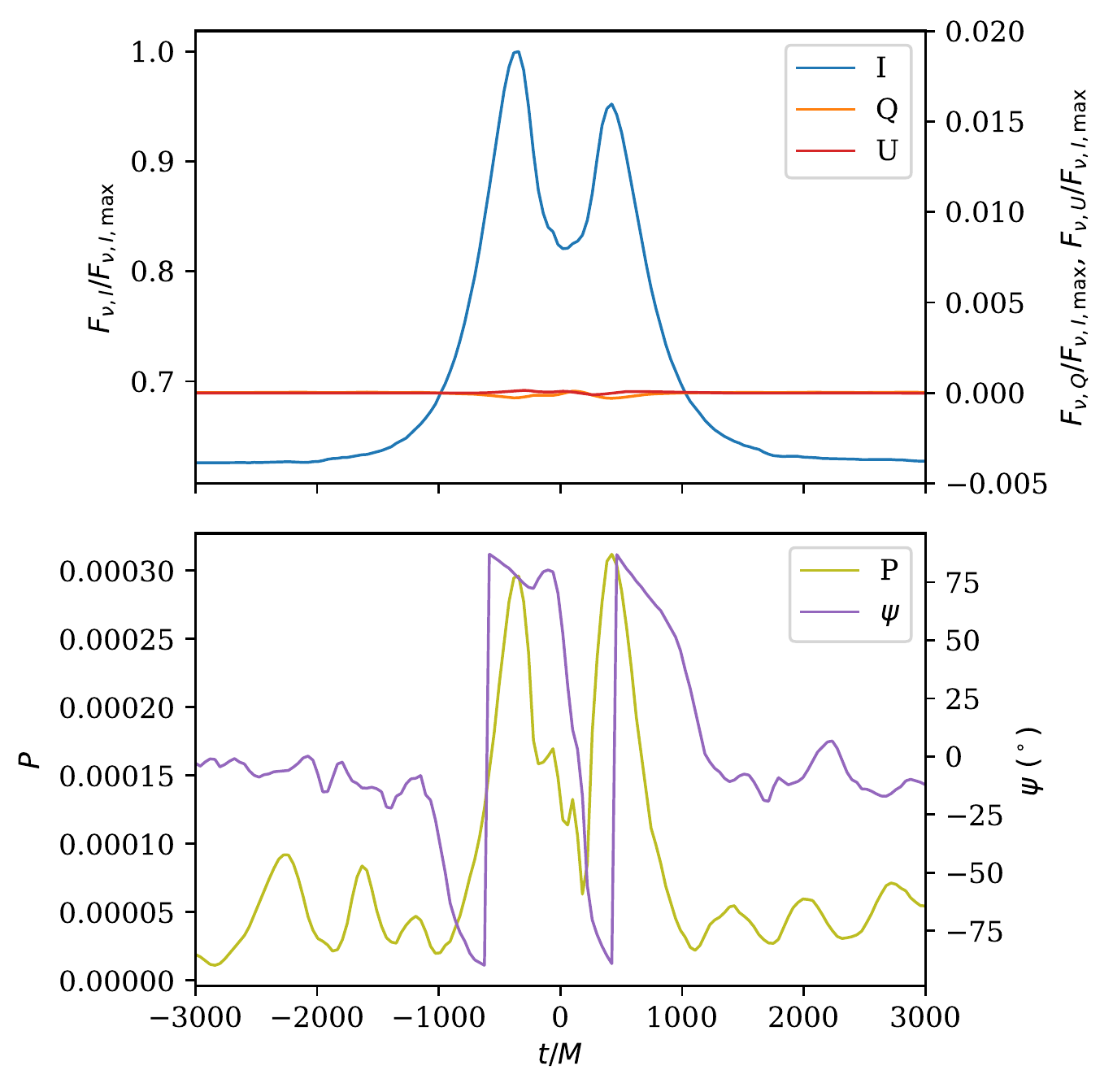}
    \caption{
        Polarized light curves at $\nu=5\times10^{16}\,\text{Hz}$ of a simulated accreting binary black
        hole system (see text), for observer inclinations of $i=60^\circ$
        (\emph{top panel})
        and $5^\circ$
        (\emph{bottom panel}).
        The horizontal time axis is shown in units of $GM_1/c^3$.
        Gray vertical lines indicate the times
        of the resolved frames shown in Figure~\ref{fig:bbh_frames}.
        Top figures show the integrated polarized fluxes of Stokes components $I$, $Q$ and
        $U$ relative to the respective maximum flux, as a function of coordinate time.
        Bottom figures show the polarization fraction $P$ and polarization angle $\psi$ as a
        function of coordinate time.
        For the bottom panel,
        the $Q$ and $U$ curves have been smoothed with a 9-point 2nd
        order Savitzky--Golay filter \citep{savitzky1964} to reduce the
        numerical noise in $P$ and $\psi$ curves caused by the very small net
        polarization.
    }
    \label{fig:bbh_lightcurve}
    \end{center}
\end{figure*}

The resulting light curves are shown in 
Figure~\ref{fig:bbh_lightcurve}, with selected resolved frames shown in
Figure~\ref{fig:bbh_frames}.
The main effect of the passing secondary is
a strong enhancement by a factor of $\sim 2$ of the observed flux from the accretion disk of
the primary, caused by gravitational lensing.
The flux curve has a clearly non-sinusoidal shape, where after the main
peak, there is a pronounced shoulder.
The double-peaked structure results from the lensing of the two main visible arcs
of the primary accretion disk. The difference is that the major peak has a larger contribution
from the Doppler boosted side of the primary accretion disk.
This asymmetry is also clearly visible in the
curves for polarization fraction $P$ and angle $\psi$, see equations \eqref{eq:poldegree} and \eqref{eq:polangle}.
The polarization
fraction curve shows a clear two-peaked shape, with a sharp peak
followed by a sharp trough. The polarization angle mirrors this
behavior, with a maximum rotation of $\sim 7^\circ$.

The light curves were also computed with a smaller value of observer
inclination of $i=5^\circ$, also shown in Figure~\ref{fig:bbh_lightcurve}.
The results show that
the double-peaked structure of the light-curve is more evident towards
$i=0$, whereas the relative amount of polarized flux grows significantly smaller.
Both effects are to be expected considering the increased symmetry when
$i\fromto 0$. The changes in degree of polarization and the polarization
angle are more pronounced as well, but due to the negligible relative amount of
polarized flux, these are unlikely to be detectable.

Over longer timescales, the recurrent lensing by the secondary produces
a periodic signal, which can be clearly observable over the baseline
brightness of the primary accretion disk, as seen from the
Figure~\ref{fig:bbh_lightcurve}. However, the signal is strongly
non-sinusoidal, which may reduce observability in periodicity searches
based on periodogram techniques. On the other hand, if a series of
accretion disk lensing events was observed, it should be possible to use
lensing mock observation simulations to obtain independent constraints
on the secondary black hole mass and the orbital parameters.

Finally, we note the interesting fact that the double-peaked light curve is
reminiscent of the light curve of the periodic binary blazar OJ~287,
which exhibits a long succession of
strongly non-sinusoidal double-peaked outbursts every $\sim 12$ years
\citep{sillanpaa1988,valtonen2008}. Many different physical mechanisms
for the outbursts have been proposed, such as
tidally enhanced accretion rate \citep{sillanpaa1988},
accretion disk impacts \citep{lehto1996,pihajoki2016} and changes in the
relativistic jet geometry \citep{katz1997,villata1998}.
Accretion disk lensing adds yet another possible outburst mechanism.

\section{Conclusions}\label{sc:conclusions}

In this paper, we have presented \am{}, a \cpp{}/Python library for
numerical computation of curves and tensor algebra in arbitrary
Riemannian and semi-Riemannian spaces. The library is designed to be easy to
extend as well as to incorporate in new or existing applications.
\am{} offers several novel
and useful features. Many of these are built around \am{}'s seamless support of
multiple simultaneous coordinate charts. For example, \am{} offers automatic
conversion of coordinates and tensors of arbitrary rank between different
charts. This conversion works even in the case where no explicit transformation
is provided between two given charts, as long as the graph formed by all the
available charts and transformations contains a path connecting the two charts.
The coordinate chart support is also used in the library to automatically pick the
numerically most appropriate chart to integrate the equations of motion for curves. 
\am{} can also be used for numerical tensor algebra, supporting all
usual tensor operations for tensors of arbitrary rank and dimension. In
addition, \am{} can parallel propagate arbitrary tensors and
user-defined quantities along curves.
In the four dimensional case,
the \am{} library contains a suite of tools designed for solving problems of
general relativistic radiative transfer using the ray-tracing approach.
These include a coordinate-invariant method for generating image planes, easy
interface for supplying user-defined fluid and radiation models and a support
for geometric objects, which can be used for example to model radiating surfaces
or define limits of computational domains.
For convenience, the library can also work with either Lorentzian metric
signature.
All of these features are thoroughly documented in the code itself, in the
documention automatically generated from the code and
via several example applications provided with the library.
In addition, the library website\footnote{\url{\amurl}} provides instructions
for installation and getting started.

In this presentation of the \am{} library we have
included a description of the internal workings of the code, as well as
numerous tests of the accuracy of the code.
The \am{} code was found to
fulfill theoretically expected convergence properties. It also
produced very similar results as an existing ray-tracing code \grtrans{},
when applied to a demanding mock observation scenario of a hot accretion
flow around a Kerr black hole. Notably, the code tests demonstrated
the critical importance of choosing the right coordinate system for the
chosen problem, and the necessity of being able to change coordinate
systems during the numerical evolution of the problem.

The code tests were followed by applications of the code to a variety of
astrophysical scenarios, showcasing the flexibility of the \am{} code.
The first example application was an
investigation of the effect of the opening angle of an optically thick
but geometrically thin accretion disk to its observable properties.
While the unpolarized flux was essentially invariant with respect to the
disk opening angle, the degree of polarization and angle of polarization
were found to significantly depend on it.

In the next application, \am{} was used to study observational
properties of hot spots on rotating neutron stars. We compared three
different commonly used models. In the AGM+Obl model, the neutron star
surface was modeled using a physical oblate shape with an exterior
space-time metric that took the oblate shape and rotation into account.
This physically accurate model was compared with two more approximate
models: Sch+Obl in which the exterior metric was changed to a Schwarzschild
metric, and Sch+Sph in which in addition the shape of the neutron star
was taken to be spherical. Our results show that the oblate shape of the
star makes a large contribution to the shape of both polarized and
unpolarized flux curves and must be taken into account. However, we also
find that in order to obtain polarized light curves with accuracies better than
$\sim1\%$ level, the Schwarzschild metric must be abandoned in favor of more
physically motivated alternatives.

Finally, we used \am{} to create mock observations of an accreting
binary black hole system, consisting of a primary black hole with an
accretion disk, together with an orbiting secondary black hole.
The application demonstrated how \am{} can
easily handle a more complex geometry where light rays can be terminated
on multiple surfaces (two event horizons and one accretion disk), some
of which may move (the secondary event horizon). We found that the
lensing caused by the orbiting secondary can produce clearly observable
changes in the observed polarized and unpolarized flux from the
accretion disk of the primary black hole. However, the changes in
polarized flux are strongly dependent on the observer inclination
due to the geometry of the simple $\alpha$-disk model we used.

In the future, we expect to use the \am{} library to build a comprehensive
radiative transfer application for investigating complex accretion flows around
compact objects.
In addition, the capabilities of the library itself will be extended. Planned
features include built-in support for outputs of other GRMHD codes besides HARM,
support for easy serialization of the code data structures and more built-in
radiation and fluid models. The coordinate chart system will also be enhanced
with support for defining domains for the charts, and improving the numerical
behavior of chart-dependent operations such as curve interpolation.

%We fully expect ray-tracing to become an even more important tool in the near
%future. This development is driven by the increasing observational capabilities,
%especially with respect to polarized light, together with the ongoing prodigious increase
%in computational resources. 
%% Peterin versio
%The ever increasing importance of ray-tracing modelling is driven by the 
%corresponding increase in observational capabilities, especially with respect to 
%polarized light, together with the ongoing prodigious increase in computational 
%resources.
% synteesi
Ray-tracing is expected to become even more important in the future, driven by
the increase in observational capabilities, especially with respect to polarized
light, together with the ongoing prodigious increase in computational resources.
We are confident that \am{} will prove to be a
highly useful and adaptable tool in this upcoming era.

\acknowledgments

PP, MM and PHJ acknowledge support from the Academy of Finland, grant
no.~274931.
This research has made use of NASA's Astrophysics Data System.

% Software acknowledgments go here
\software{
    Eigen (\url{http://eigen.tuxfamily.org/})
    Boost (\url{http://www.boost.org/})
    Pybind11 (\url{https://github.com/pybind/pybind11})
    Numpy \citep{numpy}
    Scipy (\url{https://www.scipy.org/})
    Matplotlib \citep{matplotlib}
          }

%% Appendix material should be preceded with a single \appendix command.
%% There should be a \section command for each appendix. Mark appendix
%% subsections with the same markup you use in the main body of the paper.

\appendix

\section{Differential geometry and General Relativity}\label{app:diffgeo}

The \am{} library has capabilities beyond ray-tracing and radiative transfer
in four-dimensional Lorentzian space-times. The library
offers a variety of tools for computational differential geometry
in Riemannian or semi-Riemannian manifolds of arbitrary dimension,
within the constraints of available memory and computing power. 
In the following, we give a short, self-contained review of the concepts
of differential geometry that are implemented and used in the \am{}
library. For practical reasons the
exposition is kept brief and mathematical details are omitted where
possible. The discussion is styled after a number of texts,
namely
\citet{lee2013-introduction,oneil1983-geometry,lee2006-riemannian} and
\citet{choquet1982-analysis},
in which the interested reader can find the omitted details.

\subsection{Manifolds and coordinate charts}\label{app:manifolds}

The basic building block of differential geometry is the 
\emph{manifold} $M$, which can be intuitively understood as a space which locally
`looks like' $\fR^n$, the $n$-dimensional Euclidean
space. More concretely, each manifold comes with an atlas of
\emph{charts} (coordinate systems) $\phi_i:M\supset U_i\fromto\fR^n$, defined
on open sets $U_i$ of $M$. Using a chart $\phi$, an abstract point
$p\in M$ is transformed into its coordinate representation
$(x^1(p),\ldots,x^n(p))\in\fR^n$, where $x^j = \pi^j\circ\phi$
is the projection to the $j$'th coordinate. We say that the dimension of
$M$ is $\dim(M) = n$, and use the shorthand $\phi=(x^1,\ldots,x^n)$.
A change in coordinates then corresponds to the \emph{transition map} $\phi_i \circ
\phi^{-1}_j:\fR^n\fromto\fR^n$, which changes a tuple of $n$ coordinates
to another tuple of $n$ coordinates, \emph{describing the same point}.
\am{} requires the manifold $M$ to be differentiable, meaning that the
maps $\phi_i \circ \phi^{-1}_j$ are differentiable.
However, in the following we assume \emph{smooth} (infinitely
differentiable) manifolds.

\subsection{Tensors}\label{app:tensors}

The \emph{tangent space} of $M$ at point $p$, written $T_pM$, 
is the space of all vectors tangent to $M$ at $p$. If
$\dim(M)=n$, then $T_pM$ is an $n$-dimensional real vector space.
The dual space of $T_pM$, written $T^*_pM$ and called the \emph{cotangent space}
of $M$ at $p$, is the space of all linear maps
$\omega_p:T_pM\fromto\fR$. These linear maps are called \emph{one-forms}.
For each chart $\phi=(x^1,\ldots,x^n)$, at point $p$ we can define the
vectors $\atp{\partial_i}{p} \defi \atp{\frac{\partial}{\partial x^i}}{p}$,
called \emph{coordinate vectors}. The vector $\atp{\partial_i}{p}$ 
points towards the direction where the
$i$'th coordinate increases at $p$. The complete set
of coordinate vectors, $\{\atp{\partial_1}{p},\ldots,\atp{\partial_n}{p}\}$,
forms a basis for $T_pM$, called the \emph{coordinate basis}. 
Any vector $v\in T_pM$ can be
written in terms of its components in this special basis, as 
$v = v^i\atp{\partial_i}{p}$, where now $v^i\in\fR$ are the components of
$v$ in the chart $\phi$ and the Einstein summation
convention has been assumed.
Likewise, there exists a set of one-forms, $\{\atp{\ud x^1}{p},\ldots,
\atp{\ud x^n}{p}\}$ called the \emph{coordinate one-forms}, which form a
basis for $T^*_pM$, and make it possible to write any one-form
$\theta \in T^*_pM$ in terms of its components as
$\theta=\theta_i\ud x^i$. The coordinate vectors and
one-forms obey $\atp{\ud x^i}{p}(\atp{\partial_j}{p}) = \delta^i_j$,
where $\delta^i_j$ is the Kronecker delta symbol. Together, the
coordinate vectors and one-forms are called the \emph{coordinate
frame}. The tangent and
cotangent spaces can naturally have other bases than coordinate bases as
well (see~\ref{app:local-frames}), but the coordinate basis is the
default basis used in \am{}.

A rank $(k,l)$ \emph{tensor} can then be defined as 
a multilinear map 
$T:\overbrace{T^*_pM \times \cdots\times T^*_pM}^\text{$k$ times} \times 
\overbrace{T_pM \times \cdots\times T_pM}^\text{$l$ times} \fromto \fR$.
In general, the factors $T_pM$ and $T^*_pM$ may be in any order, and the
ordering is important, but for convenience, we will use the ordering
given above. A rank $(0,0)$ corresponds to scalar quantity, and rank $(1,0)$
tensors are equivalent to vectors, and rank $(0,1)$ tensors to one-forms. 
The components of a tensor at $p$ in the coordinate basis can be directly found
from
\begin{equation}\label{eq:app-tenscomp}
    T^{i_1\cdots i_k}_{i_{k+1}\cdots+i_{k+l}} = T(
    \atp{\ud x^{i_{k+1}}}{p}, \ldots, \atp{\ud x^{i_{k+l}}}{p},
    \atp{\partial_{i_1}}{p}, \ldots, \atp{\partial_{i_k}}{p}).
\end{equation}
Using the components, a tensor can be locally defined as an expansion
\begin{equation}\label{eq:app-tensor-expansion}
    T=%\sum_{i_1=1}^{n}\cdots \sum_{i_{k+l}=1}^{n} 
    %T\ix{^{i_1\cdots i_k}_{i_{k+1}\cdots+i_{k+l}}}
    T^{i_1\cdots i_k}_{i_{k+1}\cdots+i_{k+l}}
    \atp{\partial_{i_1}}{p}\otimes\cdots\otimes \atp{\partial_{i_k}}{p} \otimes
    \atp{\ud x^{i_{k+1}}}{p} \otimes\cdots\otimes \atp{\ud x^{i_{k+l}}}{p}.
\end{equation}
If instead of the chart $\phi=(x^1,\ldots,x^n)$ we wish to use another
(overlapping) chart $\psi=(y^1,\ldots,y^n)$ to represent the tensor $T$, 
the components $T^{i_1\cdots i_k}_{i_{k+1}\cdots+i_{k+l}} \in \fR$
of a tensor must be transformed accordingly. The new components turn out
to be
\begin{equation}\label{eq:app-tensor-coord-transform}
    %T\ix{^{i_1\cdots i_k}_{i_{k+1}\cdots+i_{k+l}}} = 
    T^{i_1\cdots i_k}_{i_{k+1}\cdots+i_{k+l}} = 
    T^{j_1\cdots j_k}_{j_{k+1}\cdots+j_{k+l}}
    J^{i_1}_{j_1} \cdots J^{i_k}_{j_k}
    (J^{-1})_{i_{k+1}}^{j_{k+1}} \cdots (J^{-1})_{i_{k+l}}^{j_{k+l}},
\end{equation}
where $J^i_j = \parfrac{(y^i\circ\phi^{-1})}{x^j}$ are the components of
the Jacobian $J$ of the function $\psi\circ\phi^{-1}$ at point $p$,
and similarly for the inverse of the Jacobian $J^{-1}$.

The definitions above generalize to vector, one-form and tensor
\emph{fields}, which can be understood as functions which, for each
point $p$ of $M$, pick a specific vector, one-form or tensor,
respectively. 
%A vector field is as such a section of the \emph{tangent
%bundle} $TM$, which is the union of all tangent spaces, $TM = \cup_{p\in
%M}T_pM = TM$. Similarly a one-form field is a section of the \emph{cotangent
%bundle} and a tensor field is a section of a \emph{tensor bundle}.
%In particular, the coordinate and one-form bases generalize to fields
%collectively called \emph{coordinate frame fields}.
%Despite this generalization, it is important to note especially in a
%computational context that all operations between
%vectors, one-forms and tensors in general are only defined if these
%objects are located (defined) \emph{at the same point}.
In physics-oriented GR literature, all tensorial quantities are usually
tensor fields. In addition, a convention
called \emph{the abstract index notation} \citep{penrosebook1} is often
used. In this convention, for example, a rank $(2,3)$ tensor field $T$ can be written
as $T\ix{^{ab}_{cd}^{e}}$. The number and ordering of the upper and
lower indexes is taken only to signify the number and ordering of the
factors $T_pM$ and $T^*_pM$ in the
definition~\eqref{eq:app-tensor-expansion} of the tensor.
This approach makes it possible to write all coordinate invariant tensor
operations tersely, without specifying any underlying basis. In this
paper, we use the abstract index notation wherever possible.

\subsection{Metric}

A manifold may have a special rank $(0,2)$ tensor field called the
\emph{metric}, usually written $g_{ab}$. The metric defines the inner product
of vectors $\iprod{v^a}{w^a} = g_{ab}v^a w^b$, and consequently a norm
$\norm{v^a} = \sqrt{\abs{g_{ab}v^a v^b}}$ on each
$T_pM$. Intuitively, the metric defines the distance between nearby
points $x$ and $x+\Delta x$ as the norm of the tangent vector approximated
by $\Delta x$. In physics, the components of the metric are often
written in the form of a \emph{line element} $\ud s^2$, essentially 
an expansion in terms of coordinate basis tensors, as
$\ud s^2 = g_{ij}\, \ud x^i \otimes \ud x^j$.
The \emph{signature} of the metric is the pair $(p,q)$ of the number of
positive and negative eigenvalues of the matrix of components of
$g_{ab}$ (in any basis), respectively, assuming $p+q=n$.  If $q=0$, the metric and the
manifold are said to be Riemannian. For signatures $(n-1,1)$ or $(1,n-1)$
the metric and the manifold are said to be Lorentzian. In particular,
General Relativity is defined in terms of a four-dimensional Lorentzian
manifold. \emph{In this paper, Lorentzian metrics are always assumed to
be of type $(1,n-1)$, and in particular for GR this implies the $(+---)$
metric convention.}

A choice of metric also defines a unique metric-compatible zero-torsion
connection $\nabla$, often called the \emph{Levi--Civita connection}.
A connection in general can be roughly said to characterize which vectors of the
tangent spaces of two nearby points are to be considered equal, or
alternatively, to yield a vector field $\nabla_X Y$ describing the rate
of change of the vector field $Y$ in the direction of the vector field
$X$, called the \emph{(natural) covariant derivative of $Y$ with respect
to $X$}. Written in terms of the coordinate vector fields, the connection is
$\nabla_{\partial_i}(\partial_j) = \chr^{k}_{ij}\partial_k$, where
$\chr^k_{ij}$ are the \emph{Christoffel symbols} (of the second kind).
\emph{This work only uses the Levi--Civita connection}, the zero-torsion
property of which can be written as $\chr^k_{ij} = \chr^k_{ji}$, and the
metric compatibility as $\nabla_c g_{ab} = 0$.
In terms of the metric, the Christoffel symbols read
\begin{equation}
    \chr^{k}_{ij} = \frac{1}{2}g^{km}\left(
    \partial_i g_{jm} + \partial_j g_{im} - \partial_m g_{ij}\right).
\end{equation}

The covariant derivative can be generalized for any tensor field $T$.
Using the Christoffel symbols, the covariant derivative of $T$ with
respect to $X$ can be written in component form as
\begin{equation}\label{eq:covar-d}
    \nabla_X T
    = X^i\nabla_{\partial_i} T^{a_1\cdots a_k}_{b_1\cdots b_l}
    = X^i \left(
    \partial_i T^{a_1\cdots a_k}_{b_1\cdots b_l}
    + T^{c_1\cdots a_k}_{b_1\cdots b_l} \chr^{a_1}_{c_1 i}
    + \cdots
    + T^{a_1\cdots c_k}_{b_1\cdots b_l} \chr^{a_k}_{c_k i}
    - T^{a_1\cdots a_k}_{c_1\cdots b_l} \chr^{c_1}_{b_1 i}
    - \cdots
    - T^{a_1\cdots a_k}_{b_1\cdots c_l} \chr^{c_l}_{b_l i}
    \right).
\end{equation}
In the abstract index notation the covariant derivative is written
simply as $\nabla_X T = X^c\nabla_c T^{a_1\cdots a_k}_{b_1\cdots b_l}$.
The term covariant derivative of $T$, without additional qualifiers, is often
used to refer only to the vector field independent part 
$\nabla_c T^{a_1\cdots a_k}_{b_1\cdots b_l}$.
If $\nabla_X T = 0$, we say that $T$ is \emph{parallel (transported)}
along the vector field $X$.
This concept can be extended to the case
where $X$ is a tangent vector field of a curve $\gamma$, see below.

For semi-Riemannian spaces, the metric also divides vectors at a point
into three categories. A vector $v\in T_pM$ is said to be
\emph{timelike} if the inner product $g_{ab}v^a v^b$ is positive,
\emph{null} if it is zero, and \emph{spacelike} otherwise. The same
classification can be extended to vector fields $V$ if the sign of the
inner product is the same everywhere on $M$.

\subsection{Geodesics and other curves}\label{app:geodesics}

For ray-tracing mock observations, the notion of curves and especially
geodesics on a manifolds is essential. A curve $\gamma$ can be thought of as a map
$\gamma:\fR\supset I\fromto M$. The curve defines a vector field $u^a$
through $u(\lambda) = \ud\gamma(\lambda)/\ud\lambda$
along the curve. This vector field can be understood as the velocity vector field
of the curve. The curve itself is a solution of
\begin{equation}\label{eq:curve}
    \derfrac{u^a(\lambda)}{\lambda} = \nabla_{u^b} u^a = u^b\nabla_b u^a.
\end{equation}
If in particular $u^b\nabla_b u^a = 0$, the tangent vector of the curve is
parallel transported along the curve, and we say that the curve is a
\emph{geodesic}. Equation~\eqref{eq:curve} can also be written as
\begin{equation}\label{eq:forced-curve}
u^b\nabla_b u^a = f^a,
\end{equation}
where the vector field $f^a$ is analogous to a force, which
causes the curve to deviate from a straight path along the manifold.
For geodesics, $f^a=0$, corresponding to the
notion of geodesics as the straightest possible paths.
Written in terms of components in a specific chart,
equation~\eqref{eq:forced-curve} reads
\begin{equation}\label{eq:forced-curve-comps}
    \derfrac{u^a(\lambda)}{\lambda} = -\chr^{a}_{bc} u^b u^c + f^a.
\end{equation}
In this form the equation can be solved numerically, as long as the
solution stays within the domain of the chart.

Generic tensorial quantities can also be parallel transported along the
curve. For a rank $k+l$ tensor field $T^{a_1\cdots a_k}_{b_1\cdots b_l}$
the equation of parallel transport is
\begin{equation}\label{eq:T-partrans}
    \nabla_{u^c} T^{a_1\cdots a_k}_{b_1\cdots b_l} = u^c\nabla_c T^{a_1\cdots a_k}_{b_1\cdots b_l} = 0,
\end{equation}
which in component form can be found from equation~\eqref{eq:covar-d}
with the substitution $X = u^a$. Equation~\eqref{eq:T-partrans} can
likewise be directly solved numerically.

Finally, curves for which the tangent vector field $u^a$ is always
timelike, spacelike or null, are called timelike, spacelike or
null, respectively. This characterization is important for geodesics, for which
the tangent vector field can never change their timelike, spacelike or null
character. As such, geodesics always fall in one of these categories.

\subsection{Level hypersurfaces}\label{app:surfaces}

Manifolds may contain many kinds of submanifolds.
A particularly useful class of submanifolds are those defined by level sets
of functions, called \emph{level hypersurfaces}.
If $S:M\fromto\fR$ is a smooth map, then the sets
$S_c \defi S^{-1}(c) = \{p\in M| S(p) = c\}$ are submanifolds of $M$ for
each $c$ in the codomain of $S$ if $\atp{\ud S}{p} \neq 0$ for all 
$p\in S_c$. We can subsume the constant $c$ in the
definition of the function $S$ and take $c=0$, which is always assumed
in the \am{} code. Furthermore, with a slight abuse of notation, we use
the function $S$ defining the hypersurface to refer to the hypersurface
itself.

A level hypersurface divides the manifold $M$ into two disconnected
subsets corresponding to regions where the value of $S$ is negative or
positive. In some cases, these may be conveniently taken to be the
`outside' and the `inside' of a region bounded by the level hypersurface.

\subsection{Local frames}\label{app:local-frames}

Tensor fields can also be expressed in terms of other bases than the
coordinate basis. A \emph{local frame} defined in an open set $U\supset
M$ consists of $n$ vector fields $\{E_1,\ldots,E_n\}$ that form a basis
for $T_pM$ at each $p\in U$. For each local frame, there is a
corresponding \emph{local coframe} of one-form fields
$\{\omega^1,\ldots,\omega^n\}$ for which $\omega^i(E_j) = \kronecker_i^j$.
The components of a rank $(k,l)$ tensor $T^{a_1\cdots a_k}_{b_1\cdots b_l}$ in terms
of these bases can then be defined through (see
equation~\eqref{eq:app-tenscomp})
\begin{equation}\label{eq:app-tensframecomp}
    T^{i_1\cdots i_k}_{j_1\cdots j_l} =
    T(\omega^{i_1},\ldots,\omega^{i_k}, E_{j_1},\ldots,E_{j_l}).
\end{equation}

The transformation from the components of a tensor in a coordinate frame
to components in terms of a local frame can be found by
first writing the local (co)frame in terms of the coordinate frame as
$E_i = \sum_\alpha E^\alpha_i \partial_\alpha$ and $\omega^j =
\sum_\beta \omega^j_\beta \ud
x^\beta$. Using the multilinearity of an arbitrary $(k,l)$-tensor 
$T^{a_1\cdots a_k}_{b_1\cdots b_l}$ in
equation~\eqref{eq:app-tensframecomp} then yields
\begin{equation}\label{eq:app-tenscoordtolocal}
    T^{i_1\cdots i_k}_{j_1\cdots j_l} =
    \sum_{\beta_1}\cdots\sum_{\alpha_l}
    \omega^{i_1}_{\beta_1} \cdots \omega^{i_k}_{\beta_k}
    E_{j_1}^{\alpha_1} \cdots E_{j_l}^{\alpha_l}
    T(\ud x^{\beta_1},\ldots,\ud x^{\beta_k} , \partial_{\alpha_1},\ldots,\partial_{\alpha_l}),
\end{equation}
where $T^{\beta^1\cdots\beta^k}_{\alpha_1\cdots\alpha_l} =
T(\ud x^{\beta_1},\ldots,\ud x^{\beta_k} , \partial_{\alpha_1},\ldots,\partial_{\alpha_l})$ 
are the components of the tensor in the coordinate frame.
From a computational perspective, it is useful to note that the set of
numbers $E_i^\alpha$ can be interpreted as a matrix, for which the
corresponding set $\omega^j_\beta$ forms an inverse matrix. The
equation~\eqref{eq:app-tenscoordtolocal} is seen to resemble the
equation~\eqref{eq:app-tensor-coord-transform}, with these matrices
taking the place of the Jacobians.

Orthonormal local frames define a set of convenient projections of $T_pM$ into
orthogonal complements at each point $p$. A collection of $k$ non-null basis vectors
$\mathcal{K}=\{E_{i_j}\in T_pM|j=1,\ldots,k\}$, defines a projection onto
the subspace of $T_pM$ spanned by $\mathcal{K}$ through
\begin{equation}
    P_\parallel(E_{i_1},\ldots,E_{i_k})^a_b = 
    \sum_{j=1}^{k} \frac{ E_{i_j}^a (E_{i_j})_b }{E_{i_j}^c (E_{i_j})_c}.
\end{equation}
Similarly, a projection onto the orthogonal complement of this subspace is
defined via
\begin{equation}
    P_\perp(E_{i_1},\ldots,E_{i_k})^a_b = \delta^a_b -
    \sum_{j=1}^{k} \frac{ E_{i_j}^a (E_{i_j})_b }{ E_{i_j}^c (E_{i_j})_c }.
\end{equation}

For Lorentzian spaces, direct projections with respect to a null vector $k^a$
do not work. However, given a unit timelike vector $u^a$, a projection onto the space
orthogonal to both $u^a$ and $k^a$ can be given through
\begin{equation}\label{eq:app-screen-proj}
        P_\perp(u,k)^a_b
        = \delta^a_b - \frac{u^a u_b}{u^c u_c} - \frac{s^a s_b}{s^c s_c} 
        = \delta^a_b - \frac{u^a k_b}{K} + \frac{k^a u_b}{K} + \frac{k^a k_b}{K^2},
\end{equation}
where $s^a = P_\perp(u)^a_b k^b$ and $K = u_a k^a$. In four dimensions, the operator
$P_\perp(u,k)^a_b$ is a projection onto a two-dimensional surface on which the
observed components of polarization are defined. In this context,
$P_\perp(u,k)^a_b$ is known as the \emph{screen projection operator}
\citep{gammie2012}.

\subsubsection{Lorentz frames and observers}\label{app:lorentz_frames}

In the case of a four-dimensional Lorentzian manifold, an orthonormal
local frame is often called a \emph{(local) Lorentz frame} or a
\emph{tetrad}. The tetrad can in general consist of any permissible
combination of null, spacelike and timelike vector fields. However,
in this work we use the term Lorentz frame to mean
a mutually orthonormal combination of \emph{one
timelike vector $E_t$
and three spacelike vectors $E_x$, $E_y$ and $E_z$}, defined at a point.
Since parallel transport preserves inner products, a parallel
transported Lorentz frame is still a valid Lorentz frame.

Observers in GR are characterized by a timelike curve $\gamma$, a
\emph{world line}. An observer's rest frame can then be defined as a
Lorentz frame for which the timelike basis vector is given by the
observer's four-velocity, or $E_t = \ud \gamma(\lambda)/\ud \lambda$, and the
spatial triad $\{E_x, E_y, E_z\}$ defines the observer's choice of
spatial coordinate system.
The values of physical quantities as measured by the observer are
obtained by expressing them in the observer's rest frame basis.

For example, an important feature of GR is that angles between vectors at a
point is observer dependent. The angle $\theta$ between vectors $x^a$ and $y^a$
as measured by an observer with a four-velocity $u^a$ is then
\begin{equation}\label{eq:vecangle}
    \cos\theta(u;x,y) = \frac{ -x_\perp^a y_{\perp a} }{%
        \sqrt{\abs{ (x_\perp^b x_{\perp b})(y_\perp^c y_{\perp c}) }}
    },
\end{equation}
where $x_\perp^a = P_\perp(u)^a_b x^b$ 
and $y_\perp^a = P_\perp(u)^a_b y^b$.

\section{Radiative transfer}\label{app:radtrans}

\subsection{Geometric optics}

%For a more detailed account of the geometric optics approximation in GR,
%see e.g.\ \citet[][chap.~22]{mtw}, \citet{gammie2012} and the references
%therein. 
Briefly, a propagating monochromatic radiation front can be modeled as a
congruence of curves, each perpendicular to the surface of constant
phase. This is possible in the limit where, the wavelength of the
radiation is much smaller than the scale of variations in the radiation
front (curvature, amplitude, polarization) and much smaller than the
local `radius of curvature of the space' $\propto
\abs{R\ix{^a_{bcd}}}^{-1/2}$, where $R\ix{^a_{bcd}}$ is the Riemann curvature
tensor.

Furthermore, if the contribution of matter is insignificant, i.e.
$T^{ab}\sim 0$, the result is that the \emph{propagation of radiation can be
modeled by solving an equation of radiative transfer along null
geodesics.} If the integrated contribution of the intervening matter
is non-negligible, the radiation front normals will \emph{not be
geodesics} in
general, and there will be a matter-dependent forcing term $f^a$ in
equation~\eqref{eq:forced-curve} (see e.g.\
\citealt{broderick2003} and \citealt{broderick2004a}).

\subsection{Equation of radiative transfer}

The classical equation of radiative transfer in Cartesian coordinates is
\citep{mihalas1984}
\begin{equation}\label{eq:app-rt}
    \left(\frac{\partial}{c\,\partial t} + \vct{\hat{n}}\cdot\nabla\right)
    I_\nu(\vct{x}, t; \vct{\hat{n}}) =
    j_\nu(\vct{x}, t; \vct{\hat{n}}) -
    \alpha_\nu(\vct{x}, t; \vct{\hat{n}})
    I_\nu(\vct{x}, t; \vct{\hat{n}}),
\end{equation}
where $j_\nu$ is the total emission coefficient (emissivity) and $\alpha_\nu$ the total
absorption coefficient.
The equation~\eqref{eq:app-rt} describes the
change in specific intensity at frequency $\nu$ in the direction
$\vct{\hat{n}}$ at a point $\vct{x}$ and time $t$.
Often in astrophysical
cases, the one-dimensional time-independent case suffices, in which case
the equation~\eqref{eq:app-rt} reduces to a more commonly seen form
\begin{equation}\label{eq:app-rt1d}
    \derfrac{I_\nu(s)}{s} = j_\nu(s) - \alpha_\nu(s) I_\nu(s),
\end{equation}
where $s$ is the distance along a radiation front normal. The equations
\eqref{eq:app-rt} and \eqref{eq:app-rt1d} do not take into account
interference, quantum effects or, most importantly, polarization.

Polarization can be included by making the specific intensity $I_\nu$ vector
valued, introducing the Stokes intensities $\vct{I}_\nu = (I_\nu,
Q_\nu, U_\nu, V_\nu)$. Similarly, $j$ is replaced by the vector
\begin{equation}\label{eq:app-evec}
\evec_\nu=( j_{I,\nu}, j_{Q,\nu}, j_{U,\nu}, j_{V,\nu})
\end{equation}
of Stokes emissivities and $\alpha$ by the response (or
Müller) matrix
\begin{equation}\label{eq:app-mulmat}
\mulmat_\nu =
    \begin{pmatrix}
        \alpha_{I,\nu} & \alpha_{Q,\nu} & \alpha_{U,\nu} & \alpha_{V,\nu} \\
        \alpha_{Q,\nu} & \alpha_{I,\nu} & r_{V,\nu} & -r_{U,\nu} \\
        \alpha_{U,\nu} & -r_{V,\nu} & \alpha_{I,\nu} & r_{Q,\nu} \\
        \alpha_{V,\nu} & r_{U,\nu} & -r_{Q,\nu} & \alpha_{I,\nu} \\
    \end{pmatrix},
\end{equation}
where the $\alpha$-coefficients represent absorption effects, and the
$r$-coefficients relate to Faraday conversion and rotation.
It should be noted that different conventions for $\mulmat$ exist,
varying by the sign of $r_U$.
The one-dimensional polarized equation of
radiative transfer is then
\begin{equation}\label{eq:app-pol-rt1d}
    \derfrac{\svec_\nu(s)}{s} = \evec_\nu - \mulmat_\nu(s)\svec_\nu(s).
\end{equation}
The general relativistic generalization of equation~\eqref{eq:app-pol-rt1d} is
\citep[e.g.][]{gammie2012}
\begin{equation}\label{eq:app-pol-tens}
    \derfrac{N^{ab}}{\lambda} = J^{ab} + H^{abcd} N_{cd},
\end{equation}
where $\lambda$ is the affine parameter along the curve representing the
propagating radiation front,
$N^{ab}$ is the (complex-valued) \emph{polarization tensor}, $J^{ab}$ is the
\emph{emissivity tensor} and $H^{abcd}$ is the \emph{response tensor}.

Directly solving equation~\eqref{eq:app-pol-tens} requires integrating the 16
real independent components of $N^{ab}$.
This number can be reduced to four by parallel transporting a \emph{polarization frame}
along the geodesic. The frame consists of two orthogonal spacelike vectors also
orthogonal to the geodesic and the observer four-velocity.
Expressing all quantities in this frame using the screen projection operator
\eqref{eq:app-screen-proj}, the equation~\eqref{eq:app-pol-tens} can be written
as
\begin{equation}\label{eq:app-pol-scalar}
    \derfrac{\csvec_\nu}{\lambda} = \cevec_\nu - \cmulmat_\nu \csvec_\nu,
\end{equation}
where $\csvec_\nu = \nu^{-3}\svec_\nu$, $\cevec = C\nu^{-2} \evec_\nu$,
$\cmulmat = C\nu^{-1}\mulmat$ and
$C$ is a constant related to the
parametrization of the curve. An observer with a four-velocity $u^a$
at one end of the curve, where the tangent is $k^a$, has $C=u_a
k^a/\nu_0$ for an observed frequency $\nu_0$.

\section{Built-in manifold and chart support}\label{app:builtins}
\am{} contains a number of predefined metric spaces and space-times
together with commonly used coordinate charts for convenience.
The number of implemented spaces and charts is expected to grow, but the
selection at the time of writing is given in the following. We
list all the charts and the representations of the metric, either as a
line element or in matrix form,
in these charts for each implemented space.
All Lorentzian space-times are shown with the $(+---)$ metric convention.

\subsection{Riemannian manifolds}

\subsubsection{Two-sphere}
\paragraph{Spherical coordinates $(\theta,\phi)$}
\am{} uses two copies of the spherical chart $(\theta,\phi)$ to cover the entire
two-sphere. The metric, in either copy, is given by the usual
\begin{equation}\label{eq:app-two-sphere-metric}
\ud s^2 = \ud \theta^2 + \sin^2 \theta \ud\phi^2.
\end{equation}

\subsection{Semi-Riemannian manifolds}

\subsubsection{Minkowski space-time}
\paragraph{Cartesian coordinates $(t,x,y,z)$}
The Cartesian Minkowski coordinates are often in the literature denoted by $\eta_{ab}$, and we use
the same convention. The line element is diagonal, given by
\begin{equation}
    \ud s^2 = \ud t^2 - \ud x^2 - \ud y^2- \ud z^2.
\end{equation}
\paragraph{Spherical coordinates $(t,r,\theta,\phi)$}
\begin{equation}\label{eq:app-minkowski-sph}
\ud s^2 = \ud t^2 - \ud r^2 - r^2\left(\ud \theta^2 + \sin^2\theta\,\ud \phi^2\right)
\end{equation}

\subsubsection{Kerr space-time}\label{sc:app-kerr-space}
The Kerr space-time \citep{kerr1963} is an important solution to the Einstein
field equations, representing a rotating black hole exactly, and other rotating fluid bodies of
finite size asymptotically. The Kerr space-time is parametrized by the mass $M$ and the angular
momentum $J$. Usually $J$ is given through the \emph{normalized spin parameter} $a = J/M$, in which
case $a\in[0,M]$ or the \emph{dimensionless spin parameter} $\chi$, so that $\chi\in[0,1]$.
The solution reduces to the Schwarzschild space-time when $\chi = 0$ and further to the flat
Minkowski space-time when $M=0$. 

\paragraph{Boyer--Lindquist coordinates $(t,r,\theta,\phi)$}
Perhaps the clearest representation of the Kerr metric in the Boyer--Lindquist (BL) coordinates
is through the matrix form
\begin{equation}\label{eq:app-kerr-bl-metric}
    (g_{ab}) = \begin{pmatrix}
        1 - \frac{2Mr}{\rho^2} & 0 & 0 & \frac{2Mr a \sin^2\theta}{\rho^2} \\
        0 & -\frac{\rho^2}{\Delta} & 0 & 0 \\
        0 & 0 & -\rho^2 & 0 \\
        \frac{2Mr a \sin^2\theta}{\rho^2} & 0 & 0 & 
        -\sin^2\theta \left( r^2 + a^2 + \frac{2Mr a^2 \sin^2\theta}{\rho^2}\right),
    \end{pmatrix}
\end{equation}
where $\rho^2 = r^2 + a^2\cos^2\theta$ and $\Delta = r^2 - 2Mr + a^2$.
The BL form of the metric is singular when $\Delta = 0$ or $\rho^2 = 0$. The former
condition corresponds to $r_\pm = m^2 \pm \sqrt{m^2-a^2}$, which give the locations of the inner and outer
event horizons, where the curvature is \emph{not} singular.
The condition $\rho^2=0$ implies $r=0$ and $\cos\theta=0$, and corresponds to the curvature
singularity.

\paragraph{Cartesian Kerr--Schild coordinates $(t,x,y,z)$}
The Cartesian Kerr--Schild (KS) coordinates come in two flavors, \emph{ingoing} and \emph{outgoing},
adapted to null geodesics that move radially inwards or outwards, respectively. As such, they are a
generalization of the ingoing and outgoing Eddington--Finkelstein coordinates for a Schwarzschild
black hole.
Of the two variants, only
the ingoing form is typically seen in the literature. The metric in the ingoing KS coordinates is
most easily written as a sum
\begin{equation}\label{eq:app-KS-sum}
    g_{ab} = \eta_{ab} - F(x,y,z;a) l_a l_b,
\end{equation}
where $\eta$ is the Cartesian Minkowski metric, 
\begin{align}\label{eq:app-cart-KS-in}
    F(x,y,z;a) &= \frac{2 M r^3}{r^4 + a^2 z^2} \\
    (l_a) &= \left(1, \frac{rx + ay}{a^2+r^2}, \frac{ry - ax}{a^2+r^2}, \frac{z}{r}\right),
\end{align}
and $r$ is defined implicitly through $x^2+y^2+z^2 = r^2 + a^2(1-z^2/r^2)$.
The vector $l_a$ is null with respect to both $\eta_{ab}$ and $g_{ab}$.
The metric in the outgoing coordinates is similarly given by a sum as in
equation~\eqref{eq:app-KS-sum}, but with
\begin{align}\label{eq:app-cart-KS-out}
    (l_a) &= \left(-1, \frac{rx - ay}{a^2+r^2}, \frac{ry + ax}{a^2+r^2}, \frac{z}{r}\right).
\end{align}
When numerically computing geodesics near a Kerr black hole, it is crucial that the right form of
the KS coordinates is chosen. If the geodesic approaches the hole, the ingoing chart should be used, and
for a geodesic going away from the hole, the outgoing chart should be used. Failure to do so has
significant computational penalties, as can be seen in Section~\ref{sc:kerr-tests}.
Note that this implies that for a geodesic coming in towards a Kerr black hole, passing by it,
and then leaving, both ingoing and outgoing charts should be used with a
change of coordinates near the closest approach.

\subsection{AlGendy--Morsink (AGM) space-time}\label{app:agm}

\paragraph{AGM coordinates $(t, \barr, \theta, \phi)$}
The AGM space-time and coordinates \citep{algendy2014} actually refer to a
specific choice of a Butterworth--Ipser (BI) space-time, with the accompanying
coordinate chart \citep{butterworth1976}.
The BI space-time is a general representation of the space-time outside an
axisymmetric rotating fluid body. The AGM space-time is a special
case representing the space-time around a physically realistic oblate rotating neutron
star.
The AGM representation is accurate up to second order in the dimensionless
rotation parameter $\bar{\Omega} = \Omega R_e^{3/2} M^{-1/2}$, where $\Omega$ is
the rotational angular velocity of the star as seen by a distant observer, $R_e$
is the equatorial radius of the star and $M$ is the mass of the star. In the BI
coordinates, the generic axisymmetric metric reads
\begin{equation}
    (g_{ab}) =
    \begin{pmatrix}
        e^{2\nu} - B^2\barr^2\omega^2\sin^2\theta e^{-2\nu} & 0 & 0 & B^2 \barr^2 \omega \sin^2\theta e^{-2\nu} \\
        0 & -e^{2(\zeta-\nu)} & 0 & 0 \\
        0 & 0 & -e^{2(\zeta-\nu)}\barr^2 & 0 \\
        B^2 \barr^2 \omega \sin^2\theta e^{-2\nu} & 0 & 0 & -B^2 \barr^2 \sin^2\theta e^{-2\nu}
    \end{pmatrix},
\end{equation}
where $\nu$, $B$, $\omega$ and $\zeta$ are so-called metric functions or
potentials. The AGM metric is defined by using
\begin{align}
    \nu &= \ln \frac{1-M/(2\barr)}{1+M/(2\barr)} + \left(\frac{\beta}{3} - q P_2(\cos\theta)\right)\left(\frac{M}{\barr}\right)^3 \\
    B &= 1-\left(\frac{M}{2\barr}\right)^2 + \beta\left(\frac{M}{\barr}\right)^2 \\
    \omega &= \frac{2j}{\barr}\left(\frac{M}{\barr}\right)^2\left(
        1 - \frac{3M}{\barr}\right) \\
    \zeta &= \ln \left[1-\left(\frac{M}{2\barr}\right)^2\right]
    + \beta\frac{4 P_2(\cos\theta)-1}{3},
\end{align}
where $P_2$ is the second order Legendre polynomial.
The constants $q$ and $\beta$ are the dimensionless moments of energy density
and pressure, respectively, and $j = J/M^2$ is the dimensionless angular
momentum. \citet{algendy2014} found these constants to be well described across
various neutron star parameters and equations of state by the approximate
relations
\begin{align}
    j &= \left(1.136 - 2.53 x + 5.6 x^2\right) \bar{\Omega} \\
    q &= -0.11 x^{-2} \bar{\Omega}^2  \\
    \beta &= 0.4454 x\bar{\Omega}^2,
\end{align}
where $x = M/R_e$ is called the \emph{compactness (parameter)}.
Along with the metric potentials, \citet{algendy2014} also derived an equation for the shape of the surface of a
rotating neutron star, used in Section~\ref{sc:neutron_stars},
\begin{equation}\label{eq:app-oblate-shape}
    R(\theta) = R_e\left[
        1 + (-0.788 + 1.030x)\bar{\Omega}^2 \cos^2\theta\right].
\end{equation}
From this, the flattening, also called oblateness, of the star is given by 
\begin{equation}\label{eq:app-flattening}
    f=\frac{R(\pi/2)-R(0)}{R(\pi/2)} = (0.788 - 1.030x)\bar{\Omega}^2.
\end{equation}
It should be noted that the quantities $R_e$ and $R(\theta)$ are not defined in terms
$\barr$, but instead in terms of a radial coordinate $r = Be^{-\nu}\barr$.

\subsection{Hartle--Thorne space-time}

The Hartle--Thorne (HT) space-time \citep{hartle1968} describes the space-time around a rotating
oblate star. The original derivation was based on a perturbation of the
non-rotating Schwarzschild space-time. In \am{}, we use instead a version based
on a perturbation of the rotating Kerr space-time, given in
\citet{glampedakis2006}.

\paragraph{Glampedakis--Babak (GB) coordinates $(t,r,\theta,\phi)$}
The GB coordinates are based on the Boyer--Lindquist coordinates of the Kerr
space-time. In these coordinates, in a parametrization used by
\citet{baubock2012}, the metric is given by
\begin{equation}
    g_{ab} = g_{ab}^\text{Kerr} + \eta \chi^2 h_{ab},
\end{equation}
where $g_{ab}^\text{Kerr}$ is the Kerr metric in the BL coordinates (see
eq.~\eqref{eq:app-kerr-bl-metric}), and
\begin{align}
    h^{00} &= -(1-2M/r)^{-1}\, [(1-3\cos^2\theta)\mathcal{F}_1(r)]  &
    h^{11} &= -(1-2M/r)\, [(1-3\cos^2\theta)\mathcal{F}_1(r)] \\
    h^{22} &= r^{-2}\, [(1-3\cos^2\theta)\mathcal{F}_2(r)] &
    h^{33} &= r^{-2}\sin^{-2}\theta\, [(1-3\cos^2\theta)\mathcal{F}_2(r)].
\end{align}
Here $M$ is the mass of the star, $\chi = J/M$ is the dimensionless angular momentum
and $\eta$ parametrizes the mass quadrupole moment $q = -\chi^2(1+\eta)$, so that
$\eta = 0$ corresponds to the quadrupole moment of the Kerr space-time. The
functions $\mathcal{F}_{1,2}$ are given in \citet{glampedakis2006}.

%% The reference list follows the main body and any appendices.
%% Use LaTeX's thebibliography environment to mark up your reference list.
%% Note \begin{thebibliography} is followed by an empty set of
%% curly braces.  If you forget this, LaTeX will generate the error
%% "Perhaps a missing \item?".
%%
%% thebibliography produces citations in the text using \bibitem-\cite
%% cross-referencing. Each reference is preceded by a
%% \bibitem command that defines in curly braces the KEY that corresponds
%% to the KEY in the \cite commands (see the first section above).
%% Make sure that you provide a unique KEY for every \bibitem or else the
%% paper will not LaTeX. The square brackets should contain
%% the citation text that LaTeX will insert in
%% place of the \cite commands.

%% We have used macros to produce journal name abbreviations.
%% \aastex provides a number of these for the more frequently-cited journals.
%% See the Author Guide for a list of them.

%% Note that the style of the \bibitem labels (in []) is slightly
%% different from previous examples.  The natbib system solves a host
%% of citation expression problems, but it is necessary to clearly
%% delimit the year from the author name used in the citation.
%% See the natbib documentation for more details and options.

% BIBLIOGRAPHY
%%%%%%%%%%%%%%

\bibliographystyle{aasjournal}
\bibliography{refs} % if your bibtex file is called example.bib

%% This command is needed to show the entire author+affilation list when
%% the collaboration and author truncation commands are used.  It has to
%% go at the end of the manuscript.
%\allauthors

%% Include this line if you are using the \added, \replaced, \deleted
%% commands to see a summary list of all changes at the end of the article.
%\listofchanges

\end{document}